\documentclass[useAMS,usenatbib,onecolumn]{mn2e}

\usepackage[usenames]{color}

\newcommand{\vecr}{\mbox{\boldmath $r$}}
\newcommand{\vecv}{\mbox{\boldmath $v$}}
\newcommand{\vecf}{\mbox{\boldmath $F$}}
\newcommand{\vecg}{\mbox{\boldmath $g$}}
\newcommand{\vecb}{\mbox{\boldmath $B$}}
\newcommand{\vece}{\mbox{\boldmath $E$}}
\newcommand{\vecj}{\mbox{\boldmath $j$}}

\usepackage[dvips]{graphicx}
\title[Chemical separation of primordial L\lowercase{i}$^+$]{Chemical separation of primordial L\lowercase{i}$^+$ during structure formation caused by nanogauss magnetic field}
\author[Motohiko Kusakabe and Masahiro Kawasaki]{Motohiko Kusakabe$^{1}$\thanks{E-mail: motohiko@kau.ac.kr}\thanks{Present address: School of Liberal Arts and Science, Korea Aerospace University, Goyang 412-791, Korea, and Department of Physics, Soongsil University, Seoul 156-743, Korea} and
Masahiro Kawasaki$^{1,2}$ \\
$^{1}$Institute for Cosmic Ray Research, University of Tokyo, Kashiwa,
Chiba 277-8582, Japan\\
$^{2}$Kavli Institute for the Physics and Mathematics of the Universe (Kavli IPMU), TODIAS, \\
the University of Tokyo, 5-1-5 Kashiwanoha, Kashiwa, 277-8583, Japan}
\begin{document}

\date{Accepted xxx. Received xxx; in original form xxx}

\pagerange{\pageref{firstpage}--\pageref{lastpage}} \pubyear{20XX}

\maketitle

\label{firstpage}

\begin{abstract}
During the structure formation, charged and neutral chemical species may have separated from each other at the gravitational contraction in primordial magnetic field (PMF).   A gradient in the PMF in a direction perpendicular to the field direction leads to the Lorentz force on the charged species.  Resultantly, an ambipolar diffusion occurs, and charged species can move differently from neutral species, which collapses gravitationally during the structure formation.   We assume a gravitational contraction of neutral matter in a spherically symmetric structure, and calculate fluid motions of charged and neutral species.  It is shown that the charged fluid, i.e., proton, electron and $^7$Li$^+$, can significantly decouple from the neutral fluid depending on the field amplitude.  The charged species can, therefore, escape from the gravitational collapse.  We take the structure mass, the epoch of the gravitational collapse, and the comoving Lorenz force as parameters.  We then identify a parameter region for an effective chemical separation.  This type of chemical separation can reduce the abundance ratio of Li/H in early structures because of inefficient contraction of $^7$Li$^+$ ion.  Therefore, it may explain Li abundances of Galactic metal-poor stars which are smaller than the prediction in standard big bang nucleosynthesis model.  Amplitudes of the PMFs are controlled by a magneto-hydrodynamic turbulence.  The upper limit on the field amplitude derived from the turbulence effect is close to the value required for the chemical separation.
\end{abstract}

\begin{keywords}
atomic processes -- hydrodynamics -- magnetic fields -- plasmas -- Galaxy: abundances -- early Universe.
\end{keywords}

\section{Introduction}\label{sec1}

In the standard cosmology, abundances of light elements, i.e., hydrogen, helium, lithium, and very small amounts of other nuclides, evolve during big bang nucleosynthesis (BBN) at the redshift of $z\sim 10^9$ \citep{Fields:2011zzb}.  Lithium abundance predicted in standard BBN (SBBN) model \citep*{Coc:2011az,Coc:2013eea}, however, disagrees with that determined by spectroscopic observations of metal-poor stars (MPSs) ~\citep{Melendez:2004ni,Asplund:2005yt}.   The observational number ratio of lithium and hydrogen is $^7$Li/H$=(1-2) \times 10^{-10}$~\citep*{Spite:1982dd,Ryan:2000zz,Melendez:2004ni,Asplund:2005yt,bon2007,Shi:2006zz,Aoki:2009ce,Hernandez:2009gn,Sbordone:2010zi,Monaco:2010mm,Monaco:2011sd,Mucciarelli:2011ts}.  It is 2--4 times lower than the prediction in SBBN model with the baryon-to-photon ratio from the observation of the cosmic microwave background radiation by Wilkinson Microwave Anisotropy Probe (WMAP)~\citep{Spergel:2003cb,Spergel:2006hy,Larson:2010gs,Hinshaw:2012aka}.  

The formations of atom and molecules proceed in the redshift range of $z\la 10^4$ \citep{Saslaw1967,Peebles1968,Lepp1984,Dalgarno1987,Galli:1998dh,Vonlanthen:2009ns}.  Since lithium has a low ionization potential, it remains ionized when the recombination of hydrogen occurs \citep{Dalgarno1987}.  The relic abundance of Li$^+$ is, therefore, high \citep{Galli:1998dh}.  A recent study \citep{Vonlanthen:2009ns} shows that abundances of Li and Li$^+$ are almost equal at $z=10$. 

Magnetic fields exist in various astronomical objects, such as Sun, Galaxy, galactic cluster (see \citep{Grasso:2000wj} for a review).  Magnetic field have possibly existed in the early universe.  The origin of the magnetic fields is, however, not determined yet.  Magnetic fields can be generated through electric currents induced by a velocity difference of electrons and ions \citep{Biermann1950,Browne1968ApL}.  Such an electric current is produced in a rotating gas system because of different viscous resistances of electrons and ions \citep{Browne1968ApL}.  This current creates poloidal magnetic field.  Similarly, the drift current can be produced from gravitation working on electrons and ions, and it can generate a magnetic field \citep{Browne1968ApL,Browne1982,Browne1985}.  It has been noted \citep{Harrison1969}, however, that these batteries \citep{Biermann1950,Browne1968ApL} can not generate a large magnetic field since the time-scale of field generation is much larger than the age of the universe \citep{Spitzer1948,Hoyle1960,Harrison1969}.

The primordial magnetic field (PMF) can be generated at a couple of epochs in the early universe, i.e., the inflation, electroweak and quark-hadron transitions, and reionization \citep[see][and references therein]{Grasso:2000wj,Widrow:2002ud,Widrow:2011hs}.  The PMF generation, however, most probably occurs around the cosmological recombination epoch \citep*{Harrison1970,Matarrese:2004kq,Takahashi:2005nd,ichiki2006,Ichiki:2007hu,Takahashi:2007ds,fen2011,Maeda:2011uq}.  In the evolution of primordial density perturbation, the magnetic field can be perturbatively generated at second order through the vorticity \citep{Matarrese:2004kq} and the anisotropic stress of photon \citep{Takahashi:2005nd,ichiki2006}.  These generation processes can be calculated rather precisely with use of the cosmological perturbation theory.  Recent calculation \citep{fen2011} shows that the comoving amplitude of generated field on cluster scales, i.e., 1 Mpc, is about $3\times 10^{-29}$ G at redshift $z=0$.

Effects of PMFs on Galaxy formation have been studied \citep{Rees1972,Wasserman1978,Coles1992}.  Effects on Galactic angular momentum and Galactic magnetic fields have been also investigated utilizing magneto-hydrodynamic (MHD) equations \citep{Wasserman1978,Coles1992}.  It was found that a magnetic field can trigger a large density fluctuation with an overdensity of $\delta=1$.  Such a large fluctuation is produced in a structure with a scale $L_B$ if the comoving field amplitude measured in the present intergalactic medium (IGM) is as large as $B_0(L_B)\sim 10^{-9}(L_B/1~{\rm Mpc})$ G \citep{Wasserman1978,Coles1992}.  It has been suggested that an inhomogeneous magnetic field causes a streaming velocity of baryon relative to dark matter, and resultantly an infall of baryon in potential wells of dark matter may be inhibited.  Cosmological structure formation is thus affected by the inhomogeneous field \citep{Coles1992}.  

In this paper, we study a chemical separation of  charged and neutral species triggered by a PMF during the structure formation.  Neutral chemical species collapse gravitationally during the structure formation.  Motions of charged species can, however, decouple from that of neutral species by PMF, and an ambipolar diffusion occurs.  If the PMF has a gradient in a direction perpendicular to the field direction in the early universe, an electric current of charged species necessarily exists in the direction perpendicular to both of the field lines and the gradient direction.  The Lorentz force working on the charged species then causes a velocity difference between charged and neutral species in the direction of the field gradient.  This velocity difference enables an ambipolar diffusion.  Therefore, it is possible that $^7$Li$^+$ ions did not collapsed, while neutral $^7$Li atoms gravitationally collapsed into structures.  We suggest that the ambipolar diffusion provides a possible explanation of the small Li abundance in MPSs.

The situation of the $^7$Li$^+$ depletion due to PMFs and structure collapse studied in this paper is analogous to that of the charged grain depletion in the star-forming magnetic molecular clouds (MCs).  The chemical separation by an ambipolar diffusion has been studied for the case of the gravitational collapse in dusty interstellar MCs \citep[e.g.][]{cio1994,cio1996}.  In the MCs, the abundance of charged dust grains which is a component of their plasma is reduced since the magnetic field retards the infall of the grains while the neutral particles collapse to form a protostellar core \citep{cio1994}.  The depletion of the grain abundance by the magnetic field is a very important phenomenon since information on the star formation mechanism can in principle be obtained from the ratio of observed abundances of grains in the core and the envelope of MC \citep{cio1996}.  

The organization of this paper is as follows.  In Sec.~\ref{sec2} we describe the model of chemical separation during a gravitational collapse of a structure.  In Sec.~\ref{sec3} we introduce physical quantities used in this study, and typical numerical values relevant to the structure formation.  In Sec.~\ref{sec4} we show results of calculations of the chemical separation caused by the magnetic field.  In Sec.~\ref{sec5} we comment on the magnetic field amplitude.  In Sec.~\ref{sec6} we comment on a possible generation of a magnetic field gradient during the gravitational collapse.  In Sec.~\ref{sec7} we identify a parameter region required for a successful chemical separation.  In Sec.~\ref{sec8} we briefly mention a later epoch of the structure formation and possible reactions neglected in this study.  We suggest that the chemical separation of the $^7$Li$^{+}$ ion can reduce the abundance ratio $^7$Li/H in the early structure.  Another theoretical constraint on the magnetic field amplitude is also described.  In Sec.~\ref{sec9} we summarize this study.  In Appendix \ref{app2} we show drift velocities of protons and electrons in a structure, equations for ions and electrons which should be satisfied in equilibrium states, and typical values of variables required for an efficient chemical separation.  In Appendix \ref{app3} we show supplemental results for the calculations of the chemical separation.  
In this paper, the Boltzmann's constant ($k_{\rm B}$) and the light speed ($c$) are normalized to be unity.

\section{Model}\label{sec2}

We focus on the leaving of ionic species behind forming structures at redshift $z=\mathcal{O}(10)$.  First, let us define the initial state of the model structure.  The structure has a uniform density and parallel magnetic field.  The field amplitude has a gradient in a direction perpendicular to the field lines.  This simple condition is assumed as one example case that ionized chemical species can have bulk velocities different from that of neutral hydrogen.  It is considered that the relevant magnetic field has been generated  by motions of charged species which exist outside the structure originally.  In order to precisely follow the evolution of the spatial field distribution in the structure, the evolution of electric circuit including both inside and outside of the structure should be considered \citep[][Chaps. III and V]{alf1981}.  In this calculation, however, we do not treat the outer region, and use a boundary condition.  Second, the structure is axisymmetric in a cylindrical coordinate system $(r, \phi, z)$ with an axis of symmetry taken to be $z$-axis.  Azimuthal components of all physical quantities, therefore, do not depend on the azimuthal angle $\phi$.  The outer boundary of the structure exists at $r=r_{\rm str}$.  

\begin{figure}
\begin{center}
\includegraphics[width=84mm]{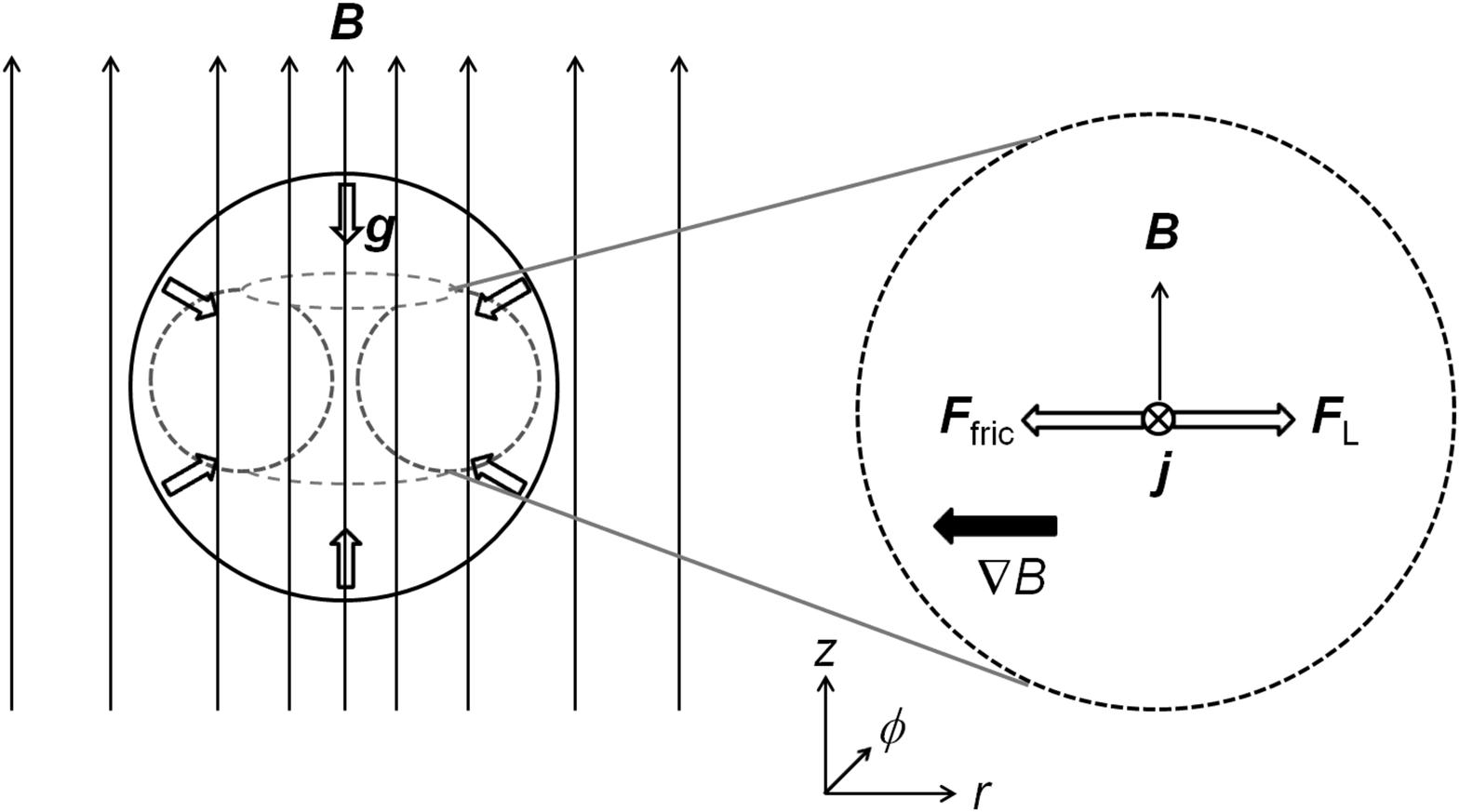}
\caption{Illustration of chemical separation in a collapsing structure.  In the left panel, the large solid circle delineates a collapsing structure, thin arrows are magnetic field lines, and open arrows are gravitational accelerations.  The structure is axisymmetric with respect to the field direction, and can be seen as a large coil indicated with a dashed lines.  The right panel shows an enlarged view of the cross section.  The three axes of the cylindrical coordinate are defined.  We assume a magnetic field along the $z$-axis (the thin arrow), and a gradient of the field amplitude in the $-r$ direction (the filled thick arrow).  There is a $\phi$ component of the $\nabla \times \vecb$ term or an azimuthal electric current (mark $\otimes$).  Charged fluid with this current in the magnetic field receives the Lorentz force ($\vecf_{\rm L}$) in the $r$ direction.  Resultantly, the charged fluid has a radial velocity relative to the neutral fluid.  The Lorentz force is then balanced with a friction force by neutral fluid ($\vecf_{\rm fric}$).\label{intro}}
\end{center}
\end{figure}

Figure \ref{intro} is an illustration of the physical concept of chemical separation.  In the left panel, the large solid circle is a boundary of collapsing structure, thin arrows are magnetic field lines, and open arrows are gravitational accelerations.  Since the axial symmetry is assumed, this structure itself can be roughly regarded as a large coil as indicated with a dashed lines.  The right panel shows an enlarged cross-sectional view of the structure.  The three axes of the cylindrical coordinate are defined in the panel.  A magnetic field exits along the $z$-axis (the thin arrow), and the field amplitude has a gradient in the $-r$ direction (the filled thick arrow).  Then, there is a $\phi$ component of the $\nabla \times \vecb$ term or an azimuthal electric current (mark $\otimes$).  The combination of this current and the magnetic field generates the Lorentz force ($\vecf_{\rm L}$) on the charged fluid in the $r$ direction.  As a result, the charged fluid has a radial velocity relative to the neutral fluid.  The Lorentz force is then balanced with a friction force by neutral fluid which depends on the radial relative velocity ($\vecf_{\rm fric}$).

\subsection{Fluid and electromagnetic equations}\label{sec2_1}

The following equations are adopted.
\begin{enumerate}
\item equation of continuity for neutral matter:
\begin{equation}
 \frac{\partial \rho_{\rm n}}{\partial t} + \nabla\cdot (\rho_{\rm n} \vecv_{\rm n})=0,
\label{eq1}
\end{equation}
where $\rho_{\rm n}$ and $\vecv_{\rm n}$ are the density and the fluid velocity of the neutral matter, respectively, and $t$ is the cosmic time.  The neutral matter is mainly composed of neutral hydrogens.

\item equation of continuity for ionized species $i$:
\begin{equation}
 \frac{\partial \rho_i}{\partial t} + \nabla\cdot (\rho_i \vecv_i)=0,
\label{eq5}
\end{equation}
where
$\rho_i$ and $\vecv_i$ are the density and the velocity of charged species $i$.
The finite differential expression in the cylindrical coordinate system is
\begin{equation}
 \frac{\Delta \rho_i}{\Delta t} =- \frac{1}{r}\frac{\Delta \left(r \rho_i v_{ir}\right)}{\Delta r} - \frac{\Delta \left(\rho_i v_{iz}\right)}{\Delta z}.
\label{eq6}
\end{equation}

\item force equations of proton and electron:

Equations of motion are given by
\begin{eqnarray}
\frac{D\vecv_p}{Dt}&=&-\frac{1}{\rho_p}\nabla P_p +\frac{e}{m_p}\left(\vece + \vecv_p \times \vecb\right) + \frac{1}{\tau_{p{\rm n}}} \left(\vecv_{\rm n}-\vecv_p\right) +\frac{1}{\tau_{pe}}\left(\vecv_e - \vecv_p\right), \label{eqb1}\\
\frac{D\vecv_e}{Dt}&=&-\frac{1}{\rho_e}\nabla P_e -\frac{e}{m_e}\left(\vece + \vecv_e \times \vecb\right) + \frac{1}{\tau_{en}} \left(\vecv_{\rm n}-\vecv_e\right) -\frac{1}{\tau_{ep}}\left(\vecv_e - \vecv_p\right),\label{eqb2}
\end{eqnarray}
where
$D/(Dt)$ is the material time-derivative, 
$\vecv_j$, $P_j$, and $m_j$ are the velocity, pressure, and particle mass of species $j$, respectively,
$e$ is the electronic charge,
$\vece$ is the electric field,
$\vecb=(B_r,~B_\phi,~B_z)$ is the magnetic field in a cylindrical coordinate, and $\tau_{ab}^{-1}$ is the energy loss rate of $a$ through the scattering with $b$, or the slowing-down rate of relative velocity of $a$ and $b$ [cf. Eq. (\ref{eq69})].
In the force equations for charged species, a term of cosmological redshift is neglected since the time-scale relevant to the redshift is much larger than those for others.  We neglect terms of pressure gradient in this paper.  
In the steady state, the force equations reduce to the form of
\begin{eqnarray}
 \vece&=&-\vecv_p\times \vecb -\frac{\rho_p}{\tau_{p{\rm n}}} \frac{(\vecv_{\rm n}-\vecv_p)}{en_p} -\frac{\rho_p}{\tau_{pe}} \frac{(\vecv_e-\vecv_p)}{en_p},\label{eqb6}\\
 \vece&=&-\vecv_e\times \vecb +\frac{\rho_e}{\tau_{e{\rm n}}} \frac{(\vecv_{\rm n}-\vecv_e)}{en_e} -\frac{\rho_e}{\tau_{ep}} \frac{(\vecv_e-\vecv_p)}{en_e}.
\label{eqb7}
\end{eqnarray}

We neglected an effect of $\nabla B$ drift since it would be small.  The force of $\nabla B$ is given by
\begin{eqnarray}
F_{\nabla B}&=&\left|-\frac{m_j v_{j \perp}^2 \nabla B}{2B}\right|\nonumber\\
&=&1.55\times 10^{-47}~{\rm GeV}^2~\left(\frac{m_j}{\rm GeV}\right) \left(\frac{v_{j\perp}}{6.59 \times 10^4~{\rm cm~s}^{-1}}\right)^2 \left(\frac{\nabla B/B}{{\rm kpc}^{-1}}\right),
\label{eqb16}
\end{eqnarray}
where
$v_{j\perp}$ is the velocity of $j$ perpendicular to the $\vecb$ direction.
On the other hand, the friction force from neutral species on charged species is given by
\begin{eqnarray}
F_{\rm fric}&=&\left|\frac{m_j \left(\vecv_{\rm n}-\vecv_j \right)}{\tau_{j{\rm n}}}\right|\nonumber\\
&=&2.01
\times 10^{-41}~{\rm GeV}^2~\left(\frac{m_j}{\rm GeV}\right) 
\frac{m_{\rm H}}{m_{\rm H} +m_j}
\left(\frac{v_{{\rm n}r}-v_{jr}}{1.61~{\rm km~s}^{-1}}\right) \left(\frac{n_{\rm H}}{5.69\times 10^{-3}~{\rm cm}^{-3}}\right) \left[\frac{\left(\sigma v\right)_{j{\rm n}}}{10^{-9}~{\rm cm}^3~{\rm s}^{-1}}\right],
\label{eqb15}
\end{eqnarray}
where $(\sigma v)_{ab}$ is the product of the momentum transfer cross section $\sigma$ and the velocity $v$ at the reaction of $a$+$b$ [cf. Eq. (\ref{eq45})].  
Since the equation $F_{\rm fric}\gg F_{\nabla B}$ holds, the effect of the field gradient is much smaller than that of the friction. 

\item Faraday's law of induction:
\begin{equation}
 \frac{\partial \vecb}{\partial t}=-\nabla\times \vece.
\label{eq2}
\end{equation}
The following equation is derived with Eqs. (\ref{eqb6}) and (\ref{eq2})
\begin{equation}
 \frac{\partial \vecb}{\partial t}=\nabla\times (\vecv_p \times \vecb) -\frac{m_p}{e} \nabla \times \left[\frac{(\vecv_p-\vecv_{\rm n})}{\tau_{p{\rm n}}} 
+\frac{(\vecv_p-\vecv_e)}{\tau_{pe}}\right].
\label{eq3}
\end{equation}
Note that the azimuthal component of magnetic field is always much smaller than $B_z$ in the setup of this study.

\item electric current density:
\begin{equation}
\vecj = e n_p \vecv_p -e n_e \vecv_e.
\label{eq75}
\end{equation}
The Lorentz force term is balanced with the friction term from neutral matter [cf. Eqs. (\ref{eqb6}), (\ref{eqb7}) and (\ref{eq75})]:
\begin{eqnarray}
\vecj \times \vecb &=& \frac{\rho_p}{\tau_{p{\rm n}}} \left(\vecv_p - \vecv_{\rm n} \right) +\frac{\rho_e}{\tau_{e{\rm n}}} \left(\vecv_e - \vecv_{\rm n} \right),
\label{eq72}
\end{eqnarray}
In the steady state, the force balance in the radial direction leads to
\begin{equation}
j_\phi B_z = e n_p \left(\alpha_{p{\rm n}} + \alpha_{e{\rm n}} \right)(v_{pr}-v_{{\rm n}r}),
\label{eq64}
\end{equation}
where we used $\alpha_{ab}=m_a/(e \tau_{ab})$ and additionally assumed the conditions, $n_p=n_e$ and $j_r=j_z=0$.  The latter condition is derived from the fact that the radial and longitudinal fluid velocities of ions and electrons are essentially the same, $v_{pr}=v_{er}$ and $v_{pz}=v_{ez}$ because of the charge neutrality of the system.

\item Ampere's law:
\begin{equation}
\nabla \times \vecb=4\pi \vecj
\label{eq76}
\end{equation}

\item Gauss's law for magnetism:
\begin{equation}
\nabla \cdot \vecb=0.
\end{equation}
The divergence of the Maxwell-Faraday equation [Eq. (\ref{eq2})] is given by
\begin{equation}
\frac{\partial (\nabla \cdot \vecb)}{\partial t} = -\nabla \cdot (\nabla \times \vece) =0.
\end{equation}
When the Gauss's law is satisfied at the initial time, it remains satisfied because of this equation.

\item force equation of neutral matter:

When the pressure gradient term is neglected, the force equation for neutral particles is given \citep{cio1993} by
\begin{equation}
 \frac{\partial\left(\rho_{\rm n}\vecv_{\rm n} \right)}{\partial t}+\nabla\cdot \left(\rho_{\rm n} \vecv_{\rm n}{\vecv_{\rm n}}\right)=\rho_{\rm n} \vecg +\frac{\rho_{\rm n}}{\tau_{{\rm n}p}}\left( \vecv_p -\vecv_{\rm n} \right) +\frac{\rho_{\rm n}}{\tau_{{\rm n}e}} \left( \vecv_e -\vecv_{\rm n} \right),
\label{eq77}
\end{equation}
where $\vecg$ is the gravitational acceleration.
The second and third terms in the right-hand side (RHS) represent frictions from protons and electrons, respectively, working on the neutral particles.
Using the balance between the Lorentz force and friction force [Eq. (\ref{eq72})], the Ampere's equation [Eq. (\ref{eq76})], and the Newton's law of action and reaction [Eq. (\ref{eq52})], we can transform the friction terms to the Lorentz-force term \citep{cio1993} as
\begin{equation}
 \frac{\partial\left(\rho_{\rm n}\vecv_{\rm n} \right)}{\partial t}+\nabla\cdot \left(\rho_{\rm n} \vecv_{\rm n}{\vecv_{\rm n}}\right)=\rho_{\rm n} \vecg +\frac{\left(\nabla\times \vecb\right)\times \vecb}{4\pi}.
\label{eqa1}
\end{equation}

When the azimuthal component of magnetic field is negligibly small, i.e., $B_\phi \simeq 0$, the second term in RHS is described as
\begin{eqnarray}
 (\nabla\times \vecb)\times \vecb=\left(\frac{\partial B_r}{\partial z}-\frac{\partial B_z}{\partial r}\right) \left(
\begin{array}{c}
B_z \\ 0 \\ -B_r
\end{array}
\right).\label{eqa2}
\end{eqnarray}

Under the spherical symmetry, the gravitational acceleration is given by
\begin{eqnarray}
 |\vecg(\vecr_{\rm sph})|=g(r_{\rm str})&=&G\frac{M(r_{\rm str})}{r_{\rm str}^2}=G\left(\frac{4\pi}{3}\right)^{2/3}~\rho_{\rm m}^{2/3}~M_{\rm str}^{1/3}\nonumber\\
&=&3.93\times 10^{-11}~{\rm cm}~{\rm s}^{-2}~\left(\frac{\rho_{\rm m}}{7.64\times 10^{-26}~{\rm g~cm}^{-3}}\right)^{2/3}~\left(\frac{M_{\rm str}}{10^{6}~M_\odot}\right)^{1/3},
\label{eqa3}
\end{eqnarray}
where
$\vecr_{\rm sph}$ is the position vector from the centre in a spherical coordinate,
$r_{\rm str}$ is the radius at the boundary of the structure,
$G$ is the gravitational constant,
$M(r_{\rm sph})$ is the mass contained inside the radius $r_{\rm sph}$,
$\rho_{\rm m}$ is the matter density, and
$M_{\rm str}$ is the mass of the structure.
In the equation the energy density is normalized to the value at the turnround ($z_{\rm tur}=16.5$; see Sec. \ref{sec3}).

On the other hand, the amplitude of the first term in RHS of Eq. (\ref{eqa1}) for gravitation is estimated as
\begin{eqnarray}
\rho_{\rm n}g&\simeq&\rho_{\rm b} g\nonumber\\
&=&
4.72\times 10^{-89}
~{\rm GeV}^5~\left(\frac{\rho_{\rm b}}{1.27\times 10^{-26}~{\rm g~cm}^{-3}}\right)~\left(\frac{\rho_{\rm m}}{7.64\times 10^{-26}~{\rm g~cm}^{-3}}\right)^{2/3}~\left(\frac{M_{\rm str}}{10^{6}~M_\odot}\right)^{1/3},
\label{eqa4}
\end{eqnarray}
where $\rho_{\rm b}$ is the baryon density.

The amplitude of the second term in RHS of Eq. (\ref{eqa1}) for the Lorentz force is estimated to be
\begin{eqnarray}
\left|\frac{\left(\nabla\times \vecb\right)\times \vecb}{4\pi}\right|&\sim&\frac{1}{4\pi}\frac{B^2}{L_B}\nonumber\\
&=&4.09
\times 10^{-89}~{\rm GeV}^5~\left(\frac{B}{10^{-7}~{\rm G}}\right)^2~\left(\frac{L_B}{597~{\rm pc}}\right)^{-1},
\label{eqa5}
\end{eqnarray}
where
$L_B$ is the length scale of coherent magnetic field.
\end{enumerate}

The ratio of the gravitational and Lorentz terms, Eqs. (\ref{eqa4}) and (\ref{eqa5}), respectively, is related to the mass-to-magnetic flux ratio for the gravitational collapse of a structure with a frozen-in magnetic field \citep{mou1976,cio1993}.  The critical mass-to-magnetic flux ratio has been determined from a numerical calculation \citep{mou1976} as
\begin{equation}
\left( \frac{M_{\rm b}}{\Phi_B} \right)_{\rm crit} =\frac{0.126}{G^{1/2}},
\label{m_phi_ratio}
\end{equation}
where
$M_{\rm b}$ is the total baryonic mass of the structure, and
$\Phi_B =\pi B r_{\rm str}^2$ is the total magnetic flux through the structure.
This ratio is invariant in comoving coordinates if the ambipolar diffusion is negligible.  Above the critical ratio a gravitational collapse can occur while below the ratio the collapse cannot.  The ratio of the gravitational and Lorentz terms can be rewritten in the form
\begin{equation}
\frac{\rho_{\rm b}g}{\left|\left(\nabla\times \vecb\right)\times \vecb/\left( 4\pi \right) \right|}= 3\pi^2 G \left( \frac{M}{M_{\rm b}}\right) \left( \frac{M_{\rm b}}{\Phi_B} \right)^2,
\label{grav_Lorentz_ratio}
\end{equation}
where
we supposed $L_B \sim r_{\rm str}$.
The factor $M/M_{\rm b}$ takes into account that not only the baryon but also the dark matter contributes to the gravitation of the system.  This factor is absent in the case of collapsing MC since effects of the dark matter mass is negligible.  The combination of the critical mass-to-magnetic flux ratio [Eq. (\ref{m_phi_ratio})] and the ratio $M/M_{\rm b}=6.03$ leads to the value of $\rho_{\rm b}g/ |(\nabla\times \vecb )\times \vecb/( 4\pi ) | =2.83$.   Therefore, roughly speaking, a gravitational collapse occurs if $M_{\rm b}/\Phi_B> (M_{\rm b}/\Phi_B)_{\rm crit}$ while a collapse does not occur if $M_{\rm b}/\Phi_B< (M_{\rm b}/\Phi_B)_{\rm crit}$.

\subsection{Galactic infall model}\label{sec2_2}
We assume that the second term in RHS of Eq. (\ref{eqa1}) is negligible, and that the initial density is exactly uniform inside a sphere.  This setup defines a toy model of collapsing structure.  Then, Eqs. (\ref{eq1}) and (\ref{eqa1}) are spherically symmetric, and Eq. (\ref{eqa1}) describes a free fall of spherical material.  A gas heating associated with virialization is neglected, and the gas temperature is assumed to evolve adiabatically after it decoupled from the temperature of the cosmic background radiation (CBR) at $z\sim 200$ \citep{pee1993}.  It is then given by $T=2.3$ K$[(1+z)(1+\delta)^{1/3}/10]^2$, where 
\begin{equation}
\delta \equiv (\rho_{\rm m}-\bar{\rho}_{\rm m})/\bar{\rho}_{\rm m}
\label{eq70}
\end{equation}
is the ratio of overdensity of matter relative to the cosmological average density $\bar{\rho}_{\rm m}$ \citep{Loeb:2003ya}.  We assume that the baryon density is proportional to the matter density.  This approximation is good as long as any radiative astrophysical objects such as first stars do not form yet.

The free fall of the sphere controlled by a self gravity is described by the Lagrangian equation of motion, i.e., $\partial^2 r_{\rm sph}/\partial t^2=-GM(r_{\rm sph})/r_{\rm sph}^2$ 
\citep{Hunter1962,pea1999}.
The radius and velocity are then related to the time $t$ as described \citep{pea1999} by
\begin{eqnarray}
r_{\rm sph}=A_{\rm G}(1-\cos\theta_{\rm G}),\label{eq7}\\
t=B_{\rm G}(\theta_{\rm G}-\sin\theta_{\rm G}),\label{eq8}\\
v_{\rm n}=\frac{\partial r_{\rm sph}}{\partial t}=\frac{A_{\rm G}}{B_{\rm G}} \frac{\sin\theta_{\rm G}}{1-\cos\theta_{\rm G}}\label{eq9},
\end{eqnarray}
where the condition $A_{\rm G}^3=GMB_{\rm G}^2$ is satisfied.  The parameters with the subscript G are used for the gravitational collapse and distinguished from parameters without the subscript.  The velocity evolution, $v_{\rm n}(t)=|\vecv_{\rm n}(t)|$, is given by this equation set.  The assumption of the initial uniform density corresponds to a constant $B_{\rm G}$ value for any $A_{\rm G}$.  Then, the velocity depends on the radius parameter $A_{\rm G}$ only.  In this case, a homologous evolution occurs, and the density is alway independent of the spatial coordinate.  Every mass shell satisfies 
\begin{eqnarray}
\rho_{\rm n}(t)=\rho_{\rm n,i}[r_{\rm sph,i}/r_{\rm sph}(t)]^3,
\label{eq78}
\end{eqnarray}
where 
$\rho_{\rm n}(t)$ and $\rho_{\rm n,i}$ are the densities at time $t$ and initial time $t_{\rm i}$, respectively, and $r_{\rm sph,i}$ is the radius at initial time.  In the present assumption, the ratio $r_{\rm sph,i}/r_{\rm sph}(t)$ is position-independent.  For a given time $t$, $\theta_{\rm G}$ and corresponding $r_{\rm sph}$ and $v_{\rm n}$ are derived.

\subsection{Velocities of charged species}\label{sec2_7}

For the system composed mainly of protons, electrons, and neutral matter, the total plasma force equation holds \citep{cio1993}:
\begin{equation}
\frac{\rho_{\rm n}}{\tau_{{\rm n}p}}\left( \vecv_p -\vecv_{\rm n} \right) +\frac{\rho_{\rm n}}{\tau_{{\rm n}e}} \left( \vecv_e -\vecv_{\rm n} \right) = \frac{\left(\nabla\times \vecb\right)\times \vecb}{4\pi}.
\label{eq73}
\end{equation}
Thus, a velocity difference of charged and neutral species is related to the Lorentz force operating on whole charged species that is mostly composed of protons and electrons.  
In general, matters in astrophysical objects are nearly complete charge-neutral.  We, therefore, assume that  fluid velocities of protons and electrons are equal as for $r$- and $z$-components.  Charged species are then considered as one component as long as motions in $r$- and $z$-directions are concerned.  Since the proton density is larger than the electron density by a factor of $m_p/m_e=1836$, the friction on proton is the predominant in the total fluid.  The following relation is derived from a balance of the friction and the Lorentz force:
\begin{equation}
\vecv_{p,(rz)}=\vecv_{{\rm n},(rz)} + \frac{\tau_{{\rm n}p}}{\left(1+\tau_{{\rm n}p}/\tau_{{\rm n}e} \right)\rho_{\rm n}} \frac{\left[\left(\nabla \times \vecb \right) \times \vecb\right]_{,(rz)}}{4\pi},
\label{eqb17}
\end{equation}
where
vector components in the $r$-$z$ plane are represented by subscript $(rz)$.
The factor $(1+\tau_{{\rm n}p}/\tau_{{\rm n}e})$ in the denominator is neglected because of $\tau_{{\rm n}p}/\tau_{{\rm n}e} \ll 1$.

The $\phi$-component of Eq. (\ref{eq73}), on the other hand, does not give an equation with $v_{p\phi}$ when the balance relation between $v_{p\phi}$ and $v_{e\phi}$ [Eq. (\ref{eqb11})] is satisfied.  In this case, terms of $v_{p\phi}$ and $v_{e\phi}$ cancel with each other [cf. Eqs. (\ref{eq69}) and (\ref{eq52})].  The equation then reduces to
\begin{equation}
\rho_{\rm n}\left(\frac{1}{\tau_{{\rm n}p}} + \frac{1}{\tau_{{\rm n}e}} \right) v_{{\rm n}\phi} + \frac{\left[\left(\nabla\times \vecb\right)\times \vecb\right]_\phi}{4\pi} =0.
\label{eq74}
\end{equation}

The proton velocity $\vecv_p$ is derived from the rotation of magnetic field, $\nabla \times \vecb$, using the Ampere's equation [Eq. (\ref{eq76})], as in studies of MCs \citep*{cio1993,cio1994,bas1994,mou2011}.  The rotation of the magnetic field is related to the electric current density.  Both physical quantities have existed from the start time of calculation (see Sec. \ref{sec5}).  Using Eqs. (\ref{eq75}) for the current density, and (\ref{eqb11}) for the velocities of protons and electrons, the $\phi$-component of the Ampere's equation gives the azimuthal proton velocity as
\begin{equation}
v_{p\phi}=\frac{\partial_z B_r -\partial_r B_z}{4\pi e n_p (1+\alpha_{p{\rm n}}/\alpha_{e{\rm n}})}.
\label{eq68}
\end{equation}
The assumption of $v_{pr}=v_{er}$ and $v_{pz}=v_{ez}$ correspond to no current density in the $r$- and $z$-directions.  Then, the Ampere's equation does not give constraints on velocities of charged species in the $r$-$z$ plane.

\subsection{Atomic mass and cross section data}\label{sec2_3}

Table~\ref{tab1} shows adopted masses of atoms and ions (H, H$^+$, Li, and Li$^+$) which are derived with atomic and electronic mass data \citep*{aud2003,wap2003}, and ionization energies of $j$ or binding energies of $j^+$ and $e^-$, BE($j^+$,$e$) \citep{nist}:  BE(H$^+$,$e$)=13.5984 eV \citep{joh1985} and BE(Li$^+$,$e$)=5.3917 eV \citep{lor1982}.

\begin{table}
 \caption{\label{tab1} Mass of chemical species.}
  \begin{tabular}{cc}
   \hline
   species & mass (GeV) \\ 
   \hline
   H      & 0.93878 \\
   H$^+$  & 0.93827 \\   
   Li     & 6.53536 \\
   Li$^+$ & 6.53485 \\
   \hline
  \end{tabular}
\end{table}

Reaction cross sections $\sigma_{i{\rm n}}$ are taken from \citet*{gla2005,sch2008} for $i=$H$^+$ and \citet{krs2009} for $^7$Li$^+$.  Linear interpolations are utilized with velocities taken as parameters.  Cross sections for energies lower than the minimum energy of data ($E_{\rm min}$) are given by the value at the energy $E=E_{\rm min}$, while those for energies larger than the maximum energy ($E_{\rm max}$) are given by the value at $E=E_{\rm max}$.

\subsection{Initial conditions}\label{sec2_6}

We take a typical comoving magnetic field value $B_{z0}$ as an input parameter.  The initial magnetic field is then assumed to be $B_{z z_{\rm i}}(r)=B_{z0}(1+z_{\rm i})^2 (1.5-r/r_{\rm str})$ for $0 \leq r \leq 1.4r_{\rm str}$ and $B_{z z_{\rm i}}(r)=0.1 B_{z0} (1+z_{\rm i})^2$ for $1.4r_{\rm str}\leq r$, where $B_{z z_{\rm i}}(r)$ is the $z$-component of magnetic field in IGM at the initial redshift $z_{\rm i}$ at radius $r$.
We note that in this calculation the ambipolar diffusion is caused by the magnetic pressure gradient [Eq. (\ref{eqb17})].  The pressure gradient does not depend on the amplitude of the magnetic field alone.  However, we fix the pattern of the gradient distribution, and take the comoving field value $B_{z0}$ as the only free parameter. 

As for initial velocities of protons and Li$^+$, the radial and $z$-components are assumed to be the same as those of hydrogens.  The $\phi$-component of proton velocity is given by Eq. (\ref{eq68}) with the initial $\vecb(\vecr)$ distribution.

Initial chemical abundances are taken from values at $z=10$ calculated in the model of homogeneous universe \citep{Vonlanthen:2009ns}:  H$^+$/H$=6.52\times 10^{-5}$, and Li$^+$/Li$=1.0$.  
We assume the Li nuclear abundance in SBBN model, Li/H$=5.2\times 10^{-10}$ \citep{Kawasaki:2012va}.  The chemical number fractions relative to hydrogen are then given by H$^+$/H$=6.52\times 10^{-5}$, Li/H$=2.6\times 10^{-10}$,
 and Li$^+$/H$=2.6\times 10^{-10}$.  A precise calculation of ionic motions should include chemical reactions coupled to the hydrodynamical calculation of the structure formation.  This is, however, beyond the scope of this paper.

\subsection{Boundary conditions}\label{sec2_4}

Boundary conditions are important to describe plasma motions since a plasma inside some region is affected by not only physical parameters inside the region but also those outside the region \citep[][Chaps. III and V]{alf1981}.  We adopt the following conditions.  Radial velocity components of any species $j$ are zero on the symmetrical axis:
\begin{eqnarray}
v_{jr}(r=0)=0.
\label{eq10}
\end{eqnarray}

The density and the recession velocity of neutral hydrogens, and number fractions of chemical species are initially given by the cosmic average values.  Outside the structure, the magnetic field is supposed to exist homogeneously in the $z$-direction.  In addition, the field amplitude evolves by redshift in the homogeneous universe.  We, however, just assume that physical variables such as ion velocities connect smoothly at the structure boundary, and do not treat the conjunction.  Calculations are performed for a contraction of material with a homogeneous overdensity of infinite size.  

As for a treatment for edges of computation domain, an origin and outer edge points are defined.  Because of the symmetry, constraints on velocities, $\vecv_j=0$ (for any $j$), always holds at the origin.  In every time step, values of $\rho_i$ (for ionic species $i$) and $\vecb$ at $r=0$ (on $z$-axis) and on the plane of $z=0$ are reset to be values calculated for the next innermost grid points, e.g. $\vecb(0, z, \phi) = \vecb(\Delta r, z, \phi)$ and $\vecb(r, 0, \phi)= \vecb(r, \Delta z, \phi)$, respectively.  Values of $\rho_i$ and $\vecb$ at the outer edge points are always given by the average value of collapsing matter.  Because of the axial symmetry, the radial and azimuthal components of the magnetic field are zero on $z$ axis.  At outer edge points of maximum $r$ and $z$ values, the densities of protons, electrons, and $^7$Li$^+$ are fixed to values derived for the homogeneous contraction.  In addition, at the edge points the magnetic field components are fixed as $B_r=0$
, $B_\phi=0$, and $B_{z z}(r_{\rm str})=0.1 B_{z0}(1+z)^2 (1+\delta)^{2/3}$.

\subsection{Calculation}\label{sec2_5}

The time step is determined so that changes in magnetic field and densities of ionized species in each step are much smaller than their amplitudes.  In the time integration of variables $A(a)$, the spatial differentiation is estimated with a finite difference method using the central difference.  The difference is derived from quantities evaluated at intermediate positions between grid points with intervals of $\Delta a$, i.e., $\partial A(a)/\partial a=[A(a+\Delta a/2)-A(a-\Delta a/2)]/\Delta a$.  The number of grid points is $260$ ($r$ direction) $\times 102$ ($z$ direction), and the spacing is $\Delta r=\Delta z=5.97$~pc. The computational region is, therefore, $0\le r \le 1.55$~kpc and $0\le z \le 0.603$~kpc.  The initial time is 9.29
 Myr, and the ending time is 474
Myr, respectively, after big bang.

In our calculation code, time evolutions of physical variables are calculated as follows.  For a time $t$, the velocity [Eq. (\ref{eq9})] and the density [Eq. (\ref{eq78})] of neutral matter, the overdensity of matter [Eq. (\ref{eq70})] and the temperature [Eq. (\ref{eq46})] are derived.  For respective reactions, the code evaluates thermal mean velocities [Eq. (\ref{eq71})] and relative fluid velocities.  Then, the friction time-scales [Eqs. (\ref{eq45}) and (\ref{eq56})] and the friction parameters [Eq. (\ref{eq69})] are derived using the law of action and reaction [Eq. (\ref{eq52})].  Besides, the electric field [Eqs. (\ref{eqb8}--\ref{eqb10})], the velocities of protons [Eqs. (\ref{eqb17}) and (\ref{eq68})], electrons [Eq. (\ref{eqb11})], and Li$^+$ [Eq. (\ref{eqb29})] are calculated.  Finally, the magnetic field [Eq. (\ref{eq3})] and the densities of charged species [Eq. (\ref{eq6})] are obtained by time integrations of their change rates.

\section{Physical quantities}\label{sec3}
\begin{enumerate}
\item cosmological parameters

The $\Lambda$CDM (dark energy $\Lambda$ and cold dark matter) model is adopted for the cosmic expansion history.  Parameter values are taken from analysis of WMAP9 CBR data ($\Lambda$CDM model \citep{Hinshaw:2012aka}) \footnote{WWW: http://lambda.gsfc.nasa.gov.}:  The Hubble parameter is $H_0=70.0\pm2.2$ km s$^{-1}$Mpc$^{-1}$, and energy density parameters of matter and baryon are $\Omega_{\rm m}=0.279\pm 0.025$ and $\Omega_{\rm b}=0.0463\pm0.0024$, respectively.  The energy density parameter is defined by $\Omega_k\equiv \rho_k/\rho_{\rm c}$, where $\rho_k$ is the density of species $k=$
m and b and $\rho_{\rm c}\equiv 3H_0^2/(8\pi G)$ is the critical density.  The present temperature of CBR is $T_{\gamma0}=2.7255$~K \citep{Fixsen:2009ug}.  The primordial abundances of hydrogen, helium, and lithium are taken from calculation of SBBN model \citep{Kawasaki:2012va} with the mean value of baryon density parameter $\Omega_{\rm b}$ described above, and the neutron lifetime $878.5 \pm 0.7_{\rm stat} \pm 0.3_{\rm sys}$~s~\citep{Serebrov:2010sg}:  mass fractions of hydrogen and helium are $X=0.753$ and $Y=0.247$, respectively, and the number ratio of lithium to hydrogen is Li/H=5.2$\times 10^{-10}$.

\item redshift ($z$) versus time ($t$) relation

\begin{eqnarray}
 a(t)&=&\frac{1}{1+z(t)} =\left(\frac{\Omega_{\rm m}}{1-\Omega_{\rm m}}\right)^{1/3} \left[\sinh \left(\frac{3\sqrt[]{\mathstrut 1-\Omega_{\rm m}}}{2}H_0 t\right)\right]^{2/3},\label{eq30}\\
t&=&\frac{2H_0^{-1}}{3\sqrt[]{\mathstrut {1-\Omega_{\rm m}}}}\sinh^{-1}\left[\left(\frac{1}{1+z}\right)^{3/2} \left(\frac{1-\Omega_{\rm m}}{\Omega_{\rm m}}\right)^{1/2}\right],
\label{eq31}
\end{eqnarray}
where
$a(t)$ is the scale factor of the universe.

\item baryon density

\begin{eqnarray}
 \rho_{\rm b}&=&\rho_{\rm c} \Omega_{\rm b}(1+z)^3 (1+\delta)\nonumber\\
&=&1.27\times 10^{-26}~{\rm g~cm}^{-3}\left(\frac{h}{0.700}\right)^2 \left(\frac{\Omega_{\rm b}}{0.0463}\right) \left(\frac{1+z}{17.5}\right)^3~\left(\frac{1+\delta}{5.55}\right),
\label{eq32}
\end{eqnarray}
where $h\equiv H_0$/(100 km s$^{-1}$ Mpc$^{-1}$) is the reduced Hubble constant, and $1+\delta=\rho_{\rm m}/\bar{\rho_{\rm m}}$ is the density normalized to the universal average value $\bar{\rho}_{\rm m}$.  It has been assumed that the baryon density is proportional to the matter density.

\item hydrogen number density

\begin{eqnarray}
 n_{\rm H}&\cong&n_{\rm b}X=\frac{\rho_{\rm b}}{m_{\rm b}}X\nonumber\\
          &=&5.69\times 10^{-3}~{\rm cm}^{-3} \left(\frac{h}{0.700}\right)^2 \left(\frac{\Omega_{\rm b}}{0.0463}\right) \left(\frac{1+z}{17.5}\right)^3 \left(\frac{X}{0.75}\right)~\left(\frac{1+\delta}{5.55}\right),
\label{eq33}
\end{eqnarray}
where
$n_{\rm b}$ is the total baryon density, and $m_{\rm b}=0.938$ GeV is the baryon mass.

\item matter density

\begin{eqnarray}
 \rho_{\rm m}&=&\rho_{\rm c} \Omega_{\rm m}(1+z)^3~\left(1+\delta\right)\nonumber\\
&=&7.64\times 10^{-26}~{\rm g~cm}^{-3}\left(\frac{h}{0.700}\right)^2 \left(\frac{\Omega_{\rm m}}{0.279}\right) \left(\frac{1+z}{17.5}\right)^3~\left(\frac{1+\delta}{5.55}\right).
\label{eq34}
\end{eqnarray}

\item spherical collapse model

The mass of the structure is $M_{\rm str}=10^{6} M_\odot$.  The collapse of the structure finishes at the redshift $z_{\rm col}=10$, or the cosmic time $t_{\rm col}=0.483$ Gyr.  The turnround then occurs at $z_{\rm tur}=16.5$, $t_{\rm tur}=t_{\rm col}/2=0.242$ Gyr.  The model structure is assumed to be a uniform density sphere with the radius at turnround of
\begin{equation}
 L=597~{\rm pc}~\left(\frac{M_{\rm str}}{10^{6}~M_\odot}\right)^{1/3} \left(\frac{h}{0.700}\right)^{-2/3} \left(\frac{\Omega_{\rm m}}{0.279}\right)^{-1/3} \left(\frac{1+z_{\rm tur}}{17.5}\right)^{-1},
\label{eq35}
\end{equation}
which derives from $M_{\rm str}=(4\pi L^3/3)\bar{\rho_{\rm m}}(1+\delta)$ with density contrast $1+\delta=9\pi^2/16$ at turnround.  The comoving length scale is $L_{0}=(1+z_{\rm tur})L=10.4$ kpc.

The parameter $A_{\rm G}$ specifies the distance from the structure centre.  When the $A_{\rm G}$ value is chosen as $2A_{\rm G}=L$ at the structure boundary at turnround, the $B_{\rm G}$ value is fixed to be
\begin{eqnarray}
 B_{\rm G}&=&\sqrt[]{\mathstrut \frac{A_{\rm G}^3}{GM_{\rm str}}}=\sqrt[]{\mathstrut \frac{1}{6\pi^3G\rho_{\rm m}}}\nonumber\\
&=&76.7~{\rm Myr}~\left(\frac{h}{0.700}\right)^{-1} \left(\frac{\Omega_{\rm m}}{0.279}\right)^{-1/2} \left(\frac{1+z_{\rm tur}}{17.5}\right)^{-3/2}. \label{eq36}
\end{eqnarray}

\item typical amplitude of magnetic field in the background universe

\begin{eqnarray}
B_z(z)&\sim &B_{z0}(1+z)^2\nonumber\\
&=&3.06\times 10^{-8}~{\rm G}~\left(\frac{B_{z0}}{10^{-10~}{\rm G}}\right)~\left(\frac{1+z}{17.5}\right)^2,~~~
\label{eq37}
\end{eqnarray}
where $B_{z0}$ is the $z$-component of the field value measured at present age, i.e., redshift $z=0$.

\item Larmor frequency of ion

\begin{eqnarray}
\Omega_i&=&\frac{Z_i eB}{m_i}\nonumber\\
&=&28.4 Z_i~{\rm yr}^{-1}~\left(\frac{B}{10^{-10}~{\rm G}}\right)~\left(\frac{m_i}{1~{\rm GeV}}\right)^{-1},
\label{eq38}
\end{eqnarray}
where
$Z_i$ is the charge number of ion $i$.

\item gyration radius of ion

\begin{eqnarray}
R_{i,{\rm g}}&=&\frac{m_i v_{i\perp}}{Z_ieB}=\frac{v_{i\perp}}{\Omega_i}\nonumber\\
&=&1.08Z_i^{-1}\times 10^{-8}~{\rm pc}~\left(\frac{v_{i\perp}}{3.00 \times 10^4~{\rm cm~s}^{-1}}\right)~\left(\frac{B}{10^{-10}~{\rm G}}\right)^{-1}~\left(\frac{m_i}{1~{\rm GeV}}\right),
\label{eq39}
\end{eqnarray}
where
$v_{i\perp}$ is the velocity of $i$ in the direction perpendicular to the magnetic field.

\item cosmic recession velocity

\begin{eqnarray}
v(r_{\rm sph},~z)&=&H(z)r_{\rm sph}\nonumber\\
&\sim&\left[H_0\Omega_{\rm m}^{1/2}\left(1+z\right)^{3/2}\right]r_{\rm sph}\nonumber\\
&=&1.61~{\rm km~s}^{-1}~\left(\frac{h}{0.700}\right) \left(\frac{\Omega_{\rm m}}{0.279}\right)^{1/2} \left(\frac{1+z}{17.5}\right)^{3/2} \left(\frac{r_{\rm sph}}{596~{\rm pc}}\right),
\label{eq40}
\end{eqnarray}
where 
$r_{\rm sph}(z)$ is the radius in a spherical coordinate at redshift $z$, and the matter dominated universe was assumed for the Hubble expansion rate at $z\ga 10$.

\item gas temperature

\begin{eqnarray}
T(z)=22~{\rm K}\left(\frac{1+z}{17.5}\right)^2~\left(\frac{1+\delta}{5.55}\right)^{2/3},
\label{eq46}
\end{eqnarray}
where
the amplitude is taken from the calculation in \citet{Loeb:2003ya}.

\item thermal average velocity of ion

\begin{eqnarray}
v_{i,{\rm th}}&=&\sqrt[]{\mathstrut \frac{8T}{\pi m_i}}\nonumber\\
&=&6.59 \times 10^4~{\rm cm~s}^{-1} \left(\frac{T}{22~{\rm K}}\right)^{1/2} \left(\frac{m_i}{1~{\rm GeV}}\right)^{-1/2}.
\label{eq47}
\end{eqnarray}

\item momentum transfer cross section of $p$+H at the relative velocity $v_{\rm rel}=1.61$ km s$^{-1}$

\begin{eqnarray}
\sigma_{p{\rm n}}=1.4\times 10^{-14}~{\rm cm}^2.
\label{eq41}
\end{eqnarray}

\item momentum transfer cross section of $^7$Li$^+$+H at $v_{\rm rel}=1.61$ km s$^{-1}$

\begin{eqnarray}
\sigma_{7{\rm n}}=1.3\times 10^{-14}~{\rm cm}^2.
\label{eq42}
\end{eqnarray}

\item elastic scattering cross section of $e$+H at $v_{\rm rel}=1.61$ km s$^{-1}$

We approximately take the elastic scattering cross section \citep{moi1962}:
\begin{eqnarray}
\sigma_{e{\rm n}}&\approx&\sigma_{{e{\rm n}},{\rm el}}=41\pi a_0^2\nonumber\\
&=&3.6\times 10^{-15}~{\rm cm}^2,
\label{eq43}
\end{eqnarray}
where
$a_0=5.29\times 10^{-9}$ cm is the Bohr radius.  This relative velocity $v_{\rm rel}=1.61$ km s$^{-1}$ corresponds to the centre of mass energy $7.37$
 $\mu$eV.  The recession velocity is smaller than the electron thermal velocity, $v_{e,{\rm th}}=29.1$
 km s$^{-1}~(T/22~{\rm K})^{1/2}$ [Eq. (\ref{eq47})].

\item Thomson scattering cross section

\begin{eqnarray}
\sigma_{e\gamma}&=&\frac{8\pi e^4}{3m_e^2}\nonumber\\
&=&6.65\times 10^{-25}~{\rm cm}^2.
\label{eq44}
\end{eqnarray}

Thomson scattering between electron and CBR is neglected since it does not occur so frequently, and its momentum transfer is negligible.  The momentum transfer rate of electrons, i.e., $n_\gamma \sigma_{e\gamma}$, multiplied by the fractional change in electron momentum at one scattering $\sim \mathcal{O}(T_\gamma/m_e)$, is much smaller than that of the $e$+H scattering.

\item momentum transfer rate of charged particles through the scattering with hydrogen
\begin{eqnarray}
\tau_{i{\rm n}}^{-1}&=& 
\frac{m_{\rm H}}{m_{\rm H} +m_i}
n_{\rm H}\left(\sigma v\right)_{i{\rm n}} \nonumber\\
&=&0.180~{\rm kyr}^{-1} 
\frac{m_{\rm H}}{m_{\rm H} +m_i}
\left(\frac{n_{\rm H}}{5.69 \times 10^{-3}~{\rm cm}^{-3}}\right) \left[\frac{\left(\sigma v\right)_{i{\rm n}}}{10^{-9}~{\rm cm}^3~{\rm s}^{-1}}\right],
\label{eq45}
\end{eqnarray}
where
$(\sigma v)_{ab}$ is the product of the cross section $\sigma$ and the velocity $v$ in the reaction of $a+b$.  In the equation, we have assumed that the reaction of $i$ with neutral matter is dominated by that of $i$+H, and neglected reactions with other neutral atoms.  
The factor $m_{\rm H}/(m_{\rm H} +m_i)$ is equal to the ratio of ionic momenta in the center of mass and laboratory systems.
The velocity is given by the larger of the hydrodynamic velocity difference, $|\vecv_a-\vecv_b|$, and the thermal mean velocity 
\begin{equation}
v_{ab,{\rm th}}= \sqrt[]{\mathstrut \frac{8T}{\pi\mu_{ab}}},
\label{eq71}
\end{equation}
where $\mu_{ab}$ is the reduced mass of the $a+b$ system.

\item friction parameter

A parameter representing the friction effect on a species $a$ from a species $b$ is defined as
\begin{equation}
\alpha_{ab}=\frac{m_a}{e\tau_{ab}}.
\label{eq69}
\end{equation}

\item energy loss rate via the Coulomb scattering

When the velocity of the incident electron measured in the rest frame of ion $i$, i.e., $w$, is much smaller than the root mean square velocity of the target ion particle, the slowing-down time (the inverse of the energy loss rate) of electrons via the scattering with ions is given \citep{spi2006} by
\begin{equation}
 \tau_{ei} = \frac{3}{4\sqrt[]{\mathstrut 2\pi}} \frac{m_e \mu_{ei} T^{3/2}}{e^4 m_i^{3/2} n_i~\ln \Lambda},
\label{eq56}
\end{equation}
where
$\mu_{ei}\sim m_e$ is the reduced mass of the $e+i$ system.
The quantity $\ln \Lambda$ is related to the cutoff scale of the scattering length, and is given by
\begin{eqnarray}
\ln \Lambda&\equiv& \ln \overline{h/p_0}=\ln \left[\frac{3}{2e^3}\left( \frac{T^3}{\pi n_e}\right)^{1/2}\right]\nonumber\\
&=& 21.5+\frac{3}{2}\ln\left(\frac{T}{22~{\rm K}}\right)-\frac{1}{2}\ln\left(\frac{n_{\rm H}}{5.69\times 10^{-3}~{\rm cm}^{-3}}\right)-\frac{1}{2}\ln\left(\frac{\chi_{{\rm H}^+}}{6.52\times 10^{-5}}\right),
\label{eq49}
\end{eqnarray}
where
$h$ is the Debye shielding distance, 
$p_0$ is the impact parameter at a scattering through which an electron is deflected by the angle of $\pi/2$, and $\chi_{{\rm H}^+}=n_{{\rm H}^+}/n_{\rm H}$ is the ionization degree of hydrogen.

The energy loss rate is then given by
\begin{equation}
\tau_{ei}^{-1} = 2.21\times 10^{-2}~{\rm s}^{-1}~\left(\frac{T}{22~{\rm K}}\right)^{-3/2}~\left(\frac{m_i}{m_p}\right)^{3/2}~\left(\frac{n_{\rm H}}{5.69 \times 10^{-3}~{\rm cm}^{-3}}\right)~\left(\frac{\chi_{{\rm H}^+}}{6.52\times 10^{-5}}\right)~\left(\frac{\ln \Lambda}{21.5}\right).
\label{eq50}
\end{equation}
The parameter $\alpha_{ei}$ is given by
\begin{equation}
 \alpha_{ei}\equiv \frac{m_e}{e \tau_{ei}}=\frac{4\sqrt[]{\mathstrut 2\pi}}{3} \frac{e^3 m_i^{3/2} Z_i^2 n_i~\ln \Lambda}{\mu_{ei} T^{3/2}}.
\label{eq51}
\end{equation}

We use the Newton's law of action and reaction \citep{cio1993}, i.e., 
\begin{equation}
 \frac{\rho_a}{\tau_{ab}}=\frac{\rho_b}{\tau_{ba}}.
\label{eq52}
\end{equation}
The following relation then holds in the case of $n_p=n_e$:
\begin{equation}
 \alpha_{pe}=\alpha_{ep}.
\label{eq53}
\end{equation}
The parameter $\alpha_{ep}=\alpha_{pe}$ is given [Eq. (\ref{eq51})] by
\begin{equation}
 \alpha_{ep}=1.26
\times 10^{-9}~{\rm G}~\left(\frac{T}{22~{\rm K}}\right)^{-3/2}~\left(\frac{n_{\rm H}}{5.69 \times 10^{-3}~{\rm cm}^{-3}}\right)~\left(\frac{\chi_{{\rm H}^+}}{6.52\times 10^{-5}}\right)~\left(\frac{\ln \Lambda}{21.5}\right).
\label{eq58}
\end{equation}

We apply Eq. (\ref{eq52}) to the $e+^7$Li$^+$ system, and derive
\begin{eqnarray}
\alpha_{7e}&=&\alpha_{e7}\frac{n_e}{n_7}=\frac{4\sqrt[]{\mathstrut 2\pi}}{3} \frac{e^3 m_7^{3/2} n_p~\ln \Lambda}{\mu_{e7} T^{3/2}}\nonumber\\
&\sim& \left(\frac{m_7}{m_p}\right)^{3/2}~\alpha_{ep}.
\label{eq54}
\end{eqnarray}

We also apply Eq. (\ref{eq52}) to the $p+^7$Li$^+$ system, and derive
\begin{eqnarray}
\alpha_{7p}&=&\alpha_{p7}\frac{n_p}{n_7}=\frac{4\sqrt[]{\mathstrut 2\pi}}{3} \frac{e^3 m_7^{3/2} n_p~\ln \Lambda}{\mu_{p7} T^{3/2}}\nonumber\\
&=&\alpha_{7e}\frac{\mu_{e7}}{\mu_{p7}} \sim  \frac{8m_e}{7m_p}\alpha_{7e} \ll \alpha_{7e}.
\label{eq55}
\end{eqnarray}
The friction from the $p+^7$Li$^+$ scattering is then neglected.

\item escape fraction of ion

The fraction of an ionic species escaping through the outer boundary of the structure during the structure formation is estimated as
\begin{eqnarray}
F_{i,{\rm esc}}(t)=\frac{\Delta M_i(t)}{M_i}=\frac{\int_{t_{\rm i}}^{t}\dot{M_i}(t')dt'}{M_i},
\label{eq59}
\end{eqnarray}
where
$M_i$ is the total mass of ion $i$ initially contained in the structure before the contraction,
$\Delta M_i(t)$ is the total mass of ion $i$ which escaped from the structure by time $t$, and
$t_{\rm i}$ is the initial time which should be larger than the time of the primordial nucleosynthesis $\sim$200~s.
We have assumed the spherical symmetry in the infall of neutral hydrogens, and the axial symmetry in the ion infall.  The mass loss rate is then given by
\begin{eqnarray}
\dot{M_i}(t)&=&2\pi r_{\rm str}^2(t)~\int_0^\pi~\sin\theta~d\theta~\rho_i(t,r_{\rm str}(t),\theta) ~v_{i,{\rm esc}}(t,r_{\rm str}(t),\theta),
\label{eq60}
\end{eqnarray}
where
$r_{\rm str}(t)$ is the structure radius in a spherical coordinate at time $t$,
$\theta=\tan^{-1}(r/z)$ is the angle between the position vector and the symmetrical $z$ axis, and
$\rho_i(t,r_{\rm str}(t),\theta)$ is the density at position ($r_{\rm str}(t),\theta$) at time $t$.
The variable $v_{i,{\rm esc}}(t,r_{\rm str}(t),\theta)$ is the escape velocity defined by
\begin{eqnarray}
 v_{i,{\rm esc}}(t,r_{\rm str}(t),\theta)&=&(\vecv_i-\vecv_{\rm n})\cdot \hat{r}\nonumber\\
&=&\sin\theta \left[v_{i{\rm r}}(t,r_{\rm str}(t),\theta)-v_{{\rm n}r}(t,r_{\rm str}(t),\theta)\right]
+\cos\theta\left[v_{iz}(t,r_{\rm str}(t),\theta)-v_{{\rm n}z}(t,r_{\rm str}(t),\theta)\right],
\label{eq61}
\end{eqnarray}
where
$\hat{r}$ is the unit vector with the direction of the position vector $\vecr$.

When we roughly assume that the density in the structure is homogeneous, the mass loss rate reduces to
\begin{eqnarray}
\dot{M_i}(t)&=&\frac{3M_i}{2r_{\rm str}(t)}~\int_0^\pi~\sin\theta~d\theta~v_{i,{\rm esc}}(t,r_{\rm str}(t),\theta).
\label{eq62}
\end{eqnarray}
The escape fraction of ion $i$ is then given by
\begin{eqnarray}
F_{i,{\rm esc}}(t)&=&\frac{3}{2}\int_{t_{\rm i}}^{t}~dt'~\frac{1}{r_{\rm str}(t')}\int_0^\pi~\sin\theta~d\theta~v_{i,{\rm esc}}(t',r_{\rm str}(t'),\theta)\nonumber\\
&=&2\int_{\ln t_{\rm i}}^{\ln t}~\frac{\langle v_{i,{\rm esc}}(t',r_{\rm str}(t')) \rangle_\mu}{H(t')r_{\rm str}(t')}~d\ln t',
\label{eq63}
\end{eqnarray}
where
\begin{equation}
\langle v_{i,{\rm esc}}(t',r_{\rm str}(t')) \rangle_\mu = \frac{1}{2} \int_{-1}^{1}~d\mu~v_{i,{\rm esc}}(t',r_{\rm str}(t'),\cos^{-1}\mu)
\end{equation}
is the average value of the escape velocity.
The recession velocity at the structure boundary, $r_{\rm sph}=r_{\rm str}(t)$, is $H(t)r_{\rm str}(t)$.   Then, in Eq. (\ref{eq63}) the time integration is dominated by the epoch when the escape velocity is a significant fraction of the recession velocity.  

We note that the escape of ions from cosmological structures is similar to that of charged dust grains from MCs.  Equations (\ref{eq60}) and (\ref{eq62}) for the time evolution of ionic mass in cosmological structures by the ambipolar diffusion during gravitational contraction is conceptually the same as equations (3a) and (3b) of \citet{cio1996} for that of the mass fraction of charged dust grains in MCs.

\end{enumerate}

\section{Result}\label{sec4}

We assume two cases of magnetic field amplitudes, 
$B_{z0}=3\times 10^{-10}$ G (Case 1) and $3\times 10^{-11}$ G (Case 2).
  The former value is so large that charged chemical species escape from a gravitational collapse of neutral atoms, while the latter is not.  The mass of the structure is $10^6$ $M_\odot$ in the both cases.  The electric current density $\vecj$ is determined from rotation of the magnetic field through the Ampere's equation [Eq. (\ref{eq76})].  The friction from inflowing neutral hydrogens determines the radial velocities of charged species through a balance between the friction and Lorentz forces.

The structure mass is chosen for the following reason.  The chemical separation of charged and neutral species proceeds when the gravitational collapse of structures enhances the matter density.  In the $\Lambda$CDM cosmological model, smaller structures form earlier.  Larger structures such as galaxies form through collisions and mergers of smaller structures.  Here we consider only structures such that they collapse at the redshift of $z=30-10$, and baryonic matter can form astrophysical high-density objects in the structures after their collapses.  Then, masses of such structures should be larger than $\sim 10^6-10^8~M_\odot$ \citep{Tegmark:1996yt}.  Significant fractions of baryonic matters in large structures which are observed today, therefore, have experience that they enhanced their densities at gravitational contractions of small structures with nearly the minimum masses.  We then assume a small structure with mass $10^6$ $M_\odot$ as a first structure.  Although the merger is a dominant cause of the formation of large structures, a part of baryonic matter is expected to have flown into the structures along filament structures (T.~Ishiyama, 2013; private communications).  It is, therefore, not to say that almost all material experienced the density enhancement at gravitational collapses of near-spherical structures.

In this section, we show results of time evolutions for average densities of chemical species, spatial distributions of the densities and azimuthal magnetic field.  Results of other physical variables are described in Appendix \ref{app3}.

\subsection{Average densities versus time}\label{sec4_1}

\begin{figure}
\begin{center}
\includegraphics[width=84mm]{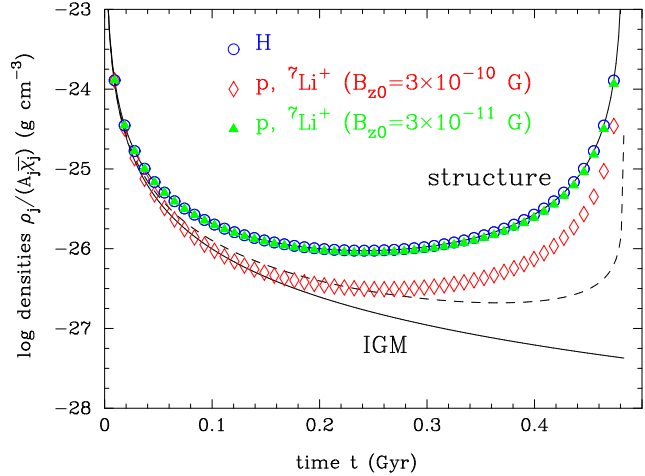}
\caption{Calculated average densities of hydrogen (open circles), proton and $^7$Li$^+$ for Case 1 (open diamonds) and Case 2 (filled triangles), respectively, in the structure as a function of cosmic time $t$.   The densities are normalized by the factor of nuclear mass number $A_j$ times the initial cosmic average value for the number ratio of $j$ and hydrogen $\bar{\chi_j}$.  Solid lines show analytical curves of hydrogen densities in the structure (upper line) and IGM (lower).  It was assumed that inside the structure, the density is determined by the gravitational free fall of matter, and that outside the density is given by the cosmic average density. The dashed line shows an analytical curve for charged species, such as proton and $^7$Li$^+$, based on the following assumption:  The species can collapse gravitationally along the axis of magnetic field, but expands across the field exactly following the cosmic average expansion. \label{pga1}}
\end{center}
\end{figure}

Figure \ref{pga1} shows densities of hydrogens (open circles), proton and $^7$Li$^+$ for Case 1(open diamonds) and Case 2 (filled triangles), respectively, averaged over the structure volume as a function of cosmic time $t$.  The densities are normalized as $\rho_j/(A_j \bar{\chi_j})$, where $A_j$ is the mass number of $j$ and $\bar{\chi_j}\equiv \overline{(n_j/n_{\rm H})}$ is the initial cosmic average value for the number ratio of $j$ to hydrogen.  Solid lines show analytical curves of hydrogen densities in the structure (upper line) and IGM (lower).  They are calculated based on the following assumption:  Outside the structure, the density is given by cosmic average density:
\begin{eqnarray}
\rho_{\rm H}^{\rm O}(t)&=&\bar{\rho_{\rm H}}(t) \nonumber \\
&=& \rho_{\rm H}^{\rm O}(t_{\rm i}) \left(\frac{1+z}{1+z_{\rm i}} \right)^3,
\label{eq80}
\end{eqnarray}
where
$z$ and $z_{\rm i}$ are redshifts corresponding to time $t$ and $t_{\rm i}$, respectively.
Inside the structure, on the other hand, the density is given by
\begin{eqnarray}
\rho_{\rm H}^{\rm I}(t)&=&\rho_{\rm b}(t_{\rm i})X \left[\frac{r_{\rm sph}(t_{\rm i})}{r_{\rm sph}(t)}\right]^3 = \frac{3}{4\pi} \frac{M_{\rm str}}{A_{\rm G}^3\left(1-\cos\theta_{\rm G}\right)^3} \left(\frac{\Omega_{\rm b}}{\Omega_{\rm m}}\right) X\nonumber\\
&=&7.57\times 10^{-26}\left(1-\cos\theta_{\rm G}\right)^{-3}~{\rm g~cm}^{-3}~\left(\frac{M_{\rm str}}{10^{6}M_\odot}\right) \left(\frac{A_{\rm G}}{298~{\rm pc}}\right)^{-3} \left(\frac{\Omega_{\rm b}}{0.0463}\right) \left(\frac{\Omega_{\rm m}}{0.279}\right)^{-1} \left(\frac{X}{0.75}\right).
\label{eq11}
\end{eqnarray}
When effects of magnetic field are small, curves of H$^+$ and $^7$Li$^+$ should be nearly the same as that of hydrogen.  The dashed line shows an analytical curve for charged species, such as proton and $^7$Li$^+$, based on the following assumption:  The species can collapse gravitationally along the axis of magnetic field ($z$-axis), and just expands across the field at the same velocity as the cosmic average expansion.  In this case, its density evolves as
\begin{equation}
\rho_i^{\rm I}(t)=\rho_i^{\rm I}(t_{\rm i}) \frac{r_{\rm sph}(t_{\rm i})}{r_{\rm sph}(t)} \left[\frac{1+z}{1+z_{\rm i}}\right]^2 =A_i \bar{\chi_i} \left[\rho_{\rm H}^{\rm I}(t)\right]^{1/3} \left[\rho_{\rm H}^{\rm O} (t)\right]^{2/3},
\label{eq12}
\end{equation}
where
it was assumed that hydrogen densities inside and outside the structure, i.e., $\rho_{\rm H}^{\rm I}(t)$ and $\rho_{\rm H}^{\rm O}(t)$, respectively, are almost equal at the initial time $t_{\rm i}(\ll t)$.  The $t_{\rm i}$ value has been taken to be enough small.

In Case 1, charged species in the structure are diluted at the intermediate phase with low densities.   This dilution can be measured as the ratio between the normalized densities of $p$ (and $^7$Li$^+$) and hydrogen.  The ratio reduces when the density becomes low around the turnround.  In the early and late phases of high densities, dilutions do not proceed effectively since the motions of charged and neutral species are strongly coupled in high density environments.  Eventually, the $^7$Li$^+$ ion is diluted in the structure by a factor of $\sim 4$ in the end of the calculation.  This dilution history is qualitatively applied to Case 2.  The dilution factor is, however, much smaller in Case 2.

\subsection{Chemical separation}\label{sec4_2}

\begin{figure*}
\begin{center}
\includegraphics[width=84mm]{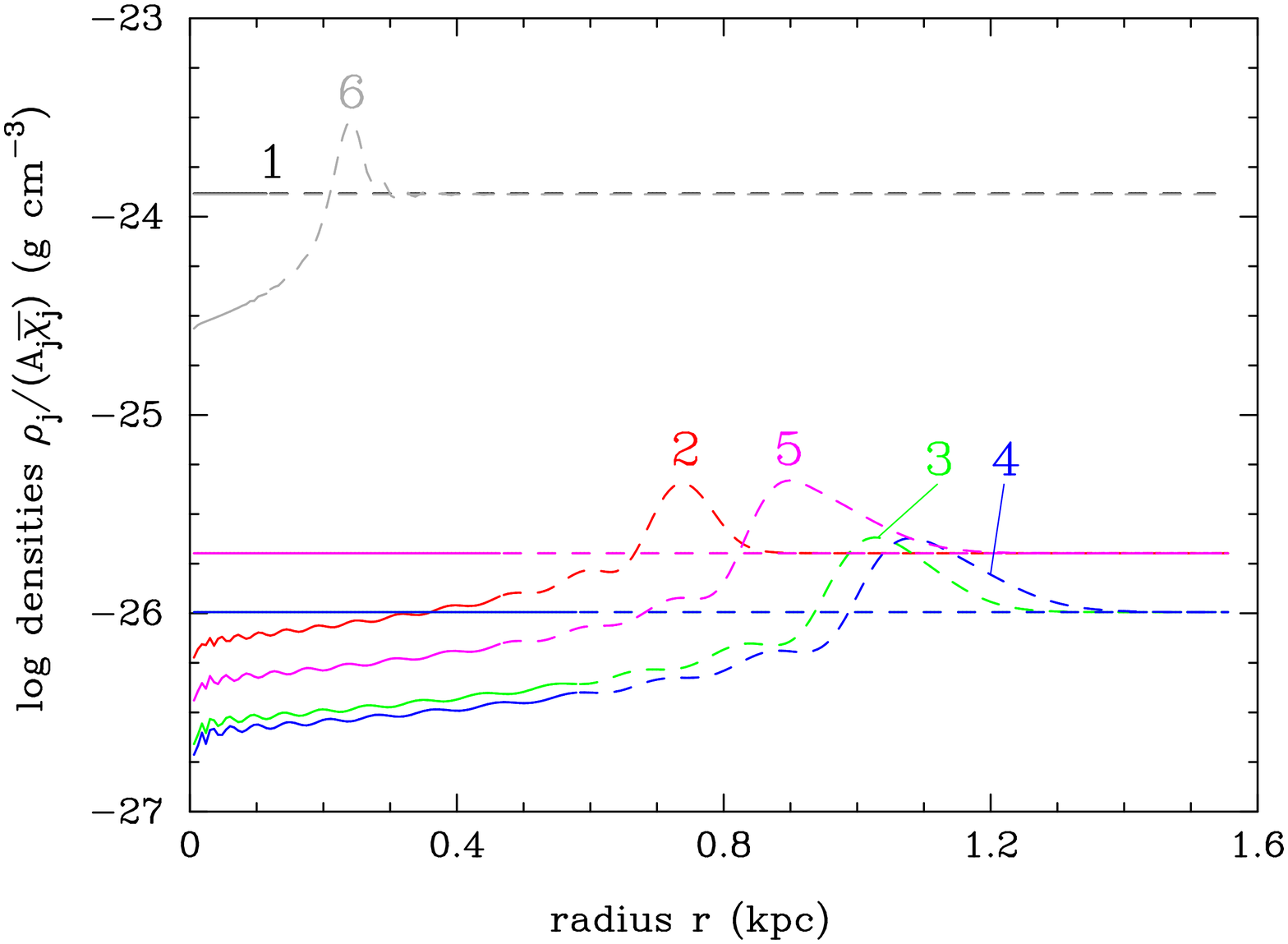}
\includegraphics[width=84mm]{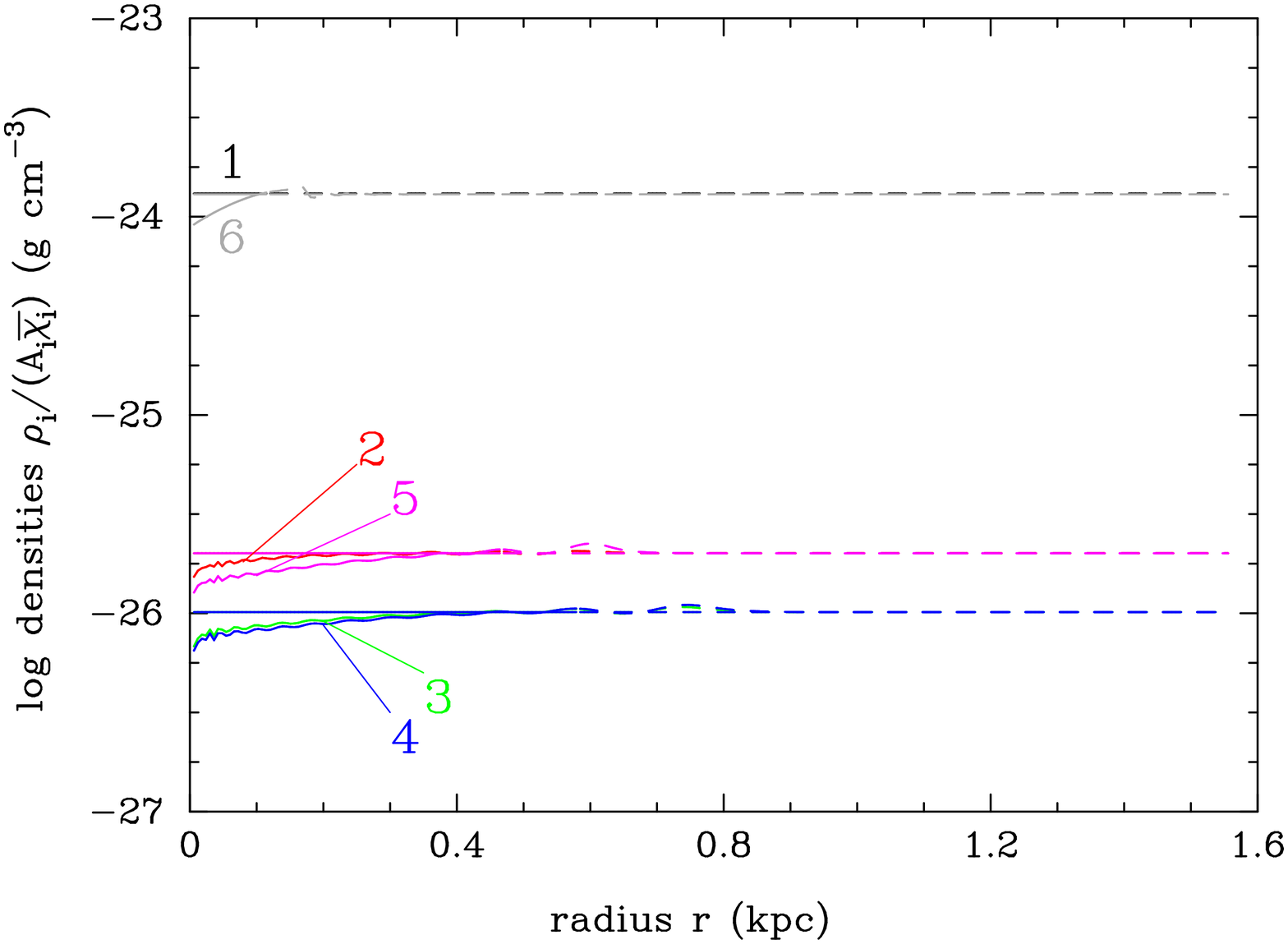}
\caption{Normalized densities of hydrogen (straight lines), proton and $^7$Li$^+$ (curves) as a function of radius from the structure centre for Case 1 (left panel) and Case 2 (right panel) at $t$=9.29 Myr (1), 102 Myr (2), 195 Myr (3), 288 Myr (4), 381 Myr (5), and 474 Myr (6).  Solid and dashed lines correspond to the regions inside and outside the structure, respectively. \label{pga2}}
\end{center}
\end{figure*}

Figure \ref{pga2} shows normalized densities of hydrogen (straight lines), proton and $^7$Li$^+$ (curves) as a function of radius from the structure centre for Case 1 (left panel) and Case 2 (right panel).  Density distributions are drawn for six different times:  $t$=9.29 Myr (denoted by number 1), 102 Myr (2), 195 Myr (3), 288 Myr (4), 381 Myr (5), and 474 Myr (6).  Times 1-3 are in an expanding phase, and times 4-6 are in a collapsing phase.  Note that structure sizes or densities of neutral hydrogen are the same at times 1 and 6, 2 and 5, and 3 and 4, respectively.  Solid and dashed lines correspond to the regions inside and outside of the structure, respectively.  The initial gradient of $B_z$ causes an expansion of the charged-species fluid.  Since the $B_z$ value is large at a small radius, the expansion is fast in the region of small $r$.  Accordingly the magnetic field amplitude and its gradient rapidly decrease in the inner region of small $r$.  Since the gradient of $B_z$ is not assumed in the outer region of large $r$, charged particles do not move in the outer region.  Then, high density shells forms at the boundaries between the inner and outer regions as seen in this figure as bumps.  The curves for the densities of charged species have oscillatory structures as well as the bumps caused by the assumed initial condition.  The charged species inside the structure are diluted more efficiently in Case 1 than in Case 2 because of the stronger magnetic field.

\subsection{Magnetic field}\label{sec4_6}

\begin{figure*}
\begin{center}
\includegraphics[width=84mm]{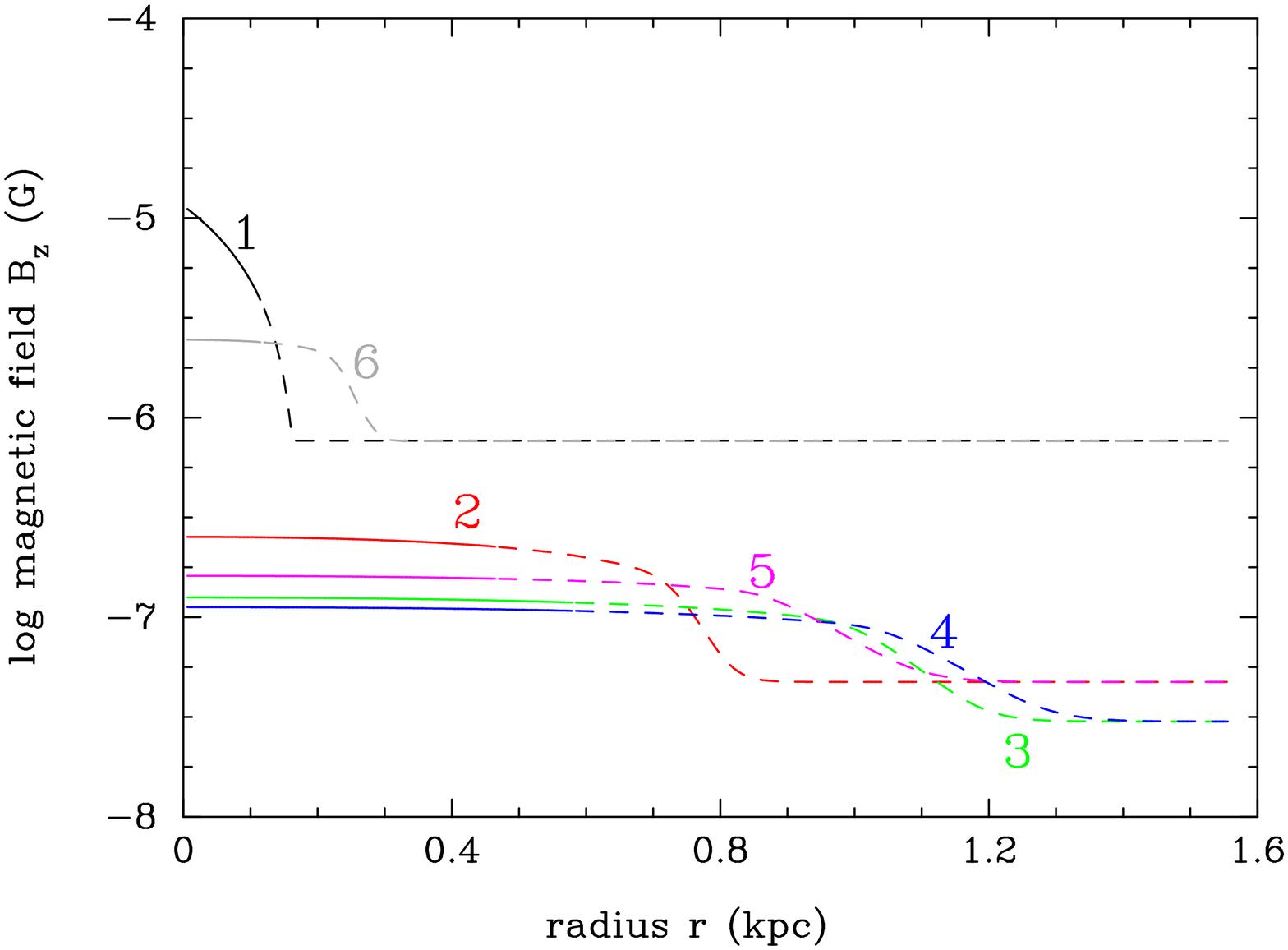}
\includegraphics[width=84mm]{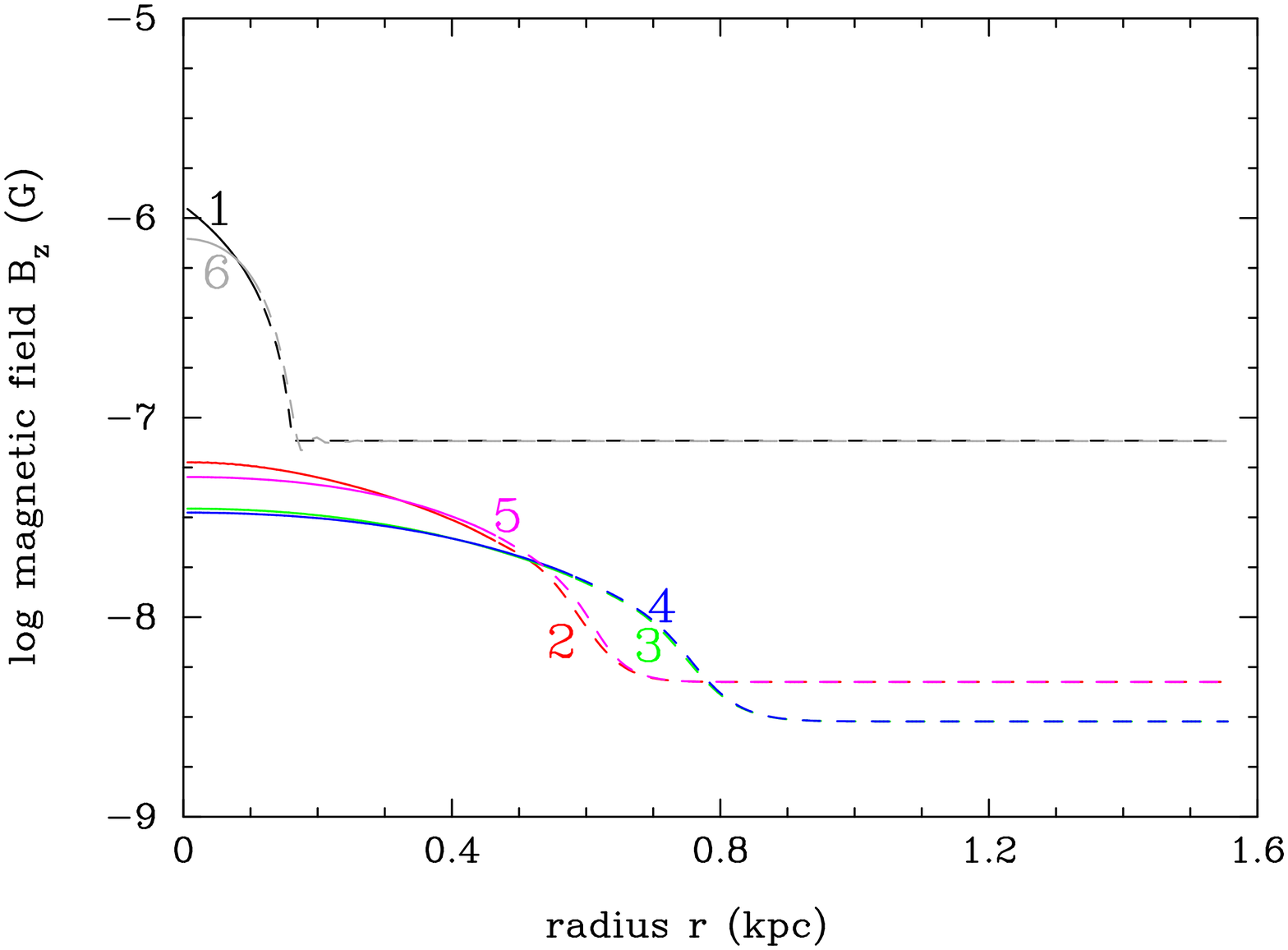}
\caption{Magnetic field ($z$-component) as a function of radius for Case 1 (left panel) and Case 2 (right panel) at $t$=9.29 Myr (1), 102 Myr (2), 195 Myr (3), 288 Myr (4), 381 Myr (5), and 474 Myr (6).  Solid and dashed lines correspond to values inside and outside the structure, respectively. \label{pga9}}
\end{center}
\end{figure*}

Figure \ref{pga9} shows the magnetic field ($z$-component) as a function of radius for Case 1 (left panel) and Case 2 (right panel).  Solid and dashed lines correspond to values inside and outside the structure, respectively.  Additionally to the effect of expansion and collapse of neutral hydrogens, weakening of magnetic field is observed in the small $r$ region.  This dilution is cased by outward movements of charged species (Fig. \ref{pga2}).  It is seen that outgoing charged species decrease the $B_z$ value in the small $r$ region more significantly in Case 1 than in Case 2.

In Case 1, the magnetic field at time 6 is almost uniform inside the collapsing structure.  The information for the initial condition of the field gradient (Sec. \ref{sec2_6}) is thus wiped out by the diffusion of the charged plasma and magnetic field through the contracting neutral matter.  It is expected that when the initial magnetic field gradient is large enough and that an ambipolar diffusion effectively operates as in Case 1, the result is possibly not so sensitive to the type or the details of the initial magnetic field gradient.  This is because if the field evolves by the ambipolar diffusion, the initial conditions will be forgotten after a certain amount of time.

\section{Magnetic field amplitude}\label{sec5}

In preceding sections, a magnetic field generation is neglected.  The generation, however, proceeds through a drift current creation in the structure, although its effect is insignificant as explained below.

The magnetic field in a structure evolves \citep{Browne1982} as
\begin{equation}
\left|\frac{d \vecb}{dt}\right|=\frac{1}{4\pi \sigma_e}\left|\nabla^2 \vecb \right|\approx \frac{B}{4\pi \sigma_e L_B^2},
\end{equation}
where
\begin{equation}
\sigma_e=\frac{n_e e^2 \tau_{ep}}{m_e}
\end{equation}
is the electron conductivity \citep{Grasso:2000wj}.  The magnetic field on a length scale $L_B$ diffuses during the early structure formation in the typical time-scale of
\begin{eqnarray}
\tau_{\rm diff}(L_B)&=&4\pi \sigma_e L_B^2 \nonumber\\
&=&6.39
\times 10^{18}~{\rm yr} \left(\frac{T}{22~{\rm K}}\right)^{3/2} \left(\frac{\ln \Lambda}{21.5}\right)^{-1} 
\left(\frac{L_B}{597~{\rm pc}}\right)^2.
\end{eqnarray}
Thus, large inductances of large astrophysical objects result in very long diffusion times.  The generation of magnetic field in the cosmological time-scale is, therefore, impossible.  The self-inductance of astrophysical objects with length scale $L_B$ is given by $L_{\rm ind}(L_B)\sim \mu_{\rm m} L_B$, where $\mu_{\rm m}$ is the magnetic permeability.  When some electromotive force is created in the structure, an electric current is produced at an approximately constant production rate.  The rate is inversely proportional to the inductance.  The magnetic energy $W$ stored in a coil, that is the structure itself in the present case, is proportional to the electric current squared, $W=L_{\rm ind}(L_B)I^2/2\sim \mu_{\rm m} L_B I^2$, where $I \sim j L_B^2$ is the electric current.  The magnetic energy per volume, $L_B^3$, is then proportional to the length scale squared:  $W/L_B^3\sim \mu_{\rm m} L_B (j L_B^2)^2/L_B^3 \sim \mu_{\rm m} j^2 L_B^2$.  The generation of magnetic field on a large scale of $L_B$, therefore, requires large amount of source energy density.  For the reason above, the magnetic field is never generated effectively by an electric current associated with dynamical friction.

The Ampere's equation [Eq. (\ref{eq76})] relates an electric current density to a magnetic field as
\begin{eqnarray}
j_\phi&\sim& \frac{B_z}{4\pi L_B} \nonumber \\
&=&2.89 \times 10^{-13}~{\rm cm}^{-2}~{\rm s}^{-1} \left(\frac{B_z}{10^{-10}~{\rm G}}\right) \left(\frac{L_B}{597~{\rm pc}}\right)^{-1}. \nonumber\\
\label{eq65}
\end{eqnarray}

If a magnetic field and an electric current density existed from the beginning of the structure formation, and the Lorentz force is enough large to realize a separation between charged and neutral species, charged species possibly do not collapse gravitationally.  The charged species, therefore, do not participate in structure formations.  Although charged species move differently from neutral species, scatterings between charged and neutral species efficiently transfer the kinetic energy of neutral species to charged species.  The velocity difference between H and proton has been assumed to be the typical cosmological recession velocity in the present case [Eq. (\ref{eq40})].  This velocity corresponds to the proton temperature of $T\sim m_p v_{\rm rel}^2/6=52.3$ 
K.  The scatterings then gradually increase the temperature of charged species as a function of time.  Resultantly, it is expected that the friction time-scale, $\tau_{ei}$ [cf. Eq. (\ref{eq50})], increases.

The equilibrium amplitude of the magnetic field is then related with the velocity difference [Eqs. (\ref{eq64}) and (\ref{eq65})]:
\begin{eqnarray}
B_z&\sim& \sqrt{\mathstrut 4\pi L_B j_\phi B_z}\sim \sqrt{\mathstrut 4\pi L_B e n_p (\alpha_{p{\rm n}}+\alpha_{e{\rm n}})(v_{pr}-v_{{\rm n}r})}\nonumber\\
&=&5.24 
\times 10^{-8}~{\rm G}~\left(\frac{n_{\rm H}}{5.69\times 10^{-3}~{\rm cm}^{-3}}\right) \left(\frac{\chi_{{\rm H}^+}}{6.52\times 10^{-5}}\right)^{1/2} \left(\frac{\Delta v_r}{1.61~{\rm km~s}^{-1}}\right)^{1/2} \left(\frac{L_B}{597~{\rm pc}}\right)^{1/2} 
\left[\frac{\left(\sigma v\right)_{p{\rm H}}}{2.3\times 10^{-9}~{\rm cm}^3~{\rm s}^{-1}}\right]^{1/2}, \nonumber \\
\label{eq16}
\end{eqnarray}
where
$\Delta v_r=v_{pr}-v_{{\rm n}r}$ is the velocity difference.  In the second line, we assumed typical physical values estimated at the gravitational turnround $z=z_{\rm tur}=16.5$, and the critical value of $\Delta v_r$ given by the cosmological recession velocity at the turnround [Eq. (\ref{eq40})].  The corresponding comoving magnetic field is $B_{z0}=B_z(z_{\rm tur})/(1+z_{\rm tur})^2= 0.171$ 
nG.  We note that the minimum amplitude of magnetic field which can support the charged species against the dynamical friction is larger when a larger structure is considered (Sec. \ref{sec5_1}).  

If the time-scale of field generation were shorter than the dynamical time of the system, an azimuthal electric current density is gradually induced by $\vecf \times \vecb$ drifts.  A poloidal magnetic field is then generated.  Magnetic fields in astronomical objects can be related to electric currents existing in their interiors.  In general, the fields are generated by electric currents which themselves are formed by motions of charged species, $\vecv_i$, in regions with finite amplitudes of magnetic fields.  This process for an amplification of magnetic field is called self-exciting dynamo, and is thought to operate in the Sun, Earth, other planets, interstellar clouds, and Galaxy \citep[][pp. 86--88]{alf1981}.  The dynamo effectively operates if a primary field exists initially, and has its origin different from the dynamo.  One of requirements for a self-exciting dynamo is an enough energy release inside the object to energize the dynamo \citep[][pp. 114--115]{alf1981}.

\subsection{Two stream instability}\label{sec5_3}

A relative motion of an electron fluid to an ion fluid can cause a micro-instability~\citep{Woods2004}.  If the relative velocity exceeds a critical value, a turbulence is triggered.  When the temperatures of electron and proton are equal, the critical values of relative velocity is $v_{\rm rel}=\mathcal{F}C_e$, where $\mathcal{F}\approx 0.604$ is a factor fixed for the maximum growth rate of instability, and $C_e\equiv (T_e/m_e)^{1/2}$ is a measure for thermal speed of electron.

However, the relative velocity is much smaller than the electron thermal velocity even when the azimuthal electric current density is so high that the radial velocity difference of protons and hydrogens is equal to the cosmic recession velocity at the turnround.  Then, the instability does not occur.  The relative velocity is given by
\begin{eqnarray}
v_{\rm rel}&=&v_{p\phi}-v_{e\phi}\approx \frac{\alpha_{p{\rm n}}}{B_z}HL\nonumber\\
&=&3.81
\times 10^{-4}~{\rm cm~s}^{-1}~\left(\frac{\chi_{{\rm H}^+}}{6.52\times 10^{-5}}\right)^{-1/2} \left(\frac{H}{2.70\times 10^3~{\rm km~s}^{-1}~{\rm Mpc}^{-1}}\right)^{1/2} \left[\frac{\left(\sigma v\right)_{p{\rm H}}}{2.3\times 10^{-9}~{\rm cm}^3~{\rm s}^{-1}}\right]^{1/2} \left(\frac{L}{L_B}\right)^{1/2}, \nonumber\\
\label{eq24}
\end{eqnarray}
where
Eqs. (\ref{eq45}), (\ref{eq69}), (\ref{eq16}), (\ref{eqb11}), and (\ref{eqb20}) were used.
On the other hand, the $C_e$ value is given by
\begin{eqnarray}
C_e&\equiv& \sqrt[]{\mathstrut T_e/m_e}\nonumber\\
&=&1.83\times 10^6~{\rm cm~s}^{-1}~\left(\frac{1+z}{17.5}\right)~\left(\frac{1+\delta}{5.55}\right)^{1/3}.
\label{eq25}
\end{eqnarray}

\subsection{Magnetic field generation in molecular cloud}\label{sec5_1}

We roughly check an amplitude of magnetic field generated through a drift current in MCs.  For this purpose, we take physical quantities at the surface of MCs in Model A of \citet{cio1994}:  $n_{\rm H}\sim 2.6\times 10^3$ cm$^{-3}$, $\chi_{{\rm H}^+}\sim 10^{-10}$, $L\sim 4.3$~pc, $B\sim 35.3~\mu$G, $|v_{{\rm n}r}|\sim 1.9\times 10^3$ cm s$^{-1}$, $\sigma_{p{\rm n}}(\Delta v_r\sim 10^3$ cm s$^{-1})\sim 1.6\times 10^{-13}$ cm $^2$.  The $\alpha_{p{\rm n}}$ value [Eqs. (\ref{eq45}) and (\ref{eq69})] is then estimated to be
\begin{eqnarray}
\alpha_{p{\rm n}}
&=&2.93
\times 10^{-12}~{\rm G}~\left(\frac{n_{\rm H}}{10^3~{\rm cm}^{-3}}\right) 
\left[\frac{\left(\sigma v\right)_{p{\rm H}}}{3.1\times 10^{-10}~{\rm cm}^3~{\rm s}^{-1}}\right].
\label{eq17}
\end{eqnarray}
The amplitude of generated field is then given [Eqs. (\ref{eq64}) and (\ref{eq65})] by
\begin{eqnarray}
\delta B_z&\sim& 4\pi j_\phi L_B = \frac{4\pi e n_p \alpha_{p{\rm n}} \Delta v_r L_B}{B_z} \nonumber\\
&=&1.82
\times 10^{-11}~{\rm G} \left(\frac{n_{\rm H}}{10^3~{\rm cm}^{-3}}\right)^2 \left(\frac{\chi_{{\rm H}^+}}{10^{-10}}\right)~\left[\frac{\left(\sigma v\right)_{p{\rm H}}}{3.1\times 10^{-10}~{\rm cm}^3~{\rm s}^{-1}}\right]~\left(\frac{B_z}{10~\mu{\rm G}}\right)^{-1}~\left(\frac{\Delta v_r}{10^3~{\rm cm}~{\rm s}^{-1}}\right) \left(\frac{L_B}{1~{\rm pc}}\right).~~~~~
\label{eq18}
\end{eqnarray}
This is much smaller than the initial magnetic field assumed in a MC, $B_{\rm eq,c0}=35.3~\mu$G.  The field generation, therefore, does not affect at all the total amplitude of magnetic field during the time evolution of the model MC.

\section{Generation of a magnetic field gradient}\label{sec6}

In the present calculation, gradients of the magnetic field in the $r$-direction are assumed in the initial conditions.  Practically, charged species of ions and electrons move outward only in special configurations of magnetic fields as in this setting.  A gradient of the magnetic field can, however, be generated through the gravitational collapse of a structure even if the initial magnetic field amplitude is coherent and homogeneous.

\begin{figure}
\begin{center}
\includegraphics[width=84mm]{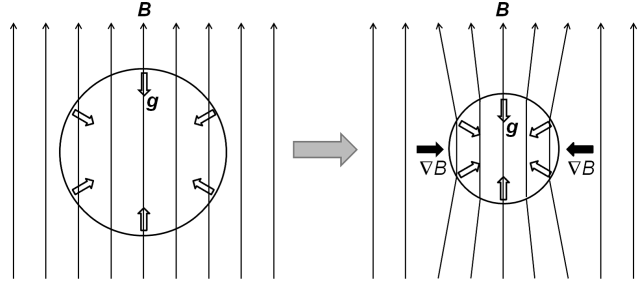}
\caption{Illustration for the creation of magnetic field gradient in the radial direction.  The upper direction is defined as the $z$-axis, and open circles correspond to boundaries of a structure at an early epoch before the gravitational contraction (left part) and at a late epoch during the contraction (right part).  Thin arrows show magnetic field lines, open thick arrows indicate directions of the gravity, and filled thick arrows indicate directions of the field gradient.  \label{fig_db}}
\end{center}
\end{figure}

Figure \ref{fig_db} shows an illustration for the creation of magnetic field gradient in the $r$-direction.  The upper direction on the plane of paper is defined as the $z$-axis, and open circles correspond to boundaries of a structure at an early epoch before the gravitational contraction (left part) and at a late epoch during the contraction (right part).  Thin arrows show magnetic field lines, open thick arrows indicate directions of the gravity, and filled thick arrows indicate directions of the field gradient.  The baryon density inside the structure increases relative to that of outside, as a function of time.  Since field lines are initially frozen into the charged plasma, the $B_z$ value increases inside the structure.  A field gradient is then generated in the $r$-direction near the boundary (right part).  Consequently, the Lorentz force is produced in the $r$-direction with its strength proportional to $[(\nabla \times \vecb)\times \vecb]_r \sim -(\partial_r B_z)B_z$ [Eq. (\ref{eqa2})].

\section{Parameter region for chemical separation}\label{sec7}

\subsection{Scales of structure and magnetic domain}\label{sec7_1}
In the $\Lambda$CDM model, large structures such as galaxies and galactic clusters are formed through collisions and mergers of smaller structures.  When we consider gravitational collapses of structures with scales smaller than that of Galaxy, effects of the magnetic field on motions of charged and neutral species are quantitatively different from that of larger structures.  For example, a smaller velocity difference is needed for charged species to escape from gravitational collapse of smaller structures at the time of turnround, i.e., $H(z_{\rm tur})r_{\rm sph}$.

We note that the typical scale of the magnetic domain in which the field direction is coherent should be larger than the system scale.  If the scale of the magnetic domain is smaller than the system scale, average radial velocities of charged species are roughly the same as that of neutral species although there are fluctuations in velocities caused by the magnetic field existing over small scales.  We, therefore, have a constraint on the comoving $L_{B0}$ value, i.e., $L_{B0}\geq L(1+z)$ [cf. Eq. (\ref{eq35})].

\subsection{Constraints}\label{sec7_2}
The condition for the gravitational collapse of neutral matter is that the gravitation [the first term in RHS of Eq. (\ref{eqa1})] is larger than the Lorentz force (the second term).  It is clear that magnetic fields on the scale larger than 600 pc with amplitude less than $\sim 10^{-7}$ G do not affect the gravitational collapse of neutral atoms [Eqs. (\ref{eqa4}) and (\ref{eqa5})].  When the amplitude and the spatial scale of magnetic field satisfy the condition, neutral species can collapse gravitationally.  Using Eqs. (\ref{eqa4}), (\ref{eqa5}), (\ref{eq32}), and (\ref{eq34}), the condition is derived:
\begin{eqnarray}
\frac{B_z^2}{L_B} < 1.93 \times 10^{-14}~{\rm G}^2~{\rm kpc}^{-1} \left(\frac{1+z_{\rm tur}}{17.5}\right)^5 \left(\frac{h}{0.700}\right)^{10/3} \left(\frac{\Omega_{\rm b}}{0.0463}\right) \left(\frac{\Omega_{\rm m}}{0.279}\right)^{2/3} \left(\frac{M_{\rm str}}{10^{6}~M_\odot}\right)^{1/3}.
\label{eqa6}
\end{eqnarray}
The comoving value is related to the proper value by $B_{z0}^2/L_{B0}=(1+z)^{-5}B_z^2/L_B$.  The condition on the comoving value is then given by
\begin{eqnarray}
\frac{B_{z0}^2}{L_{B0}}< 1.18 \times 10^{-20}~{\rm G}^2~{\rm kpc}^{-1} \left(\frac{h}{0.700}\right)^{10/3} \left(\frac{\Omega_{\rm b}}{0.0463}\right) \left(\frac{\Omega_{\rm m}}{0.279}\right)^{2/3} \left(\frac{M_{\rm str}}{10^{6}~M_\odot}\right)^{1/3}.
\label{eqa7}
\end{eqnarray}
This constraint is independent of the turnround redshift $z_{\rm tur}$.

The condition to suppress the gravitational collapse of charged species is that the Lorentz force is larger than the friction from neutral hydrogens for the velocity difference given by the cosmic recession velocity at the turnround.  In this case the equation, i.e., $v_{pr}-v_{{\rm n}r}> H(z_{\rm tur})r_{\rm sph}$, holds.  When the amplitude and the spatial scale of magnetic field satisfy the condition, charged species can get left in IGM typically.  The friction on proton is the predominant friction working on the whole charged fluid in the radial direction.  The proton velocity is then related to the velocity of neutral matter [Eq. (\ref{eqb17})].

Using Eqs. ({\ref{eqa5}}), (\ref{eq33}), (\ref{eq34}), (\ref{eq35}), (\ref{eq40}), and (\ref{eq16}), 
the condition is derived:
\begin{eqnarray}
\frac{B_z^2}{L_B}&>& 
4.59 \times 10^{-15}
~{\rm G}^2~{\rm kpc}^{-1} \left(\frac{1+z_{\rm tur}}{17.5}\right)^{13/2} \left(\frac{h}{0.700}\right)^{13/3} \left(\frac{\Omega_{\rm b}}{0.0463}\right)^2 \left(\frac{X}{0.75}\right)^2 
\left(\frac{1+\delta} {5.55}\right)^2
\left(\frac{\Omega_{\rm m}}{0.279}\right)^{1/6} \nonumber\\
&&\times \left(\frac{\chi_{{\rm H}^+}}{6.52\times 10^{-5}}\right) \left[\frac{\left(\sigma v\right)_{p{\rm H}}}{2.3\times 10^{-9}~{\rm cm}^3~{\rm s}^{-1}}\right] \left(\frac{M_{\rm str}}{10^{6}~M_\odot}\right)^{1/3}.
\label{eqb18}
\end{eqnarray}
The condition on the comoving value is also given by
\begin{eqnarray}
\frac{B_{z0}^2}{L_{B0}}&>& 
2.80 \times 10^{-21}
~{\rm G}^2~{\rm kpc}^{-1} \left(\frac{1+z_{\rm tur}}{17.5}\right)^{3/2} \left(\frac{h}{0.700}\right)^{13/3} \left(\frac{\Omega_{\rm b}}{0.0463}\right)^2 \left(\frac{X}{0.75}\right)^2 
\left(\frac{1+\delta} {5.55}\right)^2
\left(\frac{\Omega_{\rm m}}{0.279}\right)^{1/6} \nonumber\\
&&\times \left(\frac{\chi_{{\rm H}^+}}{6.52\times 10^{-5}}\right) \left[\frac{\left(\sigma v\right)_{p{\rm H}}}{2.3\times 10^{-9}~{\rm cm}^3~{\rm s}^{-1}}\right] \left(\frac{M_{\rm str}}{10^{6}~M_\odot}\right)^{1/3}.
\label{eq67}
\end{eqnarray}
In Eqs. (\ref{eqb18}) and (\ref{eq67}), the reaction rate $(\sigma v)_{p{\rm H}}$ is a function of the turnround redshift and the structure mass.  It is given by the value for the cosmic recession velocity [Eq. (\ref{eq40})] at the boundary of the structure [Eq. (\ref{eq35})].

\begin{figure}
\begin{center}
\includegraphics[width=84mm]{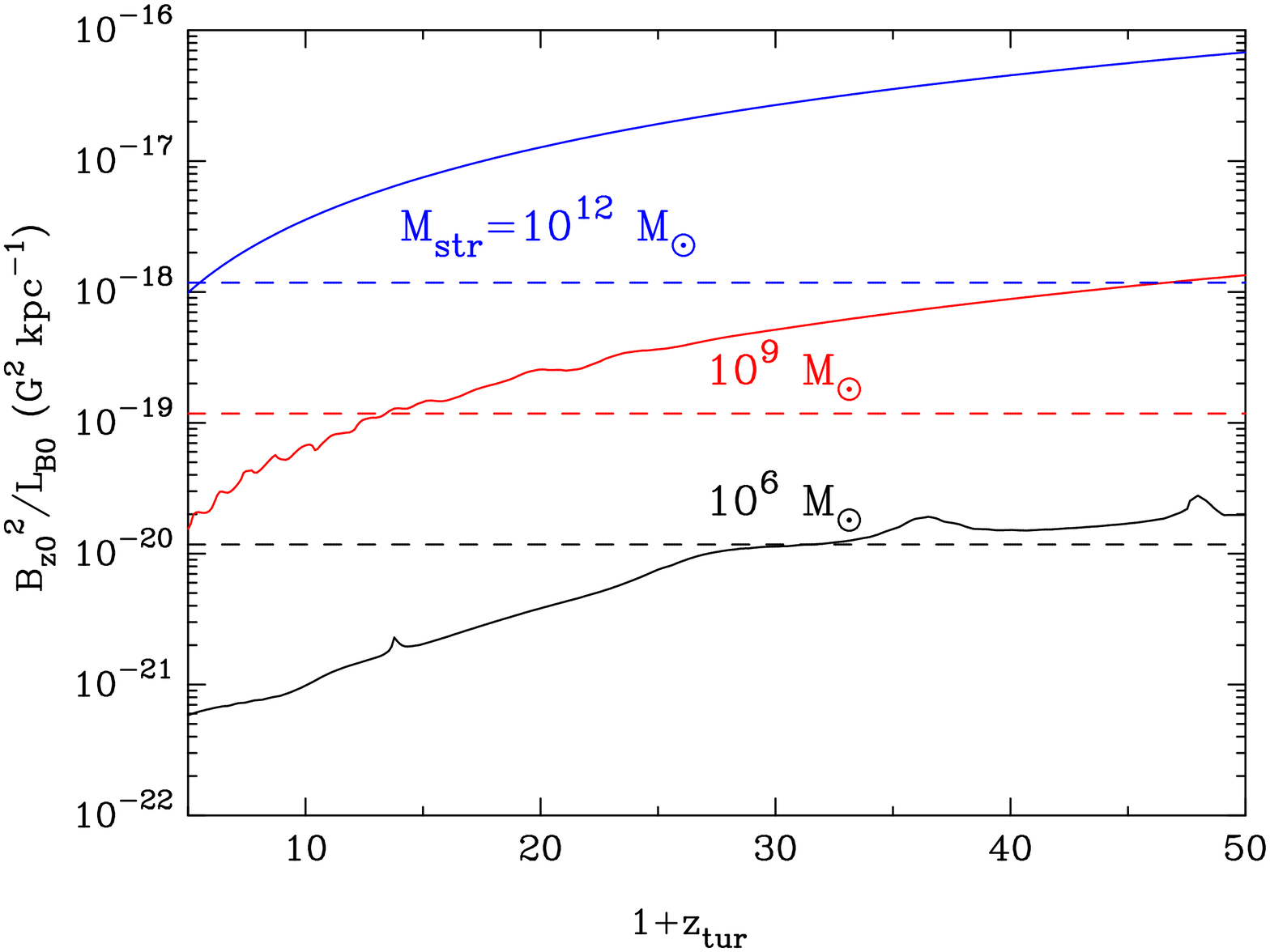}
\caption{Constraints on the comoving Lorentz force as a function of the turnround redshift.  Solid lines show lower limits from the condition that charged species do not contract along with neutral hydrogen.  Dashed lines show upper limits from the condition for the gravitational collapse of neutral hydrogen.  Lines are drawn for three cases of the structure mass, $M_{\rm str}=10^6$, $10^9$, and $10^{12}$ $M_\odot$. \label{pg13}}
\end{center}
\end{figure}

Figure \ref{pg13} shows constraints on the comoving Lorentz force $B_{z0}^2/L_{B0}$ as a function of the turnround redshift $z_{\rm tur}$.  Solid lines correspond to lower limits from the condition that charged species do not contract along with neutral hydrogen [Eq. (\ref{eq67})].  Dashed lines correspond to upper limits from the condition for the gravitational collapse of neutral hydrogens [Eq. (\ref{eqa7})].  For respective constraints, lines are shown for three cases of the structure mass, $M_{\rm str}=10^6$ (the lowest lines), $10^9$ (the middle lines), and $10^{12}$ $M_\odot$ (the highest lines).  For $M_{\rm str}=10^6$ $M_\odot$, we find a parameter region for a successful chemical separation at $B_{z0}^2/L_{B0}\la 10^{-20}$ G$^2$ kpc$^{-1}$ at redshift $1+z_{\rm tur} \la 30$.  For $M_{\rm str}=10^9$ $M_\odot$, a similar interesting parameter region exists  at $B_{z0}^2/L_{B0}\la 10^{-19}$ G$^2$ kpc$^{-1}$ and $1+z_{\rm tur} \la 15$.   For the most massive case of $M_{\rm str}=10^{12}$ $M_\odot$, no region is found at relatively high redshifts of $1+z_{\rm tur} \sim 10$.  In this way, at gravitational collapses of heavier objects, it is more difficult to separate the motions of neutral and charged particles.

\begin{figure}
\begin{center}
\includegraphics[width=84mm]{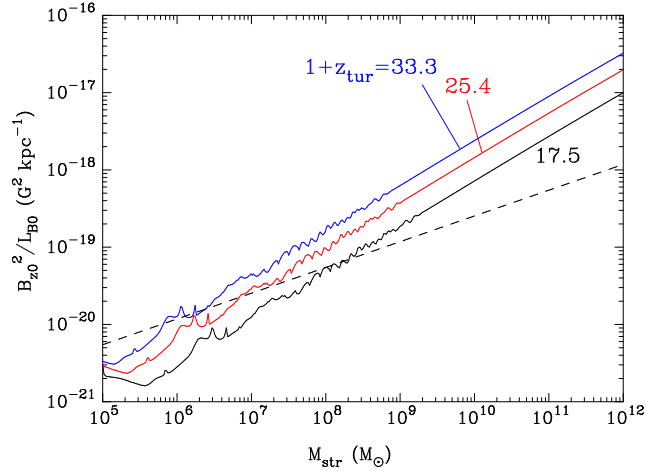}
\caption{Constraints on the comoving Lorentz force as a function of the structure mass.  Solid lines show lower limits from the condition that charged species do not contract along with neutral hydrogen.  Lines are drawn for three cases of the turnround redshift, $1+z_{\rm tur}=17.5$, 25.4, and 33.3 (corresponding to the collapse redshift $z_{\rm col}=10$, 15, and 20, respectively).  The dashed line shows the upper limit from the condition for the gravitational collapse of neutral hydrogen independently of the turnround redshift. \label{pg14}}
\end{center}
\end{figure}

Figure \ref{pg14} shows constraints on the comoving Lorentz force as a function of the structure mass $M_{\rm str}$.  Solid lines correspond to lower limits from the condition for the motion of charged species [Eq. (\ref{eq67})] for three cases of the turnround redshift, $1+z_{\rm tur}=17.5$, 25.4, and 33.3 (corresponding to the collapse redshift $z_{\rm col}=10$, 15, and 20, respectively).  The dashed line shows the upper limit from the condition for the gravitational collapse of neutral hydrogens, which are independent of the turnround redshift [Eq. (\ref{eqa7})].   It can be seen that the chemical separation is more difficult in structures which collapse earlier.  
We find parameter regions for the chemical separation at $M_{\rm str} \le {\cal O}(10^{8})$ $M_\odot$ for the latest collapse case of $1+z_{\rm tur}=17.5$, $M_{\rm str} \le {\cal O}(10^{7})$ $M_\odot$ for $1+z_{\rm tur}=25.4$, and $M_{\rm str} \le {\cal O}(10^{6})$ $M_\odot$ for $1+z_{\rm tur}=33.3$.

\section{Discussion}\label{sec8}

\subsection{Later epoch of the structure formation}\label{sec8_2}

We comment on a possibility of chemical separation in a later epoch of structure formation.  Depending on the virialization temperature of the collapsing structure, the ionization degree after the virialization can be smaller than that during the gravitational collapse because of the high density.  The baryon density in the late epoch is, on the other hand, much larger than that during the collapse.  Then, the larger friction force must be balanced by the Lorentz force originating from a larger magnetic field.  For a fixed structure mass, the gravitation term [the first term in RHS of Eq. (\ref{eqa1})] roughly scales as $\propto \rho_{\rm b}^{5/3} \propto (1+\delta)^{5/3}$ [Eq. (\ref{eqa4})].  On the other hand, the Lorentz force term (the second term) scales as $\propto B^2/L_B \propto (1+\delta)^{5/3}$ if we roughly assume adiabatic contractions of charged species and magnetic domains in the early epoch of structure formation.   Therefore, it is expected that if an ambipolar diffusion does not occur in the early structure formation epoch, it does not also in a later epoch as long as a magnetic field generation does not operate during the structure formation.

\subsection{Chemical reactions}\label{sec8_1}

Lithium atoms can be ionized by a ultraviolet (UV) photon as
\begin{equation}
 {\rm Li}+\gamma\rightarrow {\rm Li}^+ +e^-.
\label{eq28}
\end{equation}
They can be ionized also through a collision with an H$^+$ ion, which is generated by UV photons or  cosmic rays:  
\begin{equation}
 {\rm Li}+{\rm H}^+ \rightarrow {\rm Li}^+ +{\rm H}.
\label{eq29}
\end{equation}
The ionization potential of Li is $I($Li$)=5.39$ eV which corresponds to the temperature $T=2I($Li$)/3\sim 4\times 10^4$~K.  Some proportion of Li atoms can be also easily ionized by external UV sources or a gas heating at the virialization of structures.  The Li$^+$ ions produced secondarily in this way can then be trapped by magnetic field, and possibly be left out of forming structures.  Such a contribution to a resulting lithium abundance in the collapsed structure, however, operates after the gravitational collapse considered in this paper.  They are then neglected here.

\subsection{Li abundance of MPS}\label{sec8_3}
Astronomical observations indicate primordial abundances of D \citep{Pettini:2012ph}, $^3$He \citep*{Bania2002}, and $^4$He \citep*{Izotov:2010ca,Aver:2010wq} consistent with those predicted in SBBN model.  Primordial $^7$Li abundance is inferred from spectroscopic observations of metal-poor halo stars.  We adopt log($^7$Li/H)$=-12+(2.199\pm 0.086)$ determined with a 3D nonlocal thermal equilibrium model~\citep{Sbordone:2010zi}.
This estimation corresponds to the $2\sigma$ range of
\begin{equation}
1.06\times 10^{-10} < ({\rm ^7Li/H})^{\rm MPS} < 2.35\times 10^{-10}.
\label{eq27}
\end{equation}
This Li abundance level is $\sim 3$--$4$ times smaller than the SBBN prediction \citep{Coc:2011az,Coc:2013eea}, and the dispersion of observed Li abundance is small.  
Since the observed $^7$Li abundance is not so different from the SBBN prediction, it is naturally expected that SBBN model successfully describes the outline of primordial light element synthesis.  The Li abundances in MPSs can be affected by several physical processes operating after the BBN epoch.  The abundance ratio of Li and H in MPSs is then expressed as
\begin{equation}
 ({\rm Li/H})^{\rm MPS}=({\rm Li/H})^{\rm SBBN}~F^{\rm dep},
\label{eq26}
\end{equation}
where
(Li/H)$^{\rm SBBN}$ is the abundance ratio in SBBN model, and
$F^{\rm dep}$ is the depletion factor associated with 1) modified BBN models including exotic long-lived particles or changed expansion rate,
2) the structure formation as considered in this paper,
3) the virialization of the structure,
4) the formation of observed MPSs, and
5) the stellar processes in surfaces of MPSs occurring from the star formation until today.

Generally, cosmological processes change elemental abundances universally, while astrophysical processes do locally depending on physical environments of respective stars.  It is, therefore, difficult to explain the discrepancy in $^7$Li abundance with astrophysical processes which result in large dispersions in the abundance.  

The depletion factor from the chemical separation during the structure formation can be described by
\begin{eqnarray}
F^{\rm dep}&\equiv&\frac{[(n_{^7{\rm Li}}+n_{^7{\rm Li}^+})/(n_{{\rm H}}+n_{{\rm H}^+})]_{\rm str}}{[(n_{^7{\rm Li}}+n_{^7{\rm Li}^+})/(n_{\rm H}+n_{{\rm H}^+})]_{\rm uni}} \nonumber\\
&\approx & \frac{\chi_{^7{\rm Li},{\rm uni}} + \chi_{{\rm Li}^+,{\rm str}}}{\left( \chi_{^7{\rm Li}}+\chi_{^7{\rm Li}^+} \right)_{\rm uni}},
\label{eq79}
\end{eqnarray}
where
quantities with subscripts, `uni' and `str', are values of the homogeneous early universe after the cosmological recombination, and those of the collapsed structure in the late universe, respectively.   In the second line, it was assumed that the primordial ionization degree of hydrogen is negligibly small, i.e., $\chi_{{\rm H}^+} \ll 1$, and that values of the number ratio $\chi_{^7{\rm Li}}$ are equal in the homogeneous early universe and the structure.  We suppose the initial abundance ratio of $^7$Li$^+$/$^7$Li $\sim 1$ as suggested from a chemical history of homogeneous early universe \citep{Vonlanthen:2009ns}.  The chemical separation via the ambipolar diffusion can only dilute the charged $^7$Li$^+$.  The depletion factor is, therefore, 1/2 at minimum when the primordial $^7$Li$^+$ is completely expelled from the structure.  This factor would be smaller if the initial $^7$Li$^+$ abundance in the gravitational structure formation is larger for some reason.  For example, even a small intensity of ionizing photon of $^7$Li would quickly transform $^7$Li to $^7$Li$^+$ without absorption by neutral hydrogen (Sec. \ref{sec8_1}).  On the other hand, the depletion factor would be larger if the chemical separation is less efficient.

The Li abundance of MPSs may not be explained by the chemical separation only.  In that case, we need another depletion mechanism.  As an example, a rotationally induced mixing model \citep{Pinsonneault:1998nf,Pinsonneault:2001ub} for MPSs is chosen here since dispersions as well as depletion factors are predicted theoretically only in this model among stellar depletion models.  Since the predicted depletion factor is proportional to the dispersion factor, the depletion factor is constrained from observed dispersions.  Pinsonneault et al. estimated the depletion factor: `0.13 dex, with a 95 \% range extending from 0.0 to 0.5 dex' \citep{Pinsonneault:2001ub}.  This model explains a part of the Li abundance discrepancy although the complete solution by this mechanism only seems almost impossible.  The Li abundances in MPSs may, therefore, be explained by the combination of the ambipolar diffusion during the structure formation and the rotationally induced mixing in stars.

Stellar Li abundances in metal-poor globular clusters (GCs) have also been measured.  For example, GC M4 was studied using high-resolution spectra with GIRAFFE at Very Large Telescope.  The Li abundance in turn-off stars is then found to be log($^7$Li/H)$=-12+(2.30\pm 0.02 +0.10)$ \citep{Mucciarelli:2010gz}.  All Li abundances measured so far are summarized in Fig. 3 of \citet{Mucciarelli:2010gz}, and they are consistent with abundances in metal-poor halo stars at present.  If the ambipolar diffusion studied in this paper caused the small Li abundances of MPSs, however, reduction factors of MPSs can reflect respective histories of parent structure of MPSs.  In a modern model calculation for GC formation, the Galaxy formation results from a continuous process of merging and accretion which is realized in a hierarchical structure formation scenario \citep{Kravtsov:2003sm}.  In the model, GCs form at densest regions of filaments in a large-scale structure.

\subsection{Other constraint on PMF}\label{sec8_4}
Theoretical and observational constraints on the cosmic magnetic field have been summarized in \citet{Durrer:2013pga}.  The magnetic field strength in the interesting parameter region found in this study (Sec. \ref{sec7}) looks somewhat higher than the theoretical upper limit from the effect of dissipation of magnetic field through the processing by MHD turbulence.  The propagation length of Alfv\'en wave is given by $\lambda_B\sim v_A t$, where $v_A$ is the Alfv\'en speed.  This length scale corresponds to ``the size of largest processed eddies'' \citep{Durrer:2013pga} by MHD turbulence.  The Alfv\'en speed during the matter dominated epoch of the homogeneous universe is given by $v_{\rm A}=B/\sqrt[]{\mathstrut 4 \pi \rho_{\rm b}}$ with $\rho_{\rm b}\propto (1+z)^3$ the baryon density [Eq. (\ref{eq32})].  
Note that the density used in the Alfv\'en speed is that of fluid with a frozen-in magnetic field.  The density is then given by the total density if the fluid is fully ionized or if the neutral fluid is effectively coupled to the charged fluid through the collision so that the magnetic field can be considered frozen also into the neutral fluid.  The physical states considered in this paper are ones in which the matter is only weakly ionized and the coupling of the charged and neutral fluids is effective.  Although the ambipolar diffusion reduces the magnetic pressure gradient until the Lorentz force becomes comparable to the gravitation [cf. Eq. (\ref{grav_Lorentz_ratio})], the coupling is effective after then.  Therefore, the total fluid has a frozen-in magnetic field and its density is used in the Alfv\'en speed.
The distance is then given by
\begin{eqnarray}
\lambda_B &\sim& \left[\frac{B_0}{\left(4\pi \rho_{\rm b0}\right)^{1/2}} (1+z)^{1/2} \right] \left[\frac{2}{3H_0 \Omega_{\rm m}^{1/2} (1+z)^{3/2}}\right] \nonumber\\
&=&
\frac{2^{3/2} }{3^{3/2}}
\frac{B_0}{m_{\rm Pl} H_0^2 \Omega_{\rm b}^{1/2} \Omega_{\rm m}^{1/2} (1+z)},
\end{eqnarray}
where
$m_{\rm Pl}$ is the Planck mass.  Consequently, the comoving propagation length $\lambda_{B0}=\lambda_B(1+z)$ is constant.

Since the magnetic fields on scales shorter than $\lambda_{B0}(B_0)$ decay, there is a maximum amplitude of magnetic field which escapes from this decay for a given $\lambda_{B0}$.  From the above equation, an upper limit on the $B_0$ value is derived as
\begin{eqnarray}
B_{0} &\la& 
1.3 \times 10^{-10}~{\rm G}
\left(\frac{h}{0.700}\right)^2 \left(\frac{\Omega_{\rm b}}{0.0463}\right)^{1/2} \left(\frac{\Omega_{\rm m}}{0.279}\right)^{1/2} 
\left(\frac{\lambda_{B0}}{10~{\rm kpc}}\right).
\label{eq66}
\end{eqnarray}
This upper limit is lower than the field value required for the chemical separation [Eqs. (\ref{eqa7}) and (\ref{eq67})] (by a factor of $\sim$ two for $\lambda_{B0}=10$ kpc).  However, Eq (\ref{eq66}) is just a rough estimate, and realistic limits should be derived in precise calculations in future.  It is interesting that the upper limit caused by the MHD processing in the early universe is near to the interesting field strength.  It indicates that relatively large magnetic field in the early universe may have been reduced by the MHD effect to the level which is most appropriate for the chemical separation causing the lithium problem.

During the gravitational collapse of structures, the Alfv\'en speed increases as $\propto (1+\delta)^{1/6}$ if the dissipation of magnetic field is not operative.  The dissipation scale in collapsed structures which are decoupled from the cosmic expansion is then given by
\begin{equation}
\lambda_B^{\rm str}(z) \sim 
\frac{2^{3/2} }{3^{3/2}}
 \frac{B_0(1+z)^{1/2} (1+\delta)^{1/6}}{m_{\rm Pl} H_0^2 \Omega_{\rm b}^{1/2} \Omega_{\rm m}^{1/2} (1+z)^{3/2}}.
\end{equation}
The contraction increases the dissipation scale slightly.  The MHD effect then becomes significant in a large density environment.  Therefore, after the collapse, the magnetic field strength can be decreased further.

\section{Summary}\label{sec9}
We considered a possible effect of PMFs on motions of charged and neutral chemical species during the formation of first structures at redshift $z={\cal O}(10)$.  We assumed that the PMF has a gradient in a direction perpendicular to the field direction.  This gradient is realized by an electric current density in the direction perpendicular to both directions of the field lines and the gradient.  The Lorentz force on the charged species then causes a velocity difference between charged and neutral species in the direction of the field gradient.  Resultantly, a velocity of charged species can be different from that of neutral species which collapses gravitationally during the structure formation.   Therefore, $^7$Li$^+$ ions may have possibly escaped from gravitational collapse of early structures.

Calculations for fluid motions of charged and neutral species were performed through a simple estimation using fundamental fluid and electromagnetic equations.  We assumed a gravitational contraction of neutral matter in a spherically symmetric structure.  In addition, we utilized a cylindrical coordinate, and assumed a gradient of the altitudinal ($z$-component) magnetic field in the radial direction.  Related physical quantities are listed, and their typical values are given in Sec. \ref{sec3}.  Some analytical equations are introduced in Appendix \ref{app2}.

When the amplitude of magnetic field is sufficiently large, the charged fluid significantly decouples from the neutral fluid.  It is then possible that during the gravitational contraction of structure mainly composed of neutral hydrogens, contractions of protons, electrons, and $^7$Li$^+$ ions do not occur.  Although fluid motions of charged chemical species are solved for only H$^+$, $e$, and $^7$Li$^+$ in this study, other charged species are expected to have similar motions.  Because of large inductances of large astronomical structures, the generation of magnetic field is never efficient during the structure formation at $z\sim 10$.  Therefore, only PMFs which existed from the start of the structure formation can trigger the chemical separation.

The chemical separation requires the magnetic field gradient in  a direction perpendicular to the field direction.  Although such a gradient was assumed in the initial condition in this study, it may be produced associated with a density gradient during the gravitational contraction of structures without any initial field gradient.

Based on the calculated result of the chemical separation, we derived a parameter region for a successful chemical separation taking the structure mass, the turnround redshift of the gravitational collapse, and the comoving Lorenz force, i.e., $B_{z0}^2/L_{B0}$, as parameters.  It was found that the parameter region can be constrained to be very narrow.   If such a chemical separation has occurred during the structure formation, the primordial $^7$Li$^+$, which was produced via the recombination of $^7$Li$^{2+}$ but survived against its recombination during the cosmological recombination epoch, possibly does not participate in the gravitational contraction.  The abundance ratio of Li/H in early structures, which are progenitors of the Galaxy, can then be smaller than that inferred from SBBN model.  Therefore, the chemical separation may have caused the Li problem of the MPSs.  

The amplitude of the PMFs required for the chemical separation was estimated.  It is close to (somewhat smaller than) an upper limit determined from the effect of MHD turbulence on the decay of field amplitude.  This fact indicates the following possibility:  The PMF was generated via some mechanism operating in the extremely early universe.  The field amplitude was modulated by the MHD effect to the value appropriate to the chemical separation.

\appendix

\section{Solutions of variables from the force balance}\label{app2}

\subsection{Drifts in the expanding universe}\label{app2_1}

We consider two different cases of weak and strong Lorentz forces.

\subsubsection{Weak Lorentz force}\label{app2_1a}

Firstly, we suppose that a magnetic field is so weak that collisional momentum transfers from hydrogens to electrons and protons result in very small velocity differences despite the existence of the weak Lorentz force.  This case typically satisfies the condition $B_z \ll (4\pi e n_p \alpha_{p{\rm n}} H L L_B)^{1/2}$ [cf. Eqs. (\ref{eq16}) or (\ref{eqb17})].  Because of effective scatterings between protons, electrons, and neutral hydrogens, velocities of $p$ and $e$ are almost identical to that of neutral hydrogens, i.e., $\vecv_p=\vecv_e=\vecv_{\rm n}$.  The force balances for $p$ and $e$ [Eqs. (\ref{eqb1}) and (\ref{eqb2}) with an assumption $D/Dt=0$ and a neglect of $\nabla P_p$ and $\nabla P_e$ terms] give the value of electric field:
\begin{eqnarray}
\vece\cong -\vecv_{\rm n}\times \vecb=-\left(
\begin{array}{c}
0 \\
v_{{\rm n}z} B_r -v_{{\rm n}r} B_z \\
0
\end{array}
\right),
\label{eqb3}
\end{eqnarray}
where it was assumed that neither magnetic field nor neutral hydrogen velocity has an azimuthal component.

\subsubsection{Strong Lorentz force}\label{app2_1b}

Secondly, we suppose that a magnetic field is strong.  The collisional momentum transfers between charged species and hydrogens with large radial relative velocities are then counterbalanced by the Lorentz force, i.e., $B_z \sim (4\pi e n_p \alpha_{p{\rm n}} H L L_B)^{1/2}$.  The protons and electrons receive dynamical frictions from hydrogens with different amplitudes determined by momentum transfer cross sections.  In such a case, protons and electrons are promoted to start drifting in directions opposite to each other with velocities, $\vecv_{{\rm D}j}=\vecf_j \times \vecb/(Z_j eB^2)$, where $\vecf_j$ is the friction force [cf. Eqs. (\ref{eqb1}) and (\ref{eqb2})].  The both drift directions are perpendicular to the direction of the friction force.  Drift velocities of protons and electrons are given by
\begin{eqnarray}
\vecv_{{\rm D}p}&=&\frac{1}{B^2}\left(
\begin{array}{c}
\left[-\alpha_{p{\rm n}} v_{p\phi} + \alpha_{pe} \left( v_{e\phi}-v_{p\phi}\right)\right] B_z \\
\left[\alpha_{p{\rm n}} \left(v_{{\rm n}z} - v_{pz}\right)\right] B_r -\left[ \alpha_{p{\rm n}} \left( v_{{\rm n}r} - v_{pr}\right)\right] B_z \\
\left[\alpha_{p{\rm n}} v_{p\phi} - \alpha_{pe} \left( v_{e\phi}-v_{p\phi}\right)\right] B_r
\end{array}
\right), \label{eqb4}\\
\vecv_{{\rm D}e}&=&-\frac{1}{B^2}\left(
\begin{array}{c}
-\left[\alpha_{e{\rm n}} v_{e\phi} + \alpha_{ep} \left( v_{e\phi}-v_{p\phi}\right)\right] B_z \\
\left[\alpha_{e{\rm n}} \left(v_{{\rm n}z} - v_{ez}\right)\right] B_r -\left[ \alpha_{e{\rm n}} \left( v_{{\rm n}r} - v_{er}\right)\right] B_z \\
\left[\alpha_{e{\rm n}} v_{e\phi} + \alpha_{ep} \left( v_{e\phi}-v_{p\phi}\right)\right] B_r
\end{array}
\right).\label{eqb5}
\end{eqnarray}
However, these drifts never complete effectively because of large inductances of large astrophysical objects (Sec. \ref{sec5}).

\subsection{Equilibrium state}\label{app2_2}
\subsubsection{Proton and electron}\label{app2_2a}

It is expected that bulk motions of chemical species and electromagnetic fields are in equilibrium states at all times during the early epoch of structure formation.  In the equilibrium state, force balance equations for proton ($p$ or H$^+$: the dominant component of ion) and electron [Eqs. (\ref{eqb6}) and (\ref{eqb7})] include nine unknown parameters:  components of vectors $\vece$, $\vecv_p$, and $\vecv_{\rm e}$.  Because of nearly complete neutrality for local charge, fluid velocities of ions and electrons should be approximately equal as for $r$- and $z$-components.  The number of independent parameters is then reduced to be seven: $E_r$, $E_\phi$, $E_z$, $v_{pr}=v_{er}$, $v_{pz}=v_{ez}$, $v_{p\phi}$, and $v_{e\phi}$.

The following conditions have been imposed additionally: $v_{{\rm n}\phi}=0$ and $B_\phi=0$.  Then, three equations for $E_r$, $E_\phi$, and $E_z$ are obtained:
\begin{eqnarray}
 E_r=&-v_{p\phi}B_z - \alpha_{p{\rm n}} (v_{{\rm n}r}-v_{pr})
    &=-v_{e\phi}B_z + \alpha_{e{\rm n}} (v_{{\rm n}r}-v_{pr}),\label{eqb8}\\
 E_\phi=&-\left(v_{pz}B_r -v_{pr}B_z\right) + \alpha_{p{\rm n}} v_{p\phi} -\alpha_{pe} \left(v_{e\phi}-v_{p\phi}\right)
    &=-\left(v_{pz}B_r -v_{pr}B_z\right) - \alpha_{e{\rm n}} v_{e\phi} -\alpha_{pe} \left(v_{e\phi}-v_{p\phi}\right),\label{eqb9}\\
 E_z=&v_{p\phi}B_r - \alpha_{p{\rm n}} (v_{{\rm n}z}-v_{pz})
    &=v_{e\phi}B_r + \alpha_{e{\rm n}} (v_{{\rm n}z}-v_{pz}) \label{eqb10}.
\end{eqnarray}
From the second equality of Eq. (\ref{eqb9}), we instantaneously find
\begin{equation}
 v_{e\phi}=-\frac{\alpha_{p{\rm n}}}{\alpha_{e{\rm n}}}v_{p\phi}.
\label{eqb11}
\end{equation}
From the second equalities of Eqs. (\ref{eqb8}) and (\ref{eqb10}), $v_{pr}$ and $v_{pz}$ values are given, respectively:
\begin{eqnarray}
v_{pr}&=& v_{{\rm n}r} +\frac{B_z}{\alpha_{e{\rm n}}} v_{p\phi}, \label{eqb12}\\
v_{pz}&=& v_{{\rm n}z} -\frac{B_r}{\alpha_{e{\rm n}}} v_{p\phi}.\label{eqb13}
\end{eqnarray}

Insertion of Eqs. (\ref{eqb12}) and (\ref{eqb13}) in the first equality of Eq. (\ref{eqb9}) leads to an expression for the parameter $v_{p\phi}$:
\begin{eqnarray}
v_{p\phi}&=&\frac{\alpha_{e{\rm n}} \left(E_\phi +B_r v_{{\rm n}z}- B_z v_{{\rm n}r}\right)}{B^2+ \alpha_{e{\rm n}} \alpha_{p{\rm n}} +\alpha_{pe} \left(\alpha_{e{\rm n}} + \alpha_{p{\rm n}}\right)}.
\label{eqb14}
\end{eqnarray}
The $v_{p\phi}$ value is given by the rotation of $B$ field [Eq. (\ref{eq68})].  We thus have seven equations for seven variables.  Therefore, the solutions can be obtained.

\subsubsection{General singly-ionized ions}\label{app2_2c}

From the force equation for singly-ionized ions ($i=$H$^+$, Li$^+$, ...), the electric field is given by
\begin{equation}
 \vece=-\vecv_i\times \vecb -\frac{\rho_i}{\tau_{i{\rm n}}} \frac{(\vecv_{\rm n}-\vecv_i)}{en_i} - \frac{\rho_i}{\tau_{ie}} \frac{(\vecv_e-\vecv_i)}{en_i}.
\label{eqb25}
\end{equation}
The left-hand side corresponds to the term of electric field, and the first, second and third terms in the RHS correspond to the Lorentz force, the frictions from H and electrons, respectively.   The friction from protons is neglected in the force equation for $i\neq p$ since the friction parameter $\alpha_{ip}$ is $\sim m_e/m_p$ times smaller than $\alpha_{ie}$ [cf. Eq. (\ref{eq55})].  For the case of $^7$Li$^+$, the force balance equation is somewhat different from that of proton because of differences in the ion masses and the momentum transfer cross sections.

We do not assume the condition of charge neutrality since abundance of $i$ can be negligibly small (for example, the primordial number ratio of $^7$Li$^+$/H is about $2.6\times 10^{-10}$ \citep{Vonlanthen:2009ns}).  We have assumed that $v_{{\rm n}\phi}=0$ and $B_\phi=0$.  Then, three equations for $E_r$, $E_\phi$, and $E_z$ are obtained:
\begin{eqnarray}
 E_r&=&-v_{i\phi}B_z - \alpha_{i{\rm n}} \left(v_{{\rm n}r}-v_{ir}\right) -\alpha_{ie}\left(v_{er}-v_{ir}\right),\label{eqb26}\\
 E_\phi&=&-\left(v_{iz}B_r -v_{ir}B_z\right) + \alpha_{i{\rm n}} v_{i\phi}-\alpha_{ie} \left(v_{e\phi}-v_{i\phi}\right),\label{eqb27}\\
 E_z&=&v_{i\phi}B_r - \alpha_{i{\rm n}} \left(v_{{\rm n}z}-v_{iz}\right) - \alpha_{ie} \left(v_{ez}-v_{iz}\right)\label{eqb28}.
\end{eqnarray}
Using these three equations with values of $\vece$ and $\vecv_e$ derived in Appendix \ref{app2_2a}, the velocity of $i$ is solved to be
\begin{eqnarray}
\left(
\begin{array}{c}
v_{ir} \\
v_{i\phi} \\
v_{iz}
\end{array}
\right)= \frac{1}{\alpha_i \left(\alpha_i^2 +B_r^2 + B_z^2\right)} \left(
\begin{array}{ccc}
\alpha_i^2+B_r^2 & \alpha_i B_z  & B_r B_z          \\
-\alpha_i B_z    & \alpha_i^2    & \alpha_i B_r     \\
B_r B_z          & -\alpha_i B_r & \alpha_i^2+B_z^2
\end{array}
\right) \left[ \left(
\begin{array}{c}
E_r \\
E_\phi \\
E_z
\end{array}
\right) + \left(
\begin{array}{cc}
\alpha_{i{\rm n}} v_{{\rm n}r} +& \alpha_{ie} v_{er} \\
                                & \alpha_{ie} v_{e\phi} \\
\alpha_{i{\rm n}} v_{{\rm n}z} +& \alpha_{ie} v_{ez}
\end{array}
\right)\right],
\label{eqb29}
\end{eqnarray}
where
$\alpha_i\equiv \alpha_{i{\rm n}}+\alpha_{ie}$ was defined.

\subsection{Typical case for an effective chemical separation at the turnround}\label{app2_2b}

We define, as a guide, a typical case of effective chemical separation in which the radial velocity difference between charged and neutral species is exactly equal to the recession velocity at the turnround.  The velocity difference is determined from the balance between the Lorentz force and the friction from neutral hydrogens in the $r$-direction.  We assume a strong magnetic field in $z$-direction, i.e., $B_z \gg \alpha_{ab}$, $B_r$.  In this case, ions and electrons possibly do not move to the structure centre, and their radial velocities can be as large as the cosmic recession velocity.  Velocities of charged species at the radius $r_{\rm sph}$ are then matched to the cosmic recession velocity, i.e., 
\begin{equation}
v_{pr}=Hr_{\rm sph}.\label{eqb19}
\end{equation}
The proton velocities and the electric field are then determined.

Eqs. (\ref{eqb14}) and (\ref{eqb13}), respectively, give relations of 
\begin{eqnarray}
v_{p\phi}&=&\frac{\alpha_{e{\rm n}}}{B_z}\left(Hr_{\rm sph}-v_{{\rm n}r}\right), \label{eqb20}\\
v_{pz}&=&v_{{\rm n}z} -\frac{B_r}{B_z}\left(Hr_{\rm sph}-v_{{\rm n}r}\right)\sim v_{{\rm n}z}.\label{eqb21}
\end{eqnarray}

The equilibrium electric field is then derived:
\begin{eqnarray}
E_r&=&\left( \alpha_{p{\rm n}} - \alpha_{e{\rm n}}\right) \left(Hr_{\rm sph} - v_{{\rm n}r}\right), \label{eqb22}\\
E_\phi&=&\left[\frac{\alpha_{e{\rm n}} \alpha_{p{\rm n}} +\alpha_{pe}\left(\alpha_{e{\rm n}}+\alpha_{p{\rm n}}\right) + B_r^2}{B_z} \right] \left(Hr_{\rm sph} - v_{{\rm n}r}\right) 
-B_r v_{{\rm n}z} + B_z Hr_{\rm sph}, \label{eqb23}\\
E_z&=& =-\left(\alpha_{p{\rm n}}-\alpha_{e{\rm n}}\right) \frac{B_r}{B_z} \left(Hr_{\rm sph} - v_{{\rm n}r}\right).\label{eqb24}
\end{eqnarray}

Under the assumption of Eq. (\ref{eqb19}), the velocity of the singly charged ion $i$ [Eq. (\ref{eqb29})] has an approximate form of 
\begin{eqnarray}
v_{ir}&\approx& Hr_{\rm sph}, \label{eqb32}\\
v_{i\phi}&\approx& \frac{\alpha_{i{\rm n}}-\alpha_{p{\rm n}}+\alpha_{e{\rm n}}}{B_z} \left(Hr_{\rm sph}-v_{{\rm n}r} \right), \label{eqb33}\\
v_{iz}&\approx& v_{{\rm n}z}-\frac{B_r}{B_z}\left(Hr_{\rm sph}-v_{{\rm n}r}\right).\label{eqb34}
\end{eqnarray}

\section{Detailed results of physical variables}\label{app3}

In this section, we show supplemental results of calculations performed in Sec. \ref{sec4}.

\subsection{Relaxation time-scales}\label{sec4_3}

\begin{figure}
\begin{center}
\includegraphics[width=84mm]{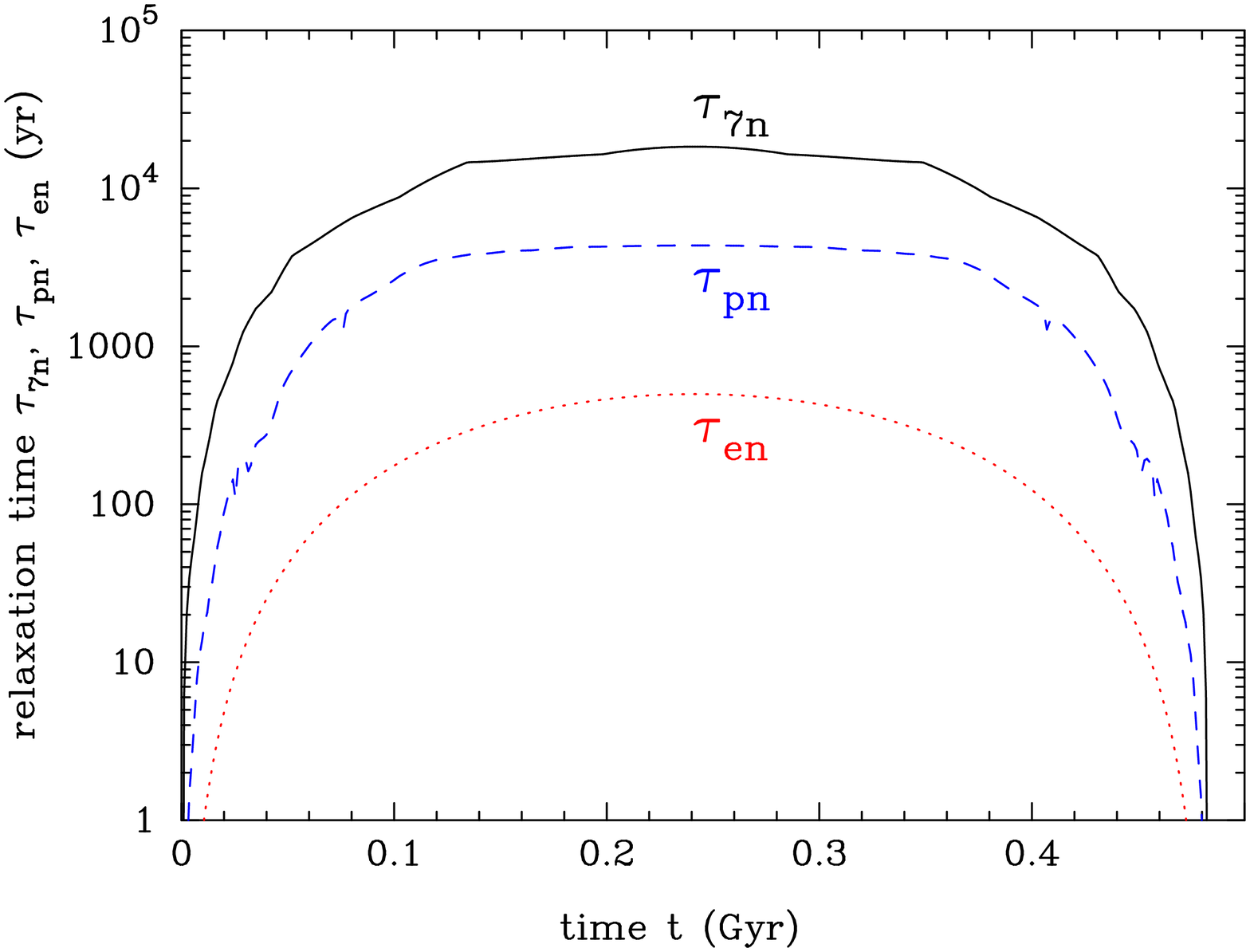}
\caption{Relaxation time-scales in collisions with hydrogen, $\tau_{7{\rm n}}$ (solid line) for $^7$Li$^+$, $\tau_{p{\rm n}}$ (dashed line) for H$^+$, and $\tau_{e{\rm n}}$ (dotted line) for $e^-$, as a function of cosmic time $t$ for both Cases 1 and 2. \label{pga13}}
\end{center}
\end{figure}

Figure \ref{pga13} shows relaxation time-scales in collisions with hydrogens, $\tau_{7{\rm n}}$ (solid line) for $^7$Li$^+$, $\tau_{p{\rm n}}$ (dashed line) for H$^+$, and $\tau_{e{\rm n}}$ (dotted line) for $e^-$, as a function of cosmic time $t$.  In this calculation, a relaxation time $\tau_{ab}$ is defined as the time it takes a species $a$ to change its velocity toward that of $b$ until a velocity difference of $a$ and $b$ becomes smaller than the thermal relative velocity of the $a$+$b$ system.  When the velocity difference, $|\vecv_a-\vecv_b|$, is smaller than the thermal velocity, the $\tau_{ab}$ value is approximated by the value evaluated at the thermal velocity.  In the present calculations for Case 1 and 2, any velocity differences were found to be smaller than thermal velocities at almost all position and time.  The relaxation time-scales are then determined only from the thermal velocity in the structure.  Therefore, they do not depend on radius.

Since the relaxation time is inversely proportional to the target number density [Eq. (\ref{eq45})], the $\tau_{ab}$ values increase during the expanding phase ($t\leq 0.242$ Gyr), while they progressively decrease during the contracting phase ($t\geq 0.242$ Gyr).  Cross sections are evaluated with linear interpolations of the adopted data.  This approximation causes a non-smooth behavior in the $\tau_{7{\rm n}}$ curve.  In addition to a similar non-smoothness, zigzags are seen in the $\tau_{p{\rm n}}$ curve, which result from fluctuations in the cross section.  The smooth shape of the $\tau_{e{\rm n}}$ curve reflects the constant cross section assumed for center of mass collision energy smaller than the lowest energy data point of $\sim 15$~meV.

\begin{figure*}
\begin{center}
\includegraphics[width=84mm]{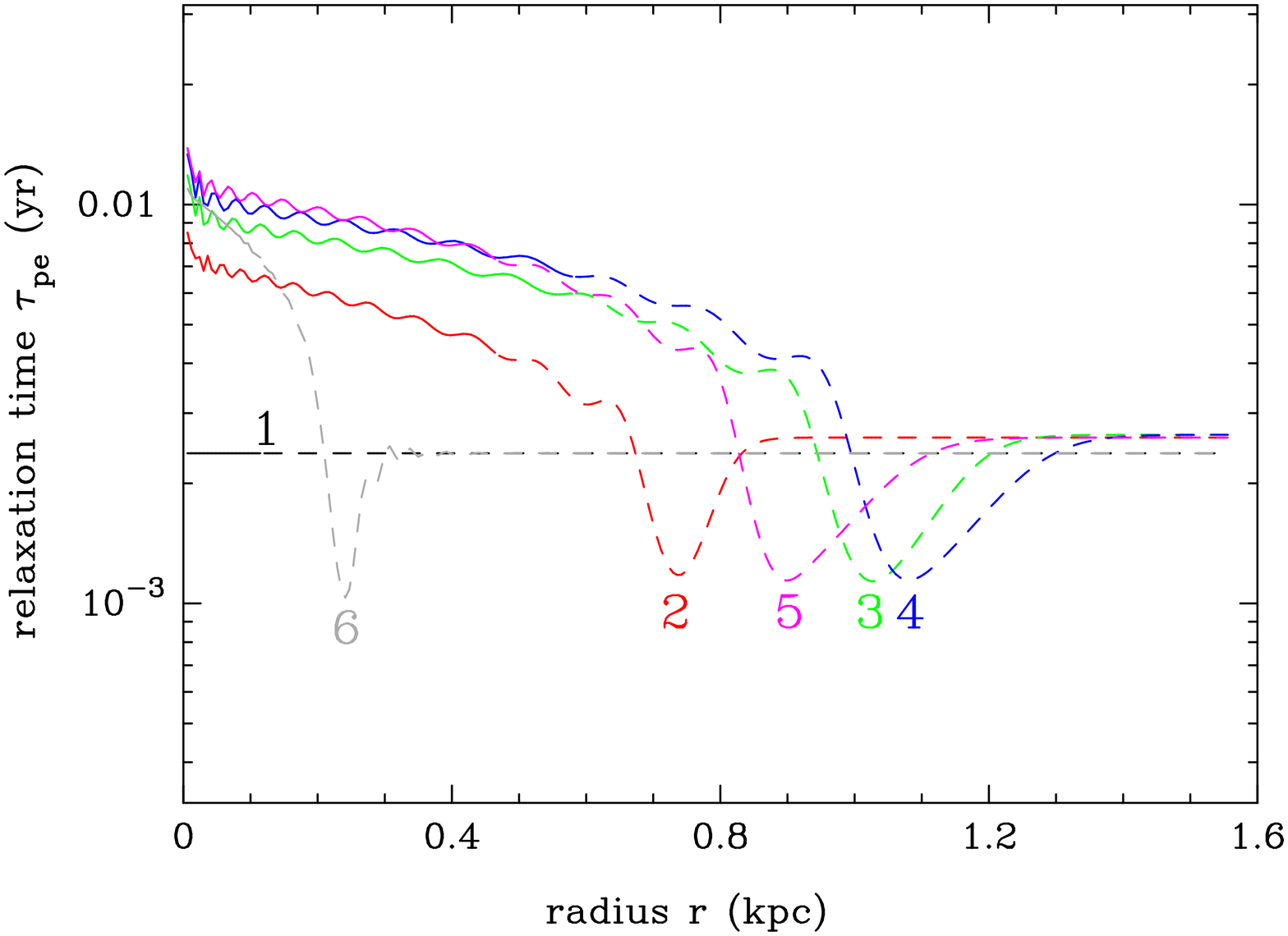}
\includegraphics[width=84mm]{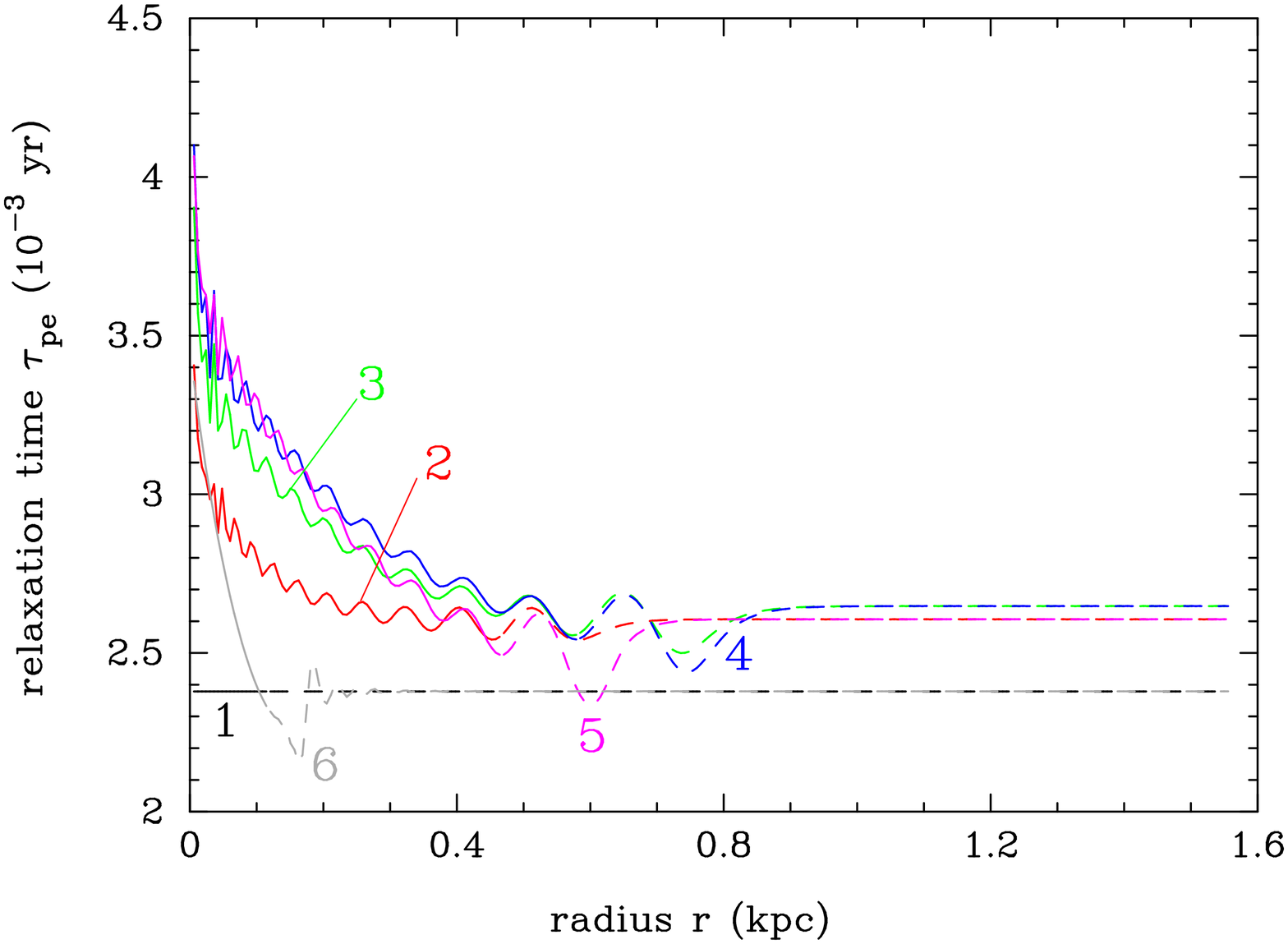}
\caption{Relaxation time $\tau_{pe}$ of H$^+$ in collisions with electron as a function of radius for Case 1 (left panel) and Case 2 (right panel) at $t$=9.29 Myr (1), 102 Myr (2), 195 Myr (3), 288 Myr (4), 381 Myr (5), and 474 Myr (6). Solid and dashed lines correspond to values inside and outside the structure, respectively. \label{pga6}}
\end{center}
\end{figure*}

Figure \ref{pga6} shows the relaxation time $\tau_{pe}$ of H$^+$ as a function of radius for Case 1 (left panel) and Case 2 (right panel).  Solid and dashed lines correspond to values inside and outside the structure, respectively.  In the whole calculation time, the relative velocity of proton and electron is given by the thermal velocity since the relative fluid velocity is smaller than the thermal velocity.  The $\tau_{pe}$ values at large radii are almost constant with time.  The reason comes from a constant reaction rate for a given thermal velocity, and a constant density of electrons.  The thermal velocity is assumed to be proportional to the square root of gas temperature $T\propto \rho_{\rm n}^{2/3}$ [Eq. (\ref{eq46})], while the proton and electron number densities scale as $n_{p,e}\propto \rho_{p,e}$.  The relaxation time is then given by $\tau_{pe}=(m_p/m_e)\tau_{ep}\propto T^{3/2}/n_p\propto \rho_{\rm n}/\rho_p$ [cf. Eqs. (\ref{eq69}), (\ref{eq50}), and (\ref{eq53})].  Inside the structure, this quantity increases with time since the number abundances of proton and other charged species decrease (Fig. \ref{pga2}).

\subsection{Relative velocities}\label{sec4_4}

\begin{figure*}
\begin{center}
\includegraphics[width=84mm]{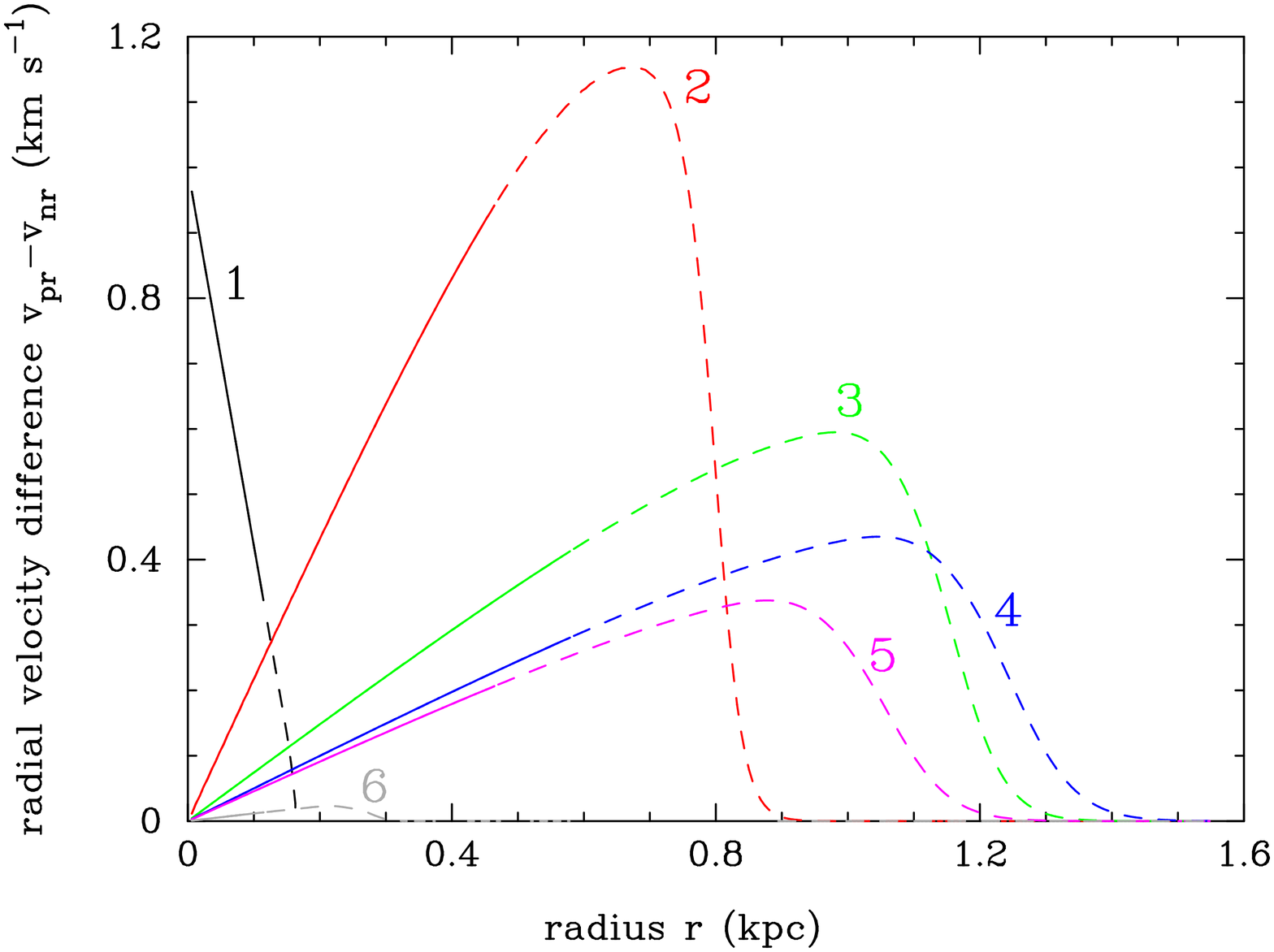}
\includegraphics[width=84mm]{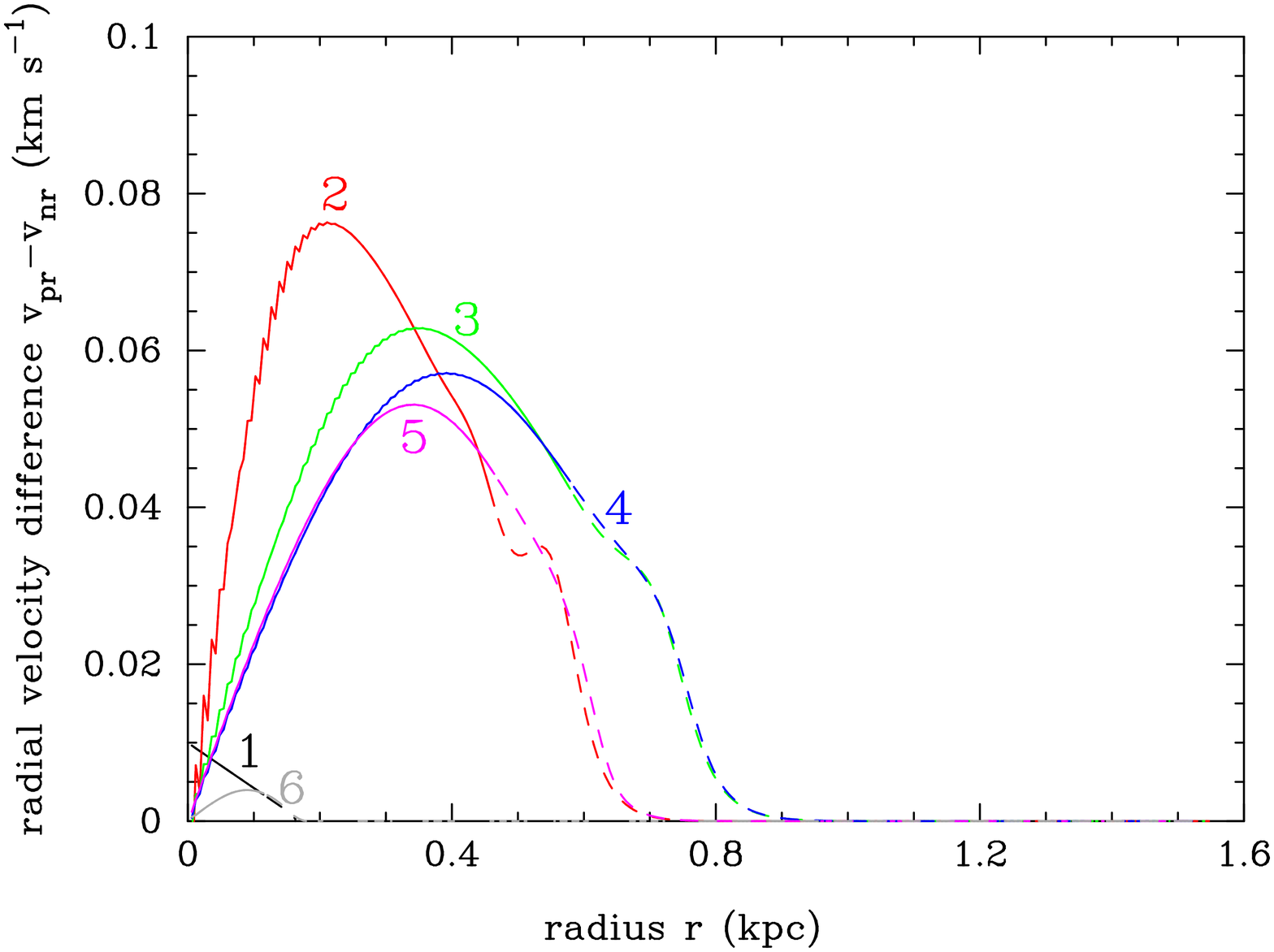}
\caption{Radial velocity difference between H$^+$ and H, $v_{pr}-v_{{\rm n}r}$, as a function of radius for Case 1 (left panel) and Case 2 (right panel) at $t$=9.29 Myr (1), 102 Myr (2), 195 Myr (3), 288 Myr (4), 381 Myr (5), and 474 Myr (6). Solid and dashed lines correspond to values inside and outside the structure, respectively. \label{pga7}}
\end{center}
\end{figure*}

Figure \ref{pga7} shows the radial velocity difference of H$^+$ and H, i.e., $v_{pr}-v_{{\rm n}r}$, as a function of radius for Case 1 (left panel) and Case 2 (right panel).  Solid and dashed lines correspond to values inside and outside the structure, respectively.  The velocity difference of $^7$Li$^+$ and H, i.e., $v_{7r}-v_{{\rm n}r}$, is the same as that of H$^+$ and H.  The conditions, $B_z \gg \alpha_{ie} \gg \alpha_{i{\rm n}}, B_r$ (for $i=p$ and $^7$Li$^+$) 
[cf. Eqs. (\ref{eq37}), (\ref{eq45}), (\ref{eq69}), (\ref{eq58}), and (\ref{eq54})]
and $v_{jr}, v_{jz} \gg v_{j\phi}$ (for any species $j$), are satisfied in the present case.  Under these conditions, the approximate relation $v_{7r}\sim v_{pr}$ is satisfied [see Eq. (\ref{eqb32})].  Strictly, ions such as the $^7$Li$^+$ ion have radial velocities very slightly different from that of proton under an electric field in which radial motions of protons and electrons are balanced \footnote{Inhomogeneous chemical abundances in cosmic plasma including solar flares have been considered \citep[][pp. 82--84]{alf1981}.  One of their mechanisms is the mass dependent gravitational drift resulting in isotope separation of single element.  In this study, we treat species-dependent dynamical frictions operating in low density plasma with a very small ionization degree of hydrogen, as realized in the early universe.  The frictions induce a chemical separation of various singly charged ionic species as one kind of separations.  Its effect is, however, negligible because of a strong electric coupling of positively charged ions and negatively charged electron.}.  The velocity difference is larger in Case 1 than in Case 2.  The time evolution is also different since the movement of charged species is larger and the $B_z$ value in small radii evolves more significantly for the larger magnetic field in Case 1.

One can intuitively understand that the radial velocity of the lithium ions is essentially the same as that of the protons as follows.  Under the conditions of $B_z \gg \alpha_{ie} \gg \alpha_{i{\rm n}}, B_r$, the Hall parameter of lithium ions is  much larger than 1.  The Hall parameter is the dimensionless ratio between the Larmor frequency and the collision rate given by
\begin{equation}
 \beta_i = \frac{\Omega_i} {\sum_j \tau_{ij}^{-1}},
\end{equation}
where
$j$ is the target particle at the collision, and $j=$n and $e$ for the case of $i$=$^7$Li.
Since $\beta_i \ga 1$ is satisfied in the present situation, the $^7$Li ions gyrate about a magnetic field line more than several times before being knocked off the line by colliding with a neutral particle.  The lithium ions, therefore, move with the magnetic field and the proton-electron plasma.

\begin{figure*}
\begin{center}
\includegraphics[width=84mm]{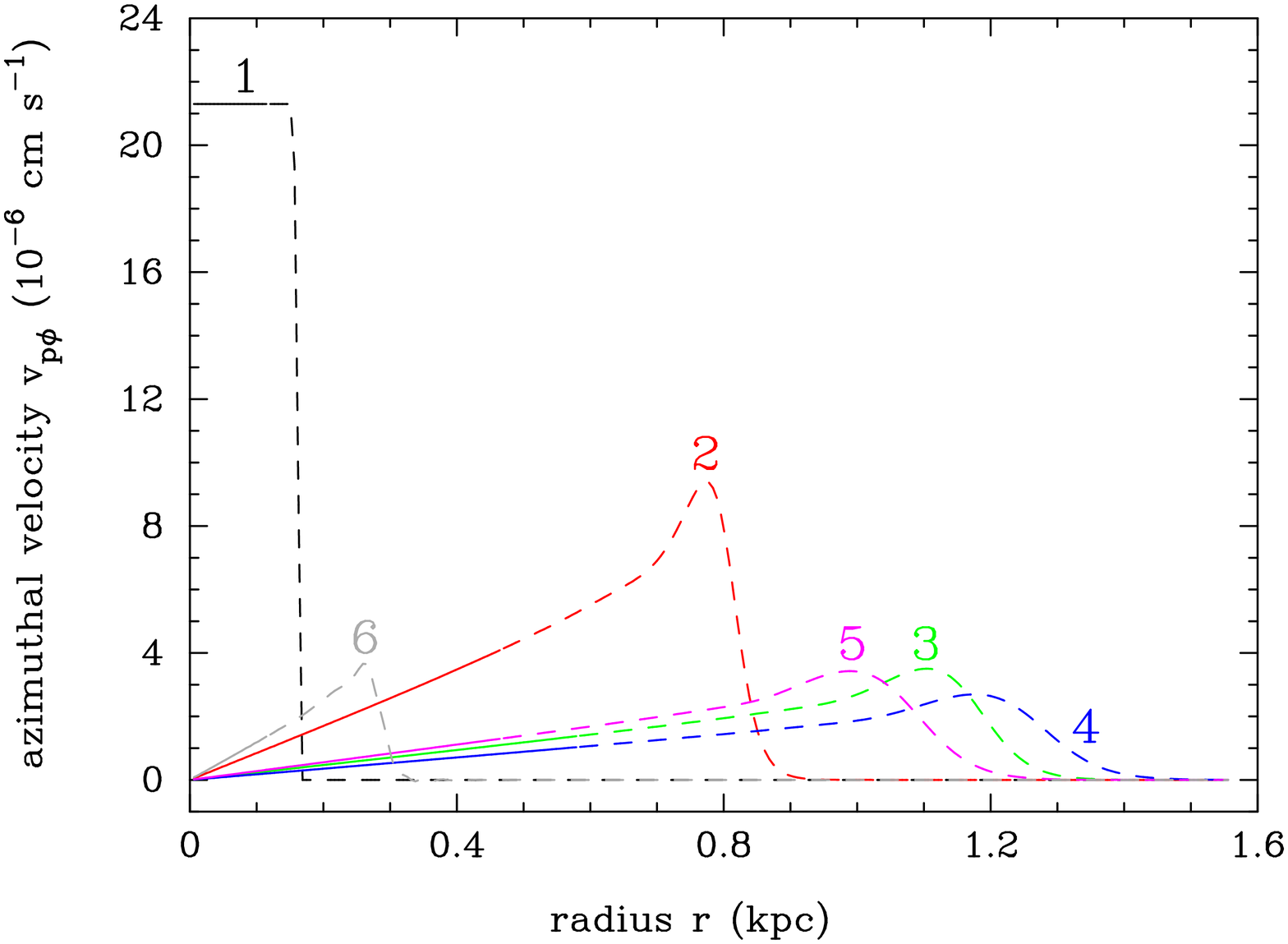}
\includegraphics[width=84mm]{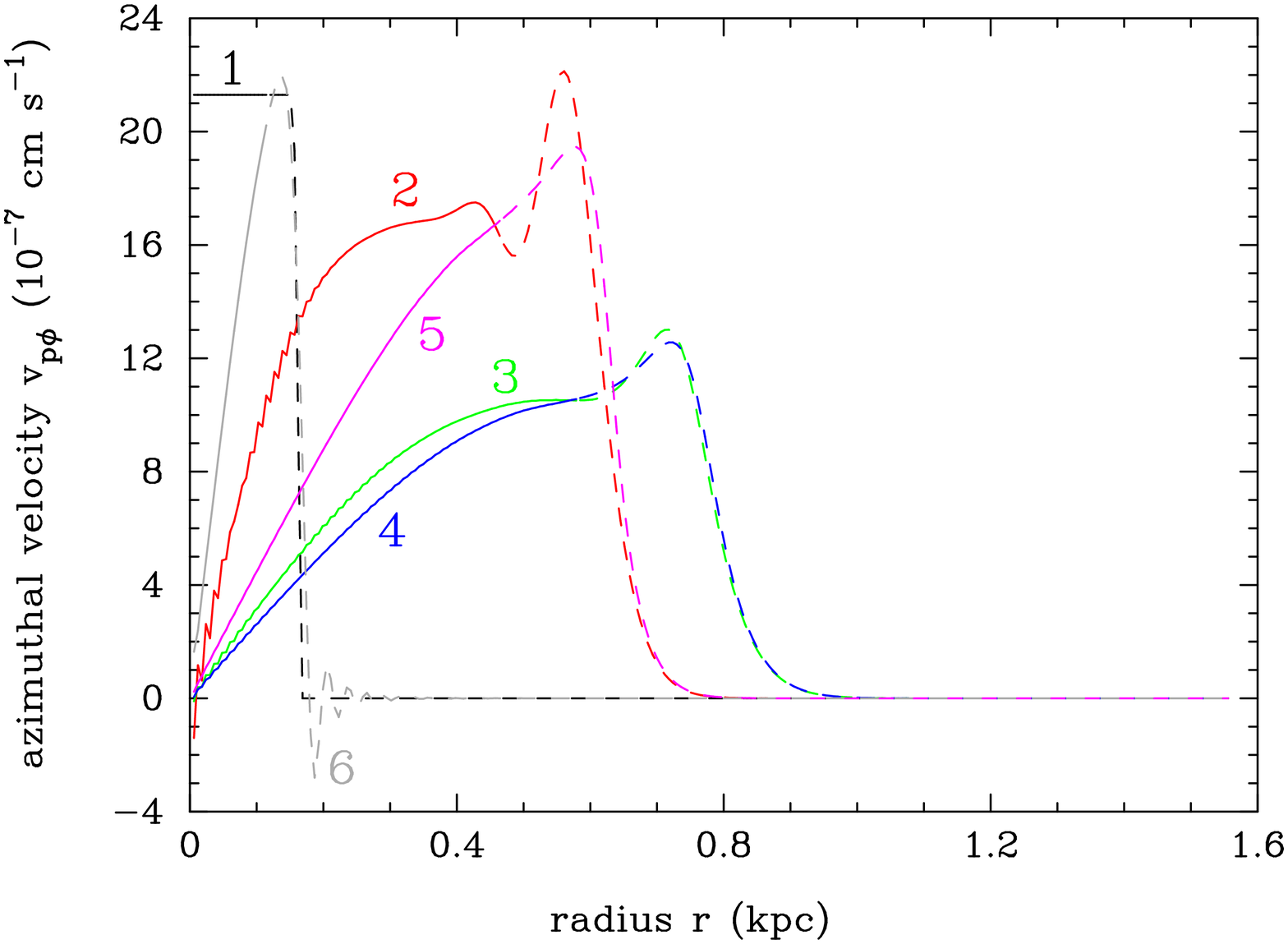}
\caption{Azimuthal velocity of proton $v_{p\phi}$ as a function of radius for Case 1 (left panel) and Case 2 (right panel) at $t$=9.29 Myr (1), 102 Myr (2), 195 Myr (3), 288 Myr (4), 381 Myr (5), and 474 Myr (6).  Solid and dashed lines correspond to values inside and outside the structure, respectively. \label{pga8}}
\end{center}
\end{figure*}

Figure \ref{pga8} shows the azimuthal velocity of proton $v_{p\phi}$ as a function of radius for Case 1 (left panel) and Case 2 (right panel).  Solid and dashed lines correspond to values inside and outside the structure, respectively.  This quantity scales as $( \nabla \times \vecb )_\phi$ [Eq. (\ref{eq68})].  Velocities at small radii are larger since a gradient of $B_z$ is assumed at small radii in the initial condition.  The velocity is roughly independent of time or hydrogen density if there is no weakening of magnetic field (cf. Fig. \ref{pga9}).  This is because $v_{p\phi}\propto (B_z/L_B)/n_p \propto 1/(L_B^3 n_p)$ is nearly constant.  The movement of charge species relative to neutral hydrogen, however,  gradually reduces the field gradient $\partial_r B_z$.  This effect reduces the velocity as a function of time.  Shapes of the curves are similar between Cases 1 and 2.  Amplitudes and the time evolutions of the velocities are, however, different for the same reason described for Fig. \ref{pga7}.

\begin{figure*}
\begin{center}
\includegraphics[width=84mm]{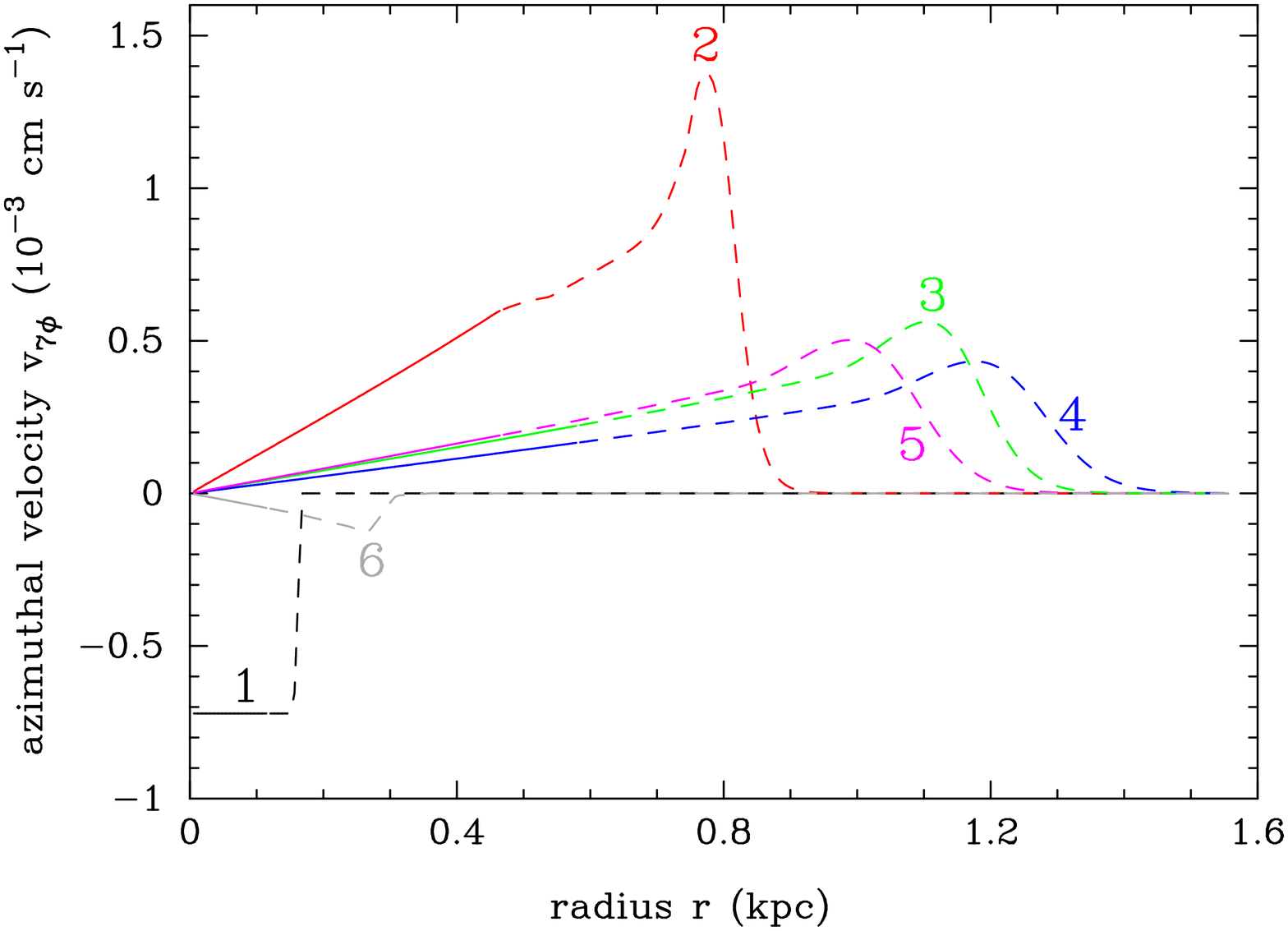}
\includegraphics[width=84mm]{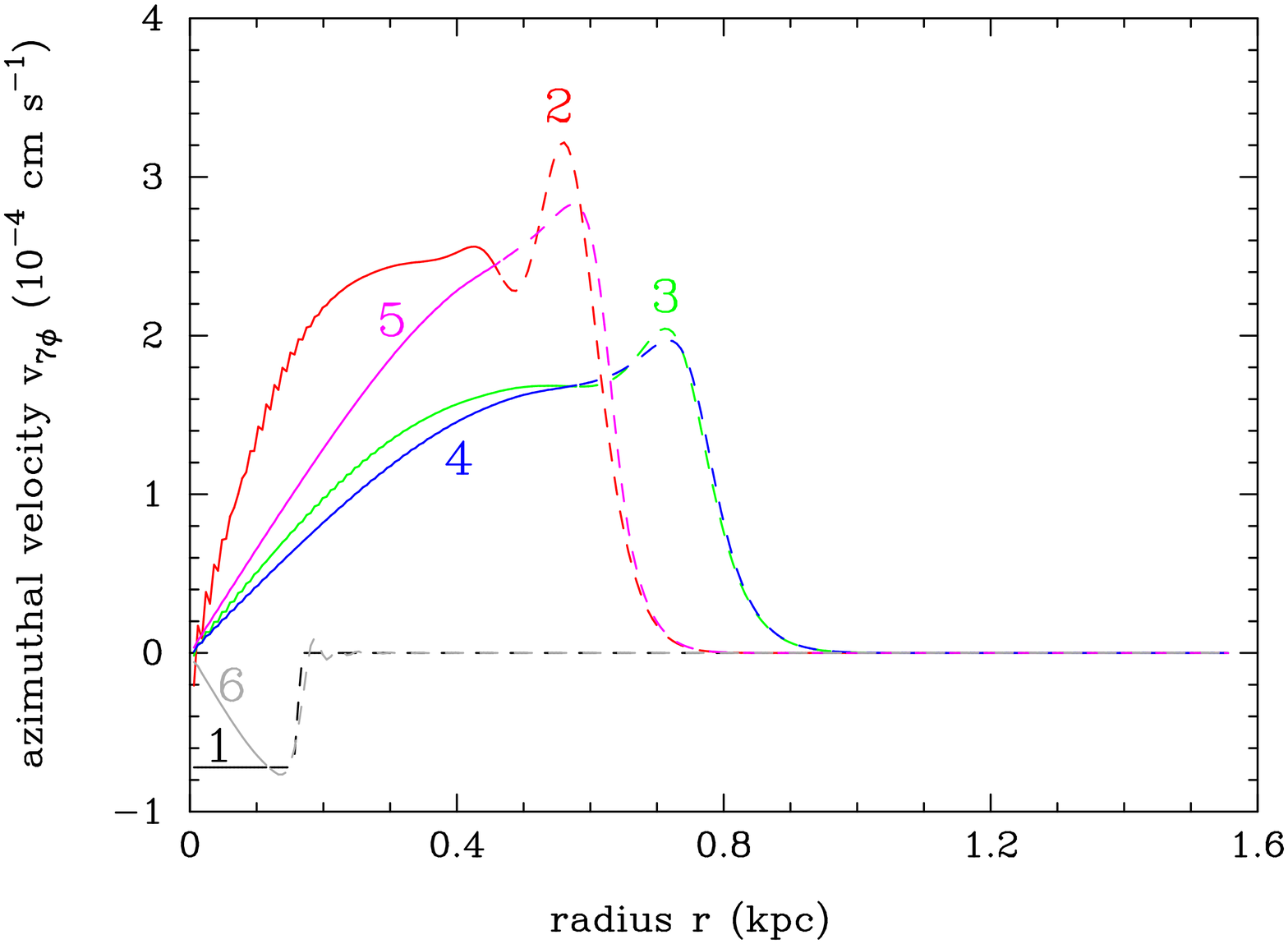}
\caption{Azimuthal velocity of $^7$Li$^+$ $v_{7\phi}$ as a function of radius for Case 1 (left panel) and Case 2 (right panel) at $t$=9.29 Myr (1), 102 Myr (2), 195 Myr (3), 288 Myr (4), 381 Myr (5), and 474 Myr (6).  Solid and dashed lines correspond to values inside and outside the structure, respectively. \label{pga14}}
\end{center}
\end{figure*}

Figure \ref{pga14} shows the azimuthal velocity of $^7$Li$^+$ $v_{7\phi}$ as a function of radius for Case 1 (left panel) and Case 2 (right panel).  Solid and dashed lines correspond to values inside and outside the structure, respectively.  This quantity reflects the force balance as described in Eq. (\ref{eqb29}).  Because of the conditions, $B_z \gg \alpha_{ie} \gg \alpha_{i{\rm n}}, B_r$ (for $i=p$ and $^7$Li$^+$) and $v_{jr}, v_{jz} \gg v_{j\phi}$ (for any species $j$), the azimuthal velocity has a limit value of $v_{7\phi}=[(\alpha_{p{\rm n}}+\alpha_{i{\rm n}})/B_z](v_{pr}-v_{{\rm n}r})$, which is different from both of $v_{p\phi}$ and $v_{e\phi}$.

\subsection{Electric field}\label{sec4_7}

\begin{figure*}
\begin{center}
\includegraphics[width=84mm]{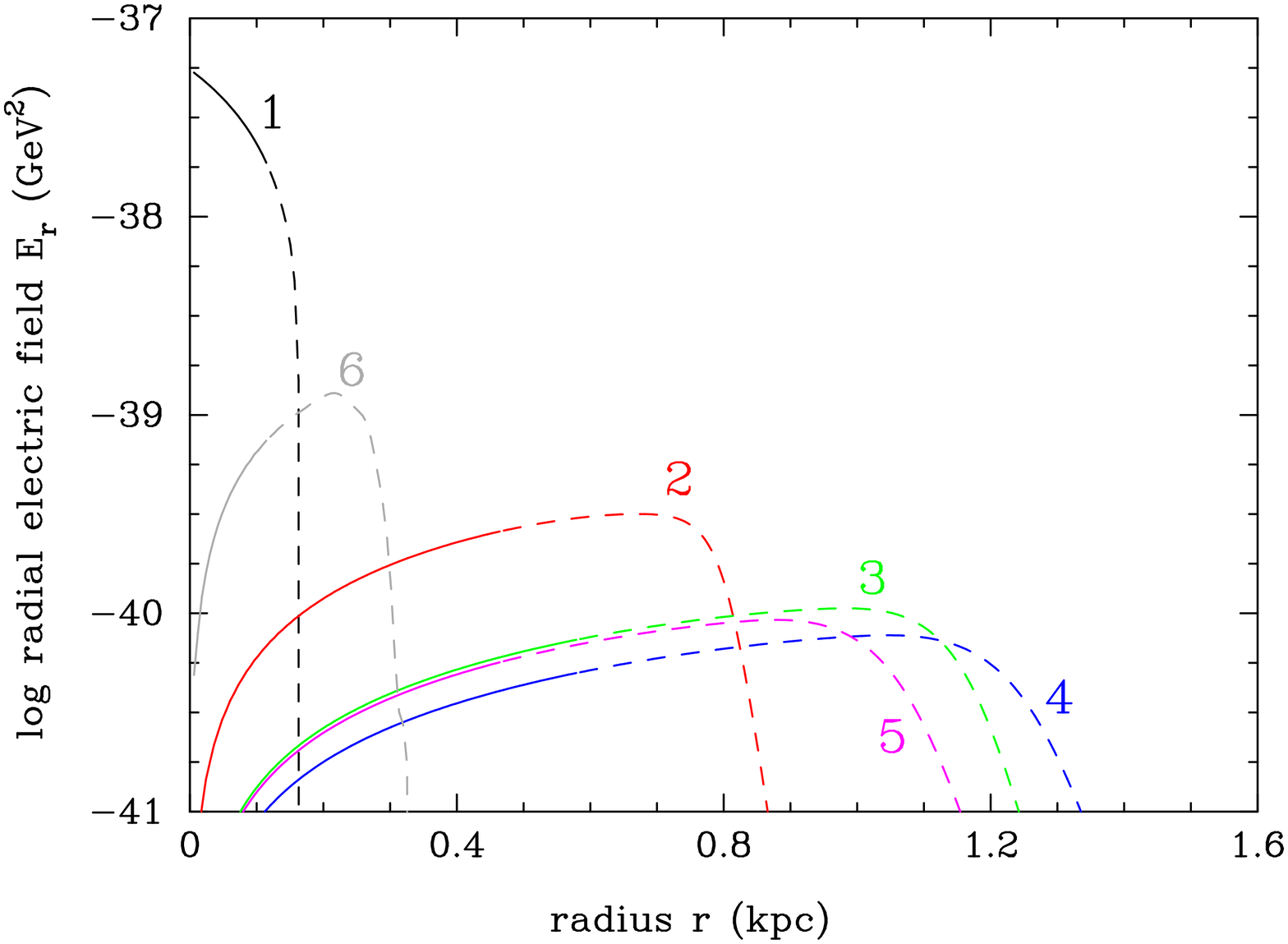}
\includegraphics[width=84mm]{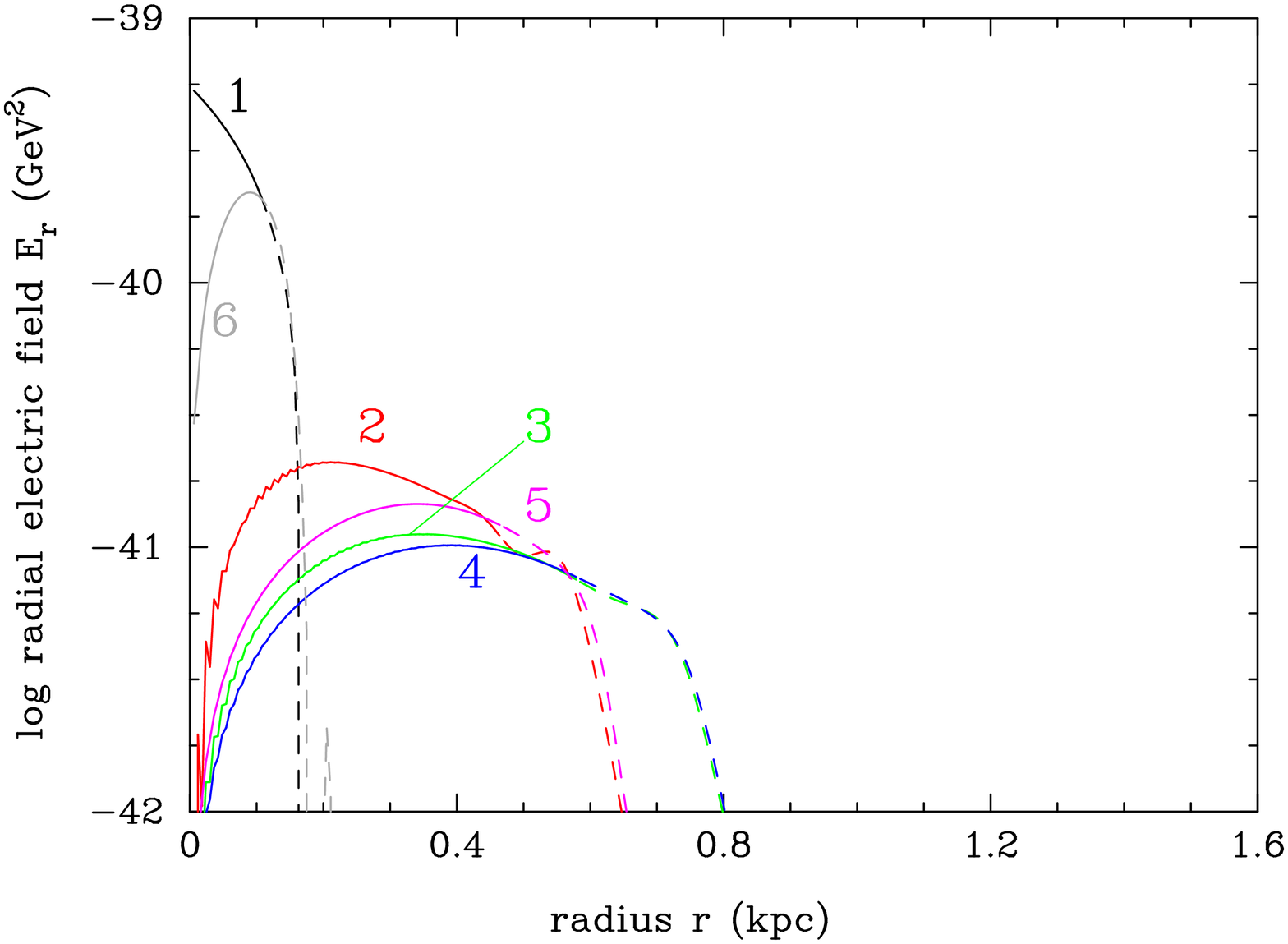}
\caption{Radial electric field as a function of radius for Case 1 (left panel) and Case 2 (right panel) at $t$=9.29 Myr (1), 102 Myr (2), 195 Myr (3), 288 Myr (4), 381 Myr (5), and 474 Myr (6).  Solid and dashed lines correspond to values inside and outside the structure, respectively. \label{pga10}}
\end{center}
\end{figure*}

Figure \ref{pga10} shows the radial component of electric field $E_r$ as a function of radius for Case 1 (left panel) and Case 2 (right panel).  Solid and dashed lines correspond to values inside and outside the structure, respectively.  The $E_r$ value is given by
\begin{eqnarray}
 E_r &=& \left( \alpha_{p{\rm n}} -\alpha_{e{\rm n}} \right) \left(v_{pr} -v_{{\rm n}r} \right) \nonumber\\
&=& \frac{1}{e} \left( \frac{m_p}{\tau_{p{\rm n}}} - \frac{m_e}{\tau_{e{\rm n}}}\right) \frac{\tau_{{\rm n}p}}{\rho_{\rm n}} \frac{\partial_z B_r - \partial_r B_z}{4\pi} B_z \nonumber\\
&\propto&\frac{-(\partial_r B_z) B_z}{\rho_{\rm n}},
\end{eqnarray}
where
Eqs. (\ref{eqb8}) and (\ref{eqb12}) were used in the first equality, and
Eqs. (\ref{eqa2}) and  (\ref{eqb17}) were used in the second equality.
It thus scales as $\propto (\partial_r B_z)B_z /\rho_{\rm n}$, and decreases during the expanding phase (curves 1--3), while it increases during the contracting phase (curves 4--6).

\begin{figure*}
\begin{center}
\includegraphics[width=84mm]{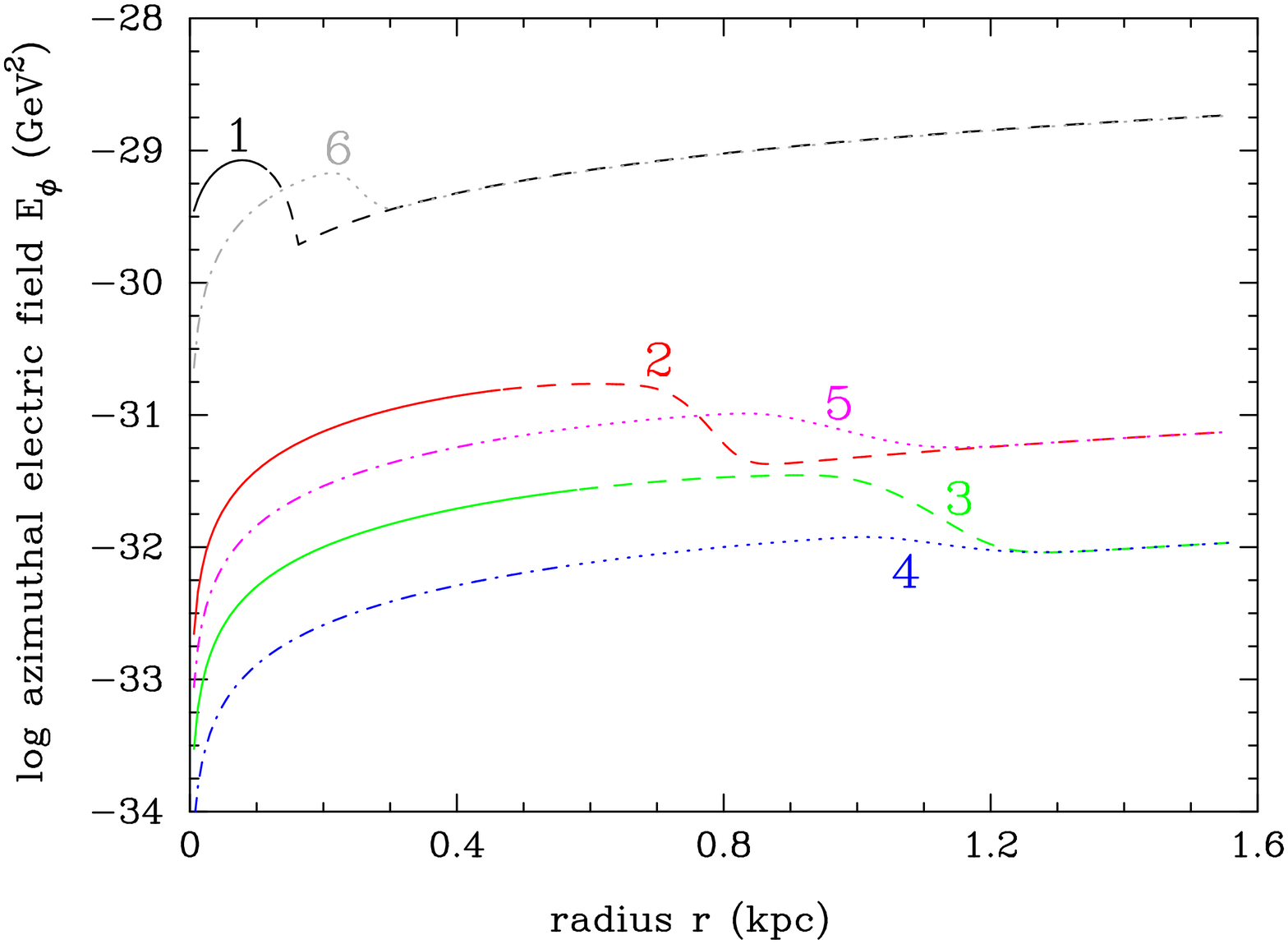}
\includegraphics[width=84mm]{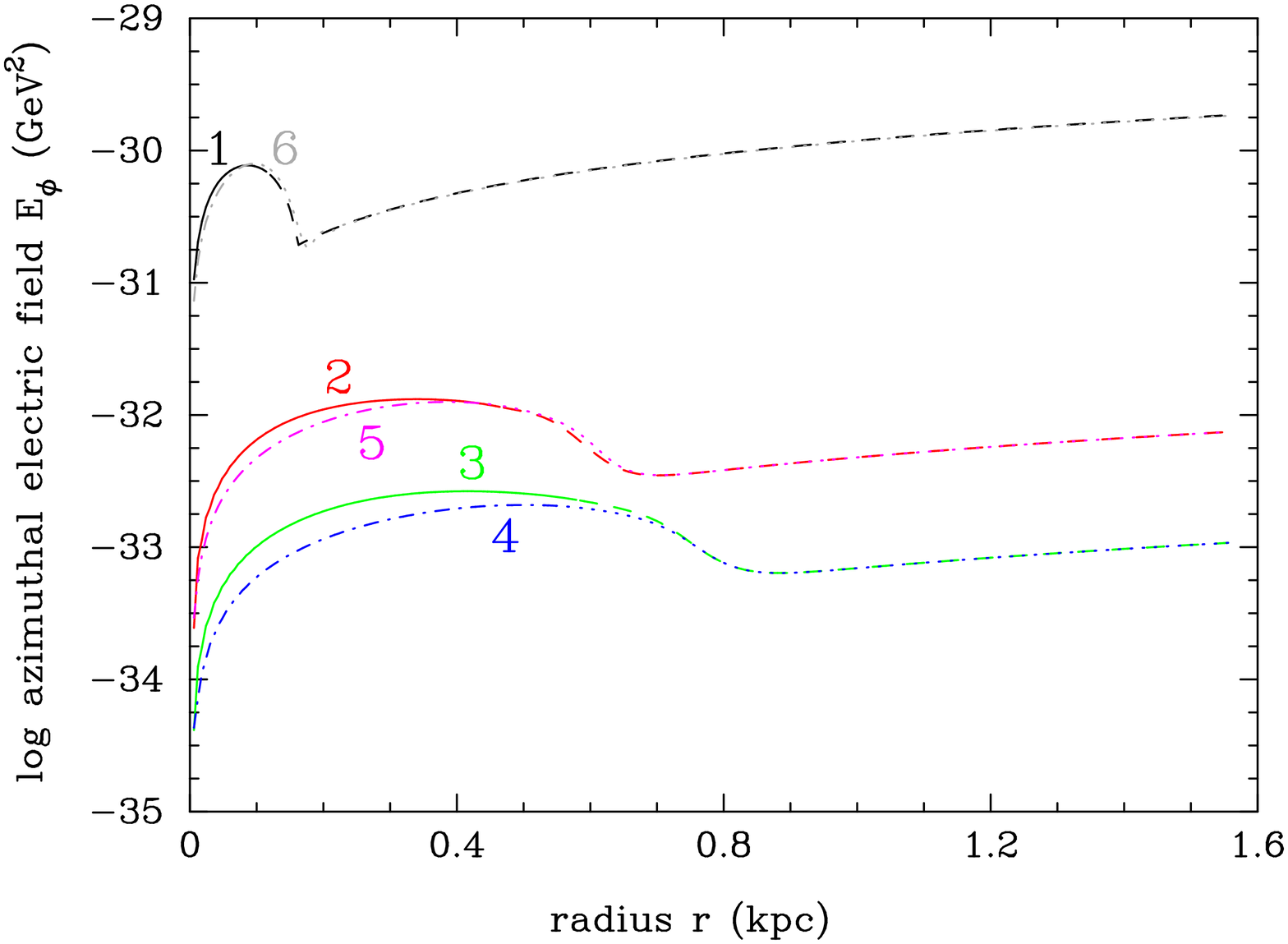}
\caption{Azimuthal electric field as a function of radius for Case 1 (left panel) and Case 2 (right panel) at $t$=9.29 Myr (1), 102 Myr (2), 195 Myr (3), 288 Myr (4), 381 Myr (5), and 474 Myr (6).  Solid and dashed lines correspond to positive values inside and outside the structure, respectively.  Dot-dashed and dotted lines correspond to negative values inside and outside the structure, respectively. \label{pga11}}
\end{center}
\end{figure*}

Figure \ref{pga11} shows the azimuthal component of electric field $E_\phi$ as a function of radius for Case 1 (left panel) and Case 2 (right panel).  Solid and dashed lines correspond to positive values inside and outside the structure, respectively.  Dot-dashed and dotted lines correspond to negative values inside and outside the structure, respectively.  In the case of strong magnetic field of $B_z \gg \alpha_{ab},~B_r$, the relation $E_\phi \propto v_{pr} B_z$ holds in this calculation [Eq. (\ref{eqb9})].  According to the relation, the $E_\phi$ value decreases with time.  The radial velocity is large in the early epoch (curve 1), which is approximately equal to the cosmic expansion velocity.  The velocity is negative and its amplitude is large in the late epoch (curve 6), which is given by free fall velocity of the structure.

\section*{Acknowledgments}
We are indebted to the referee, Glenn E. Ciolek, for instructive comments on astrophysical plasma physics.  We are grateful to Tomoaki Ishiyama for instructive information on structure formation.  This work has been supported by Grant-in-Aid for Scientific Research from the Ministry of Education, Science, Sports, and Culture (MEXT), Japan,
No.25400248 and No.21111006, and by World Premier International Research Center Initiative (WPI Initiative), MEXT, Japan.


\bsp

\label{lastpage}

\end{document}